\documentclass[modern]{aastex61}

\newcommand{\msun}{$M_{\odot}$}
\newcommand{\teff}{T_{eff}}
\usepackage{subfigure}
\usepackage{multirow}
\usepackage{array,graphicx}
\usepackage{booktabs}
\usepackage{pifont}

\newcommand*\rot{\rotatebox{90}}

\received{}
\revised{}
\accepted{\today}

\submitjournal{ApJS}

\shorttitle{M7$-$L5 Dwarf 25\,pc Sample}
\shortauthors{Bardalez Gagliuffi et al.}

\begin{document}

\title{The Ultracool SpeXtroscopic Survey. I. Volume-Limited Spectroscopic Sample and Luminosity Function of M7$-$L5 Ultracool Dwarfs.}

\correspondingauthor{Daniella C. Bardalez Gagliuffi}
\email{dbardalezgagliuffi@amnh.org}

\author[0000-0001-8170-7072]{Daniella C. Bardalez Gagliuffi}
\altaffiliation{AMNH Kalbfleisch Fellow.}
\affil{Department of Astrophysics, American Museum of Natural History, Central Park West at 79th St, New York, NY 10024, USA}
\affil{Center for Astrophysics and Space Sciences, University of California, San Diego, 9500 Gilman Dr., Mail Code 0424, La Jolla, CA 92093, USA}

\author{Adam J. Burgasser}
\affil{Center for Astrophysics and Space Sciences, University of California, San Diego, 9500 Gilman Dr., Mail Code 0424, La Jolla, CA 92093, USA}

\author{Sarah J. Schmidt}
\affiliation{Leibniz-Institute for Astrophysics Potsdam (AIP), An der Sternwarte 16, D-14482, Potsdam, Germany}

\author{Christopher Theissen}
\affil{Center for Astrophysics and Space Sciences, University of California, San Diego, 9500 Gilman Dr., Mail Code 0424, La Jolla, CA 92093, USA}

\author[0000-0002-2592-9612]{Jonathan Gagn\'e}
\affil{Institute for Research on Exoplanets, Universit\'e de Montr\'eal, D\'epartement de Physique, C.P.~6128 Succ. Centre-ville, Montr\'eal, QC H3C~3J7, Canada}

\author{Michael Gillon}
\affil{1 Space sciences, Technologies and Astrophysics Research (STAR) Institute, Universit\'{e} de Li\`{e}ge\\
All\`{e}e du 6 Ao\^{u}t 17, Bat. B5C, 4000 Li\`{e}ge, Belgium} 

\author{Johannes Sahlmann}
\affil{Space Telescope Science Institute (STScI), 3700 San Martin Drive, Baltimore, MD 21218, USA}

\author{Jacqueline K. Faherty}
\affil{Department of Astrophysics, American Museum of Natural History, Central Park West at 79th St, New York, NY 10024, USA}

\author{Christopher Gelino}
\affil{Infrared Processing and Analysis Center (IPAC), California Institute of Technology, 117 Morrisroe Astroscience Lab, Mail Code 100-22, Pasadena CA 91125, USA}
\affil{NASA Exoplanet Science Institute (NExSci), California Institute of Technology, Mail Code 100-22, 1200 East California Blvd., Pasadena, CA 91125, USA}

\author[0000-0002-1821-0650]{Kelle L. Cruz}
\affil{Department of Astrophysics, American Museum of Natural History, Central Park West at 79th St, New York, NY 10024, USA}
\affil{Hunter College, City University of New York, 695 Park Ave, New York, NY 10065, USA}
\affil{Department of Physics, Graduate Center, City University of New York, 365 5th Ave, New York, NY 10016, USA}
\affil{Center for Computational Astrophysics, Flatiron Institute, 162 5th Avenue, New York, NY 10010 USA}

\author{Nathalie Skrzypek}
\affil{Blackett Laboratory, Imperial College, Prince Consort Rd, Kensington, London SW7 2AZ, UK}

\author{Dagny Looper}
\affil{CBS Studios, 4024 Radford Ave, Studio City, CA 91604}

\begin{abstract}
We present a volume-limited, spectroscopically-verified sample of M7$-$L5 ultracool dwarfs within 25\,pc. The sample contains 410 sources, of which $93\%$ have trigonometric distance measurements ($80\%$ from \textit{Gaia} DR2), and $81\%$ have low-resolution ($R\sim120$), near-infrared (NIR) spectroscopy. We also present an additional list of 60 sources which may be M7$-$L5 dwarfs within 25\,pc when distance or spectral type uncertainties are taken into account. The spectra provide NIR spectral and gravity classifications, and we use these to identify young sources, red and blue $J-K_S$ color outliers, and spectral binaries. We measure very low gravity and intermediate gravity fractions of $2.1^{+0.9}_{-0.8}\%$ and $7.8^{+1.7}_{-1.5}\%$, respectively; fractions of red and blue color outliers of $1.4^{+0.6}_{-0.5}$\% and $3.6^{+1.0}_{-0.9}$\%, respectively; and a spectral binary fraction of $1.6^{+0.5}_{-0.5}\%$. We present an updated luminosity function for M7$-$L5 dwarfs continuous across the hydrogen burning limit that agrees with previous studies.  We estimate our completeness to range between $69-80\%$ when compared to an isotropic model. However, we find that the literature late-M sample is severely incomplete compared to L dwarfs, with completeness of $62^{+8}_{-7}\%$ and $83^{+10}_{-9}\%$, respectively. This incompleteness can be addressed with astrometric-based searches of ultracool dwarfs with \textit{Gaia} to identify objects previously missed by color- and magnitude-limited surveys. 
\end{abstract}


\section{Introduction} \label{sec:intro}

Ultracool dwarfs (UCDs) are the lowest-mass, coldest, and faintest products of star formation, encompassing objects with masses $M\lesssim0.1$\,\msun, effective temperatures $\leq2700$\,K, and spectral types M7 and later~\citep{1991ApJS...77..417K}. UCDs include both very low-mass (VLM) stars that slowly fuse hydrogen for up to a trillion years~\citep{1997ApJ...482..420L}; and brown dwarfs, which have insufficient mass to sustain hydrogen fusion in their cores ($M_{BD}\lesssim0.072\,M_{\odot}$ for solar metallicity;~\citealt{1963ApJ...137.1121K,1963PThPh..30..460H}).  Brown dwarfs never reach thermal equilibrium as they are supported by electron degeneracy pressure, and thus continue to cool and dim over time across spectral types M, L, T, and Y~(\citealt{1999ApJ...519..802K,2002PhDT........27B}, and~\citealt{2011ApJ...743...50C}, respectively). The absence of an internal energy generation mechanism results in a degeneracy between mass, age and luminosity (and its proxies, effective temperature, absolute magnitude, and spectral type). As a consequence, the characterization of isolated brown dwarfs is challenging, but the population can be evaluated statistically~(e.g.~\citealt{2004ApJS..155..191B,2005ApJ...625..385A, 2008ApJ...676.1281M, 2010MNRAS.406.1885B, 2010AandA...522A.112R, 2013MNRAS.430.1171D, 2019ApJS..240...19K}).

UCDs are yardsticks of Galactic chemical evolution, as their minimal core fusion mostly preserves their natal compositions. Their interiors are fully convective, allowing measurement of both composition and products of fusion (i.e. Li depletion) from their atmospheres. UCDs are ubiquitous, and include some of the closest neighbors to the Sun, such as the L/T transition and flux reversal binary Luhman 16AB~\citep{2013ApJ...767L...1L}, and the coldest known brown dwarf, the $\gtrsim$Y2 WISE J085510.83$-$071442.5~($\teff\sim250\,$K;~\citealt{2014ApJ...786L..18L}), both at a distance of 2\,pc. UCDs can host disks (e.g.,~\citealt{2014ApJ...791...20R,2016AandA...593A.111T}) and exoplanets (e.g., TRAPPIST-1,~\citealt{2016Natur.533..221G,2017Natur.542..456G}; OGLE-2012-BLG-0358Lb,~\citealt{2013ApJ...778...38H}); are found in binary and higher-order multiple systems (e.g.,~\citealt{2007prpl.conf..427B,2012ApJ...757..110B}), and in young clusters and associations~(e.g.,~\citealt{2015ApJ...808L..20G,2000Sci...290..103Z}); they are members of the Galactic halo (e.g.,~\citealt{2003ApJ...592.1186B,2014ApJ...783..122K,2017MNRAS.468..261Z}); and have a broad range of magnetic activity~(\citealt{2015AJ....149..158S,2000AJ....120.1085G}) including high levels of radio emission (e.g.,~\citealt{2018ApJS..237...25K,2006ApJ...648..629B}); among other distinct properties. Finally, while UCDs represent the low-mass tail of the stellar initial mass function (IMF; e.g.,~\citealt{2005ASSL..327...41C}), their formation mechanisms remain poorly understood, since the Jeans mass in typical molecular clouds favors the production of objects with masses $M\sim$0.5\,\msun~\citep{1902RSPTA.199....1J}. The dense regions that are necessary to produce UCDs are difficult to model~(e.g.,~\citealt{2012MNRAS.419.3115B}).


Large area surveys in optical, NIR and mid-infrared (MIR) bands have been crucial to the discovery and population characterization of UCDs. These include the Two-Micron All Sky Survey (2MASS;~\citealt{2006AJ....131.1163S}), the Sloan Digital Sky Survey (SDSS;~\citealt{2000AJ....120.1579Y}), the UKIRT Infrared Deep Sky Survey (UKIDSS;~\citealt{2007MNRAS.379.1599L}), the Deep Near Infrared Survey of the Southern Sky (DENIS;~\citealt{1994ExA.....3...73E}), the Canada France Brown Dwarf Survey (CFBDS;~\citealt{2008AandA...484..469D}), and the \textit{Wide-Field Infrared Survey Explorer} (\textit{WISE};~\citealt{2010AJ....140.1868W}).~\textit{Gaia}~\citep{2016AandA...595A...1G}, whose second data release~(DR2;~\citealt{2018AandA...616A...1G}) has delivered 5-parameter astrometric solutions for 1.3 billion sources, has further uncovered and characterized nearby UCDs~(\citealt{2018AandA...616A...1G,2018arXiv180908244R}).

A homogeneous and unbiased sample is key to understanding the essential mechanisms, physical processes, and environmental conditions favorable to UCD formation and evolution. The IMF is a consequence of formation, and ultracool IMF studies indicate there are fewer brown dwarfs than stars (e.g.,~\citealt{2000ApJ...540.1016L,2005ASSL..327...41C}). The incidence of rare subpopulations such as color outliers, young, and metal-poor sources, and binary and higher order systems, all probe formation and evolution mechanisms. The Solar neighborhood is the ideal region to measure these statistics. Bearing in mind the location and motion of the Sun with respect to the Galactic center, and the distinct kinematics and metallicity distributions of the thin disk, thick disk and halo populations~\citep{1983MNRAS.202.1025G}, the local volume can be treated as broadly representative of the Milky Way. Since brown dwarfs are intrinsically faint ($M_K\gtrsim10\,$mag;~\citealt{2013AJ....145....2F}), collecting data on the nearest sources is particularly advantageous to building a well-characterized sample. Spectroscopy, broad-band spectral energy distributions, kinematics, multiplicity, magnetic activity, and excesses and variability attributable to weather, magnetic activity, and presence of disks are best investigated with the nearest stars and brown dwarfs.


Previous studies of the nearby UCD population have already revealed some of the statistical properties of these low-mass objects.~\citet{2003AJ....125..354R} compiled the northern sample of systems within 8\,pc of the Sun in $V$-band magnitude, including 142 main sequence stars, 3 brown dwarfs, and 8 white dwarfs, and estimated $\sim15\%$ incompleteness.~\citet{2003AJ....126.2421C} compiled a volume-limited sample of 186 M7$-$L6 dwarfs within 20\,pc using a NIR photometric color and magnitude selection in 2MASS. Subsequently,~\citet{2007AJ....133..439C} built the first UCD NIR luminosity function, finding number densities of $n=4.9\times10^{-3}$\,pc$^{-3}$ for M7$-$M9.5 and a lower limit of $n\geq3.8\times10^{-3}$\,pc$^{-3}$ for L dwarfs\footnote{This study also converted an earlier luminosity function of the 8\,pc sample in $V$-band from~\citet{2003AJ....125..354R} into $J$-band magnitudes.}.  Using the sixth data release of SDSS,~\citet{2010AJ....139.2679B} compiled luminosity and mass functions of field low-mass stars spanning the M dwarf spectral class. Other studies have focused on the coldest brown dwarfs, to eventually obtain the low-mass end of the substellar mass function.~\citet{2008ApJ...676.1281M} measured a T dwarf number density of $n = (7.0^{+3.2}_{-3.0})\times10^{-3}$\,pc$^{-3}$ based on the detection of 15 T dwarfs in 2099\,$deg^2$ sampled by 2MASS and SDSS.~\citet{2010AandA...522A.112R} measured a late-L dwarf density of $n = (2.0^{+0.8}_{-0.7})\times10^{-3}$\,pc$^{-3}$, and T dwarf densities of $n = (1.4^{+0.3}_{-0.2})\times10^{-3}$\,pc$^{-3}$ for T0.5-T5.5 dwarfs and $n = (5.3^{+3.1}_{-2.2})\times10^{-3}$\,pc$^{-3}$ for T6-T8 dwarfs in CFBDS. Recently,~\citet{2019ApJS..240...19K} used a 20\,pc volume limited sample of sources T6 and later and estimated a number density of $0.97\times10^{-3}\, pc^{-3}$ for objects with temperatures 900-1050\,K or roughly T6 dwarfs, increasing to $3.26\times10^{-3}\,pc^{-3}$ for objects with temperatures in the 300-450\,K range, roughly corresponding to Y dwarfs.

Despite these concerted efforts, source identification and follow-up has been inhomogeneous for the local 25\,pc sample, as evidenced by ongoing nearby discoveries. The M7 dwarf 2MASS~J154043.42$-$510135.7 at 5\,pc~\citep{2014AandA...567A...6P}, the M9.5+T5 binary system WISE~J072003.20$-$084651.2 (\citealt{2014AandA...561A.113S,2015AJ....149..104B}), the L/T transition binary WISE~J104915.57$-$531906.1~\citep{2013ApJ...767L...1L}, and the 250\,K WISE~J085510.83$-$071442.5~\citep{2014ApJ...786L..18L}, all at distances of 6\,pc or less, show that the nearby sample remains incomplete. Given the availability of abundant multi-epoch survey data and astrometry from \textit{Gaia}, it is time to revisit the compilation of UCDs in the local volume.


 

In this paper we present a new volume-limited sample of M7$-$L5 ultracool dwarfs within 25\,pc, accompanied by NIR spectra homogeneously acquired with the SpeX spectrograph~\citep{2003PASP..115..362R} on the NASA Infrared Telescope Facility (IRTF). We follow a similar analysis to those of~\citet{2003AJ....126.2421C} and~\citet{2008AJ....136.1290R} by creating an unbiased, homogeneous, NIR spectroscopic sample of M7$-$L5 dwarfs selected from multiple sources in the literature. Section~\ref{sec:sampleconstruction} describes the sample selection and construction of our 25\,pc and $+1\sigma$ samples. Section~\ref{sec:specsamp} describes the construction of the spectral sample, which is analyzed in Section~\ref{sec:characterization}, for spectral and gravity classifications, color outliers, low gravity sources, spectral binaries, and resolved binaries and higher order multiples previously identified in the literature. In Section~\ref{sec:selfunc}, we estimate our biases, the completeness of the observed sample, and compute its selection function through a population simulation. We present an updated infrared luminosity function of ultracool dwarfs and compare it to previous work. Conclusions are summarized in Section~\ref{sec:conclusions}.

\section{Literature Sample Construction}\label{sec:sampleconstruction}

\subsection{Compilation of UCD Targets from the Literature}\label{sec:litsources}

Targets for the sample were drawn from a number of literature sources, including surveys and previous compilations, each designed for its own scientific purposes and with a variety of follow-up. We attempt to average over the various biases from the original surveys by compiling as many sources as possible. Some of the known biases include a red $J-K_S$ color bias~(e.g.,~\citealt{2003AJ....126.2421C,2013AJ....145..102L}, identified by~\citealt{2015AJ....149..158S}); incomplete compilations~(e.g.,~\citealt{2015ApJ...798...73G}) or partial sky coverage, e.g. Sloan Digital Sky Survey (SDSS;~\citealt{2012ApJS..203...21A,2015ApJS..219...12A}), Deep Near-Infrared Southern Sky Survey (DENIS;~\citealt{1994ExA.....3...73E}), UKIRT Infrared Deep Sky Survey (UKIDSS;~\citealt{2007MNRAS.379.1599L}); and targeted surveys~(e.g., young objects,~\citealt{2009ApJ...699..649S};  wide binaries,~\citealt{2014ApJ...792..119D}; high proper motion surveys, i.e.~SUPERBLINK,~\citealt{2011AJ....142..138L}). We believe biases due to proper motion selection are negligible due to the completeness of the photometric selection surveys. While proper motion surveys tend to be more incomplete, they also are less likely to scatter distant objects into the sample. Table~\ref{tab:litsources} lists the literature sources used to consolidate a database of $\sim16,000$ candidate nearby UCDs. Table~\ref{tab:cuts} summarizes the sequence of cuts leading to our final samples.

Duplicate sources were removed with TOPCAT~\citep{2005ASPC..347...29T} through an internal match that organized sources in near-neighbor groups with a matching radius of $15''$, large enough to catch binary components before deletion. This step reduced the number of entries to $\sim12,700$. We applied a spectral type cut requiring optical or NIR spectral types or photometric spectral type estimates (e.g.,~\citealt{2015AandA...574A..78S,2017AJ....153...92T}) to be in the M7$-$L5 range, shrinking the database to $\sim6,200$ sources. A rough distance cut eliminating objects farther than 50\,pc, trimmed this list to 1,664 sources.

Galaxies, giants, T-Tauri stars and other non-UCD sources as reported in the literature were identified using SIMBAD and removed, reducing the sample to 1,571 sources. After compiling photometric and astrometric data and recalculating spectrophotometric distances (see below), another distance cut at 30\,pc was applied for those sources with astrometric parallaxes, yielding 833 sources. 

\begin{deluxetable*}{ll}
\tabletypesize{\scriptsize}
\tablewidth{0pt}
\tablenum{2}
\tablecolumns{2}
\tablecaption{Cuts leading to the final sample\label{tab:cuts}}
\tablehead{
\colhead{Cut} & 
\colhead{Targets remaining}}
\startdata
Initial compilation & 16,322\\
Deletion of duplicates & 12,711\\
Optical, NIR or ``photometric'' spectral type between M7$-$L5 & 6,226\\
Estimated distance $\leq50$\,pc & 1,664\\
\emph{Compilation of photometry, recalculation of spectrophotometric distances} & \\
Deletion of non-stellar sources, giants, compact and young stellar objects & 1,571\\
Estimated Distance $\leq30$\,pc & 833\\
\hline
\emph{Compilation of \emph{Gaia} astrometry, recalculation of trigonometric distances} & \\
Objects with literature optical, NIR, or SpeX spectral type within M7$-$L5 (including photo-types only) & 595\\
Objects with trigonometric or NIR spectrophotometric distance $\leq25$\,pc & $435^{+21}_{-20}$\tablenotemark{a}\\
Objects with trigonometric or NIR spectrophotometric distance $\leq25$\,pc$+1\sigma$ & $470^{+22}_{-21}$\tablenotemark{a}\\
\hline
\emph{Final samples} & \\
$25\,pc$ sample of M7$-$L5 dwarfs  & $410^{+21}_{-20}$\tablenotemark{a}\\
$25\,pc$ plus $1\sigma$ sample of M7$-$L5 dwarfs & $470^{+22}_{-21}$\tablenotemark{a}\\
\enddata
\tablenotetext{a}{Uncertainties based on Poisson statistics.}
\end{deluxetable*}



\subsection{Photometric and Astrometric Data}\label{sec:photdata}

Photometry from the 2MASS~\citep{2006AJ....131.1163S}, SDSS DR9~\citep{2012ApJS..203...21A}, \textit{AllWISE}~(\citealt{2010AJ....140.1868W,2011ApJ...743..156M}), UKIDSS-LAS~\citep{2007MNRAS.379.1599L}, and \textit{Gaia} DR2~\citep{2018AandA...616A...1G} catalogs were collected for all sources, selecting the closest match up to 15$''$ through the VizieR interface to account for objects with large proper motions, using a custom routine\footnote{Available at \url{https://github.com/daniellabardalezgagliuffi/M7L5_download_phot}} built with the \texttt{Astroquery} Python package~\citep{astroquery}. We obtained coordinates, epochs, identifiers, and $GrizJHK_sKW1W2W3$ magnitudes from~\textit{Gaia}, SDSS, 2MASS, UKIDSS, and  \textit{AllWISE}. Spectral types from SDSS spectroscopy were obtained when available. In addition to these surveys, we also obtained $rizJHK_s$ magnitudes and uncertainties, spectral type, object type, and proper motions from SIMBAD with the same search radius.
 
Table~\ref{tab:photometry} provides the photometry data for the sample. All sources in our final $25\,pc$ sample (See Table~\ref{tab:cuts}) have NIR magnitudes\footnote{Except for GJ~1116B, where only unresolved photometry for the system was available~\citep{2014AJ....147...20N}. Several companions and close binaries do not have magnitudes in all three 2MASS bands (e.g., Gl~779~B, LSPM~J1314+1320AB, LHS~1901AB).}, 88\% have MIR magnitudes from AllWISE, and 39\% have optical magnitudes from SDSS. Resolved NIR photometry on the Mauna Kea Observatory (MKO) filter system~\citep{2002PASP..114..180T} was obtained from the literature (e.g.~\citealt{2012ApJS..201...19D,2018ApJS..234....1B}) and selected compilations\footnote{\url{https://jgagneastro.wordpress.com/list-of-ultracool-dwarfs/}}, particularly for closely-separated components of binary systems. We adopted 2MASS $JHK_s$ magnitudes as the standard, and use MKO $JHK$ magnitudes if those were the only NIR ones available. 

\textit{AllWISE} includes a crossmatch with the 2MASS catalog that we used to check for mismatches. We compared the $JHK_s$ magnitudes from the 2MASS and \textit{AllWISE} catalogs and kept the 2MASS magnitudes when the difference was within 0.05\,mag (typical magnitude uncertainty for 2MASS $JHK_s$). Objects whose magnitude differences were $>0.05\,$mag were flagged for visual examination in multi-wavelength finder charts, and comparison of SIMBAD and VizieR data sets. The mismatches between \textit{AllWISE} and 2MASS $JHK_s$ magnitudes were typically caused by the blending of a bright and faint source ($\Delta m \sim3$\,mag) in the larger \textit{AllWISE} pixels. In these cases, we assigned the 2MASS $JHK_s$ magnitudes to the source, and replaced the \textit{AllWISE} $W1W2W3$ magnitudes with null entries. The same procedure was followed to consolidate $JHK$ magnitudes from UKIDSS, and literature sources. While UKIDSS uses MKO filters, we keep these measurements separate because the quantum efficiency of the various NIR detectors may differ. 

\begin{figure*}[!t]
\figurenum{1}
\centering
\includegraphics[width=6in]{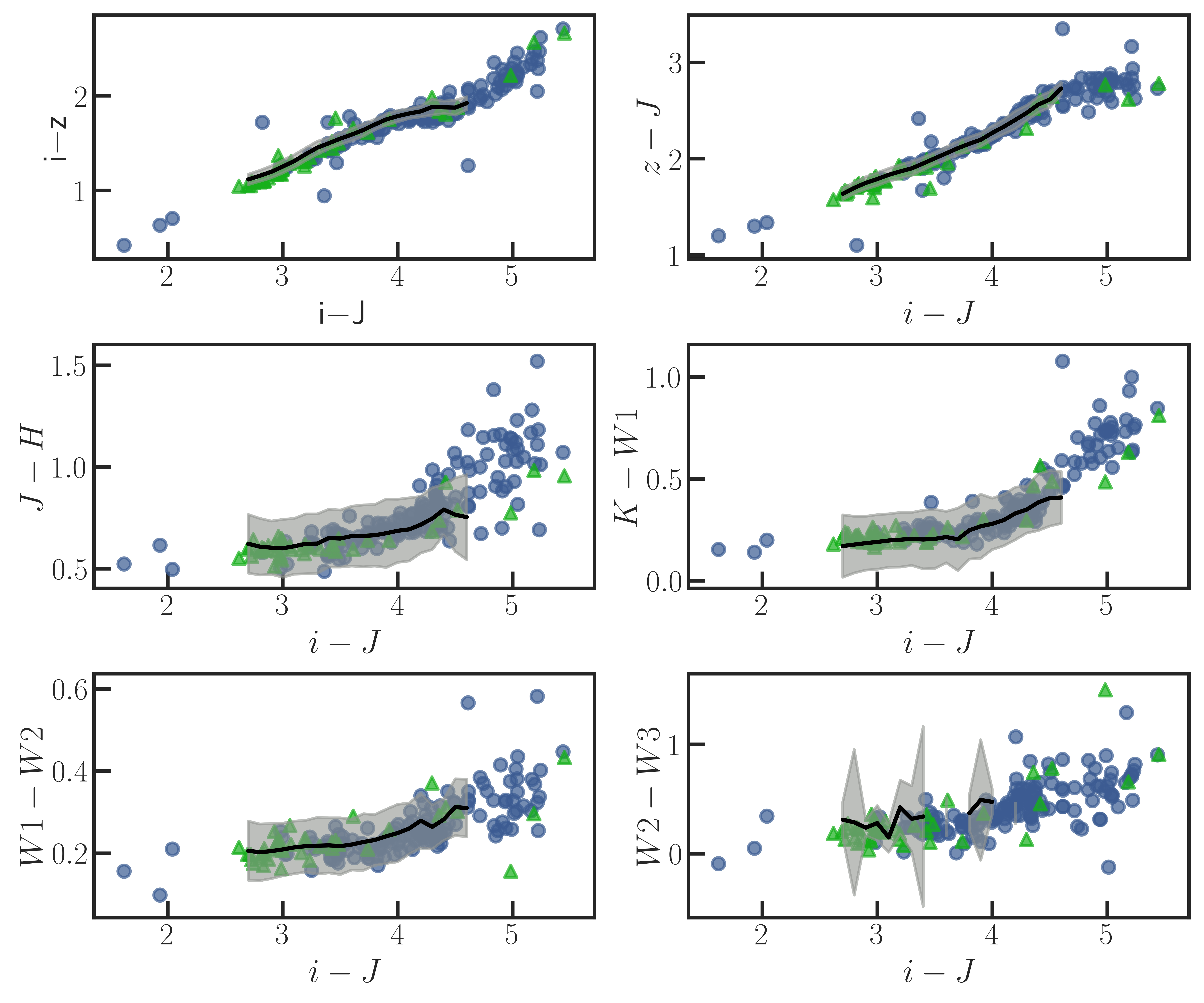}
\caption{Color locus of the known M7$-$L5 25\,pc sample in SDSS, 2MASS, and \textit{WISE} colors as a function of $i-J$~\citep{2015AJ....149..158S}. Blue circles are members of the $25\,pc$ sample, green triangles are members of the extended $1\sigma$ sample. The black line represents mean colors from~\citet{2015AJ....149..158S} (complete between M7$-$L2), with the extent of their uncertainties shaded in light gray. 
\label{fig:sch15plots}}
\end{figure*}

Further inspection on mismatched photometry between SDSS, 2MASS and AllWISE was done with color-color diagrams, as shown in Figure~\ref{fig:sch15plots}, and corrected by visual inspection using finder charts. Figure~\ref{fig:sch15plots} illustrates the color loci of M7$-$L5 dwarfs from~\citet{2015AJ....149..158S}. The most discriminating colors (e.g., $z-J$) use filters across surveys. Mismatches were corrected in a similar way as described above, using multi-wavelength finder charts and comparing magnitudes.

Astrometric data (positions, proper motions, and parallaxes) and radial velocities were drawn from SIMBAD when available. The sample was also crossmatched against the astrometric samples of~\citet{2012ApJS..201...19D} and~\citet{2016AJ....152...24W}. Upon the release of \textit{Gaia} DR2~\citep{2018AandA...616A...1G}, we crossmatched our preliminary sample against this dataset to obtain 5-parameter astrometric solutions. We used the following Astronomical Data Query Language (ADQL) query through the \texttt{astroquery.Gaia} package.

\begin{flushleft}
\texttt{SELECT g.*, t.*}\\
\texttt{FROM gaiadr1.tmass\_original\_valid AS t}\\
\texttt{LEFT OUTER JOIN gaiadr2.tmass\_neighbourhood AS xt ON xt.tmass\_oid = t.tmass\_oid}\\
\texttt{LEFT OUTER JOIN gaiadr2.gaia\_source AS g ON xt.source\_id = g.source\_id}\\
\texttt{where 1=CONTAINS(POINT(`ICRS', t.ra, t.dec),CIRCLE(`ICRS', \{\}, \{\}, 5./3600))}\\
\end{flushleft}

The \textit{Gaia} crossmatch was done in two steps. First, we crossmatched the sample with the 2MASS-\textit{Gaia} DR2 crossmatch table (\texttt{gaiadr2.tmass\_neighbourhood}) within a radius of $5\farcs0$ using 2MASS coordinates from our sample. Second, we joined this crossmatch with the \textit{Gaia} DR2 source table. We obtained 843 matches in 2MASS (10 objects with 2 matches each), 715 matched \textit{Gaia} DR2 with a $G$ magnitude, and 705 with parallaxes. To check the validity of our matches, we examined a color magnitude diagram of $G-RP$ versus absolute $G$ magnitude. We considered sources as outliers if $G-RP\leq1.25$, and if $M_G \leq 5$ to avoid giant stars. The 36 sources that failed our color/magnitude constraints were examined for crossmatch accuracy, and we found 22 mismatches of true UCDs with erroneous \textit{Gaia} data. The remaining 14 sources were dropped from the sample due to their small \textit{Gaia} parallaxes ($\vec{\omega} \ll 10~mas$), resulting in 825 sources.

\begin{figure*}
\figurenum{2}
\centering
\includegraphics[scale=0.7]{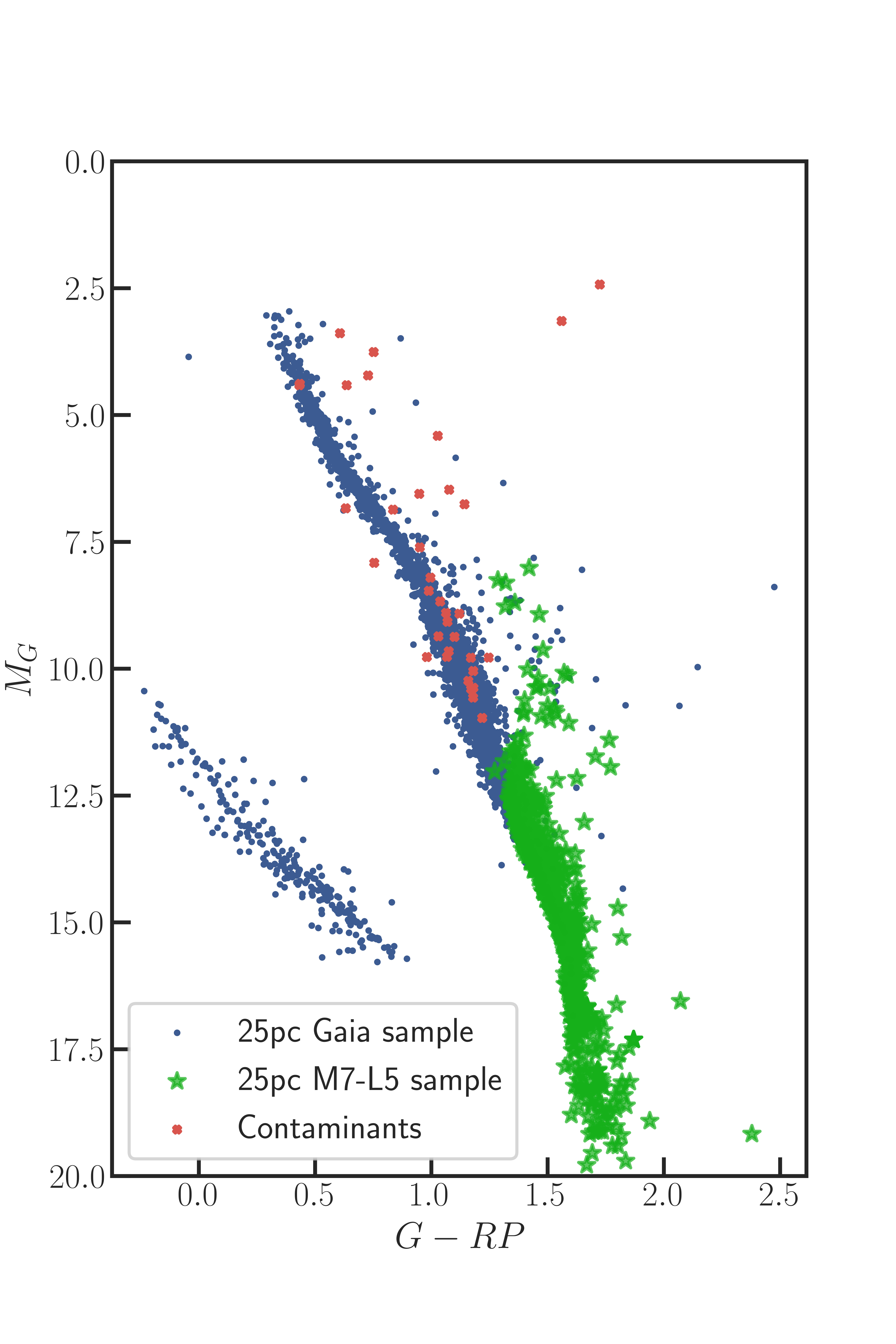}
\caption{\emph{Gaia} Hertzprung-Russell Diagram of the 25\,pc sample of M7$-$L5 dwarfs superimposed on the full 25\,pc sample from \emph{Gaia}. ~\emph{Gaia} sources are shown as blue points, and sources from the M7$-$L5 dwarf 25\,pc sample with valid \emph{Gaia} matches are shown as green stars. Sources in orange correspond to \emph{Gaia} mismatches. \label{fig:HRDgaia_xmatch}}
\end{figure*}


\subsubsection{Spectral Types}\label{sec:spts}
Most catalogs provide information on optical or NIR spectral classification, or classification estimates from photometry (\citealt{2015AandA...574A..78S,2017AJ....153...92T}). Given variations in classification schemes and intrinsic differences between optical and NIR classification (particularly for L dwarfs), we required at least one optical, NIR or photometric type belonging to the M7$-$L5 range for sources to be included in the sample. Adopted literature spectral types were chosen by prioritizing optical, NIR and photo-types, in that order. In the final $25\,pc$ sample, the adopted spectral type is optical for 334 objects, NIR for 73, and photometric for 4. The objects whose adopted literature type is photometric have SpeX observations~(see Section~\ref{sec:specsamp}) confirming their status as M7$-$L5 dwarfs. Figure~\ref{fig:spthist} shows the distribution of adopted literature spectral types color-coded by the nature of their measurement. 

\begin{figure*}
\figurenum{3}
\centering
\includegraphics[width=\textwidth]{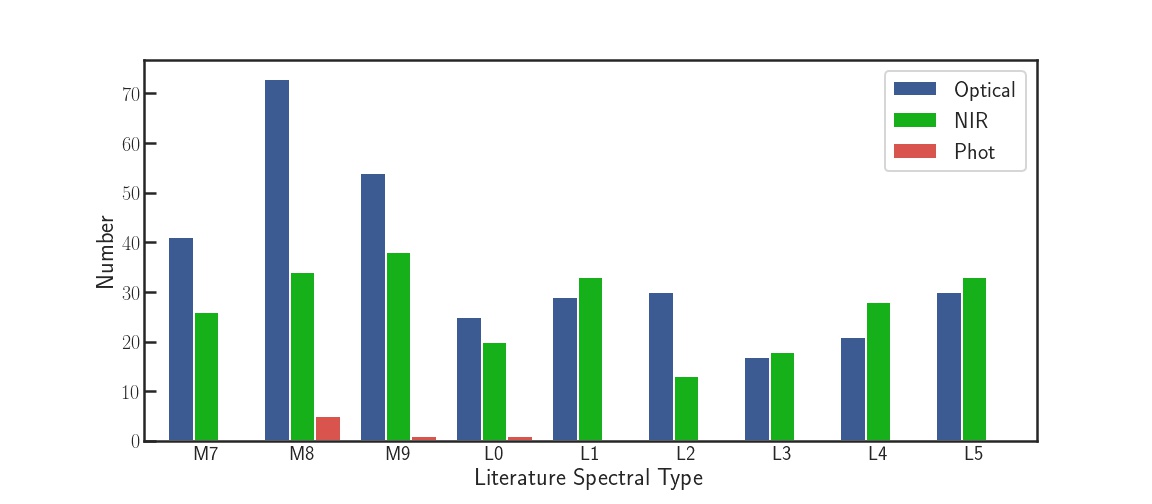}
\caption{Adopted literature spectral type for the M7$-$L5 25\,pc sample, broken down by optical (blue), NIR (green), and photometric types (red). \label{fig:spthist}}
\end{figure*}

\begin{figure*}
\figurenum{4}
\centering
\includegraphics[scale=0.5]{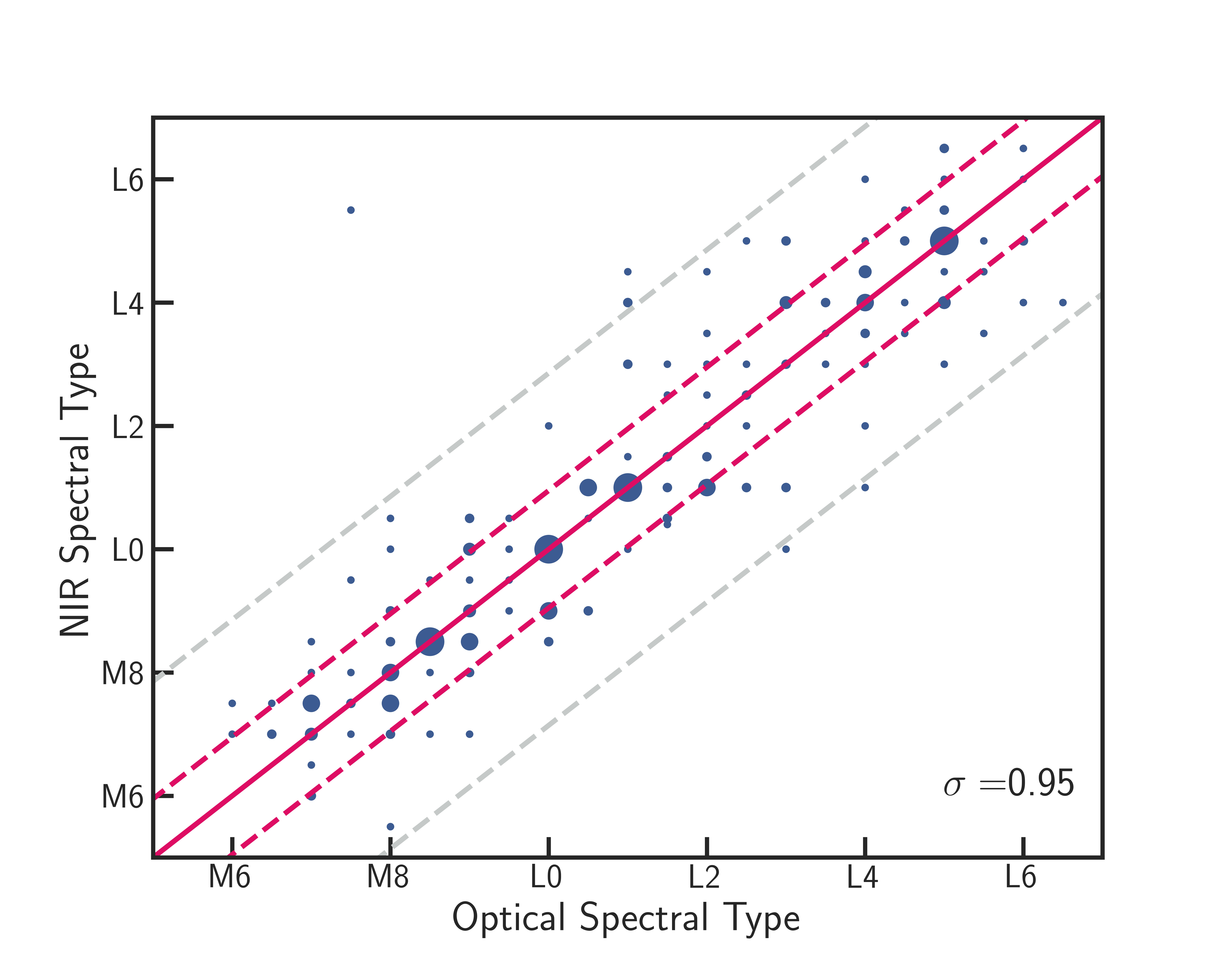}
\caption{Comparison of optical and NIR spectral types from the literature for the M7$-$L5 25\,pc sample. The size of the circles scales as the cube of the number of repeated points. The solid line marks where the slope equals one, while the dashed lines encompass the $1\sigma$ and $3\sigma$ limits in magenta and light grey, respectively.\label{fig:optvsnir}}
\end{figure*}

One hundred and eighty-nine objects have both optical and NIR measurements from the literature. With our SpeX observations (see Section~\ref{sec:specsamp}), we have added 109 NIR classifications (see Section~\ref{sec:sptclass}). Figure~\ref{fig:optvsnir} shows a comparison between literature optical and NIR spectral types. The size of each circle is proportional to the number of overlapping sources. The scatter between spectral types is 0.95 subtypes; the $3\sigma$ boundaries are delineated by the dashed light grey lines.  

\subsubsection{Distances}\label{sec:dists}

Trigonometric and spectrophotometric distances were calculated from parallaxes and from spectrophotometric empirical relations in the NIR, respectively. \textit{Gaia} DR2 provided most of the parallaxes in the sample, 80\% of the total or 327 in our $25\,pc$ sample. Distances from \textit{Gaia} were calculated simply as $d = 1000/\omega~(mas)$ , rather than using a likelihood function with Bayesian probabilities (e.g.,~\citealt{2018arXiv180410121B}), since we are concerned with sources with large parallaxes $(\omega \geq 35~mas$ or $d\leq 28.5\,pc$ to account for uncertainties beyond $d = 25$\,pc) with small relative errors of the order of $0.04\%-4\%$. Trigonometric distances from parallaxes predating \textit{Gaia} DR2 were calculated in the same way. We also calculated trigonometric distances from \emph{WISE} following the prescription of~\citet{2018ApJ...862..173T} for 16 sources. 

\begin{figure*}
\figurenum{5}
\centering
\includegraphics[width=1\textwidth]{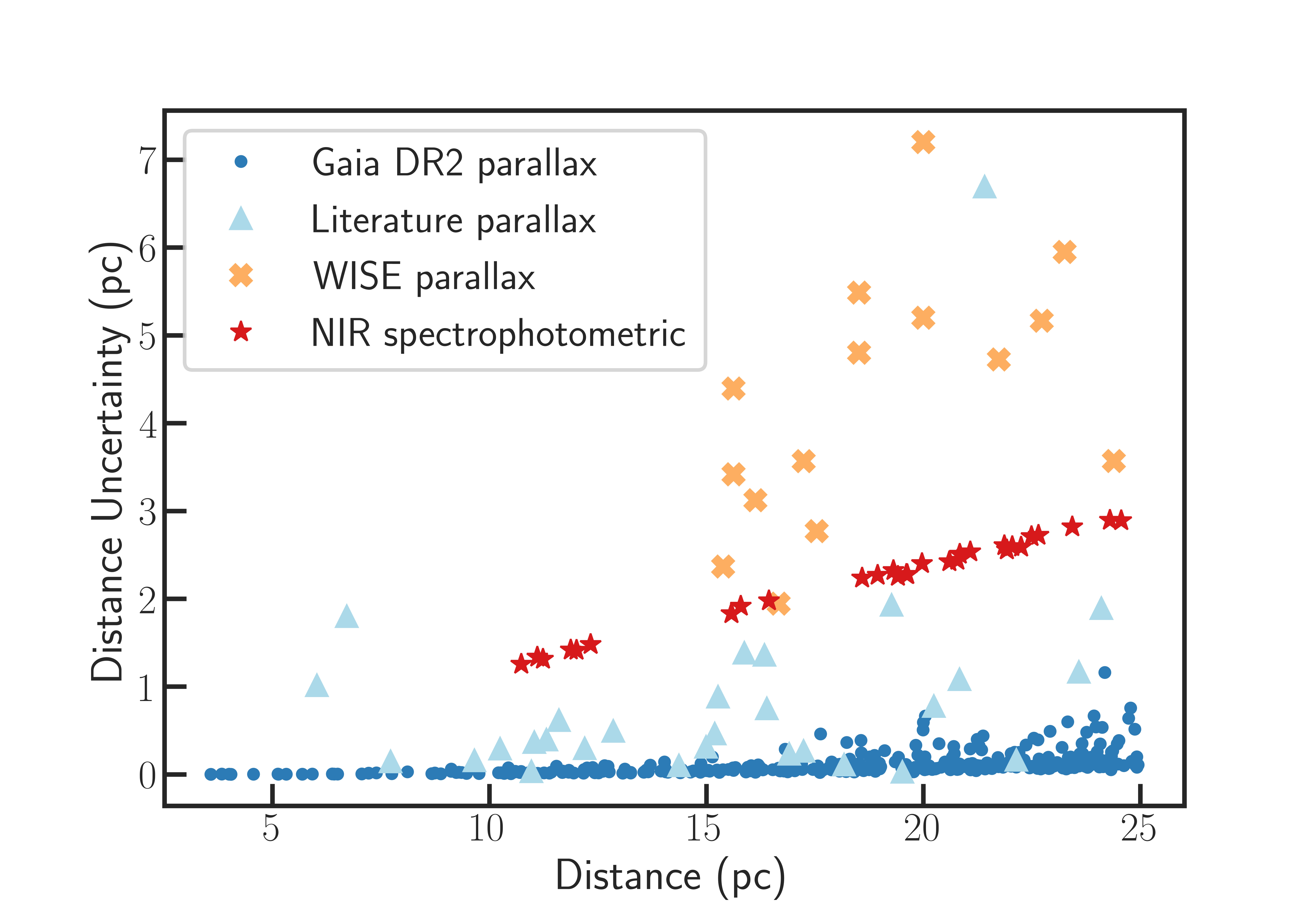}
\caption{Comparison of distance values and uncertainties. The most precise distances are those found through \textit{Gaia} parallaxes shown as blue dots. Distances found through parallaxes from the literature (i.e. SIMBAD) are plotted as light blue triangles, and show a large scatter since they come from a variety of studies with different systematics. Parallaxes obtained through \emph{WISE}~\citep{2018ApJ...862..173T} are shown as orange crosses and have the largest uncertainties. NIR spectrophotometric distance estimates are shown as red stars also with large uncertainties, and growing as a function of distance.\label{fig:distunc}}
\end{figure*}

\begin{figure*}
\figurenum{6}
\centering
\includegraphics[width=1\textwidth]{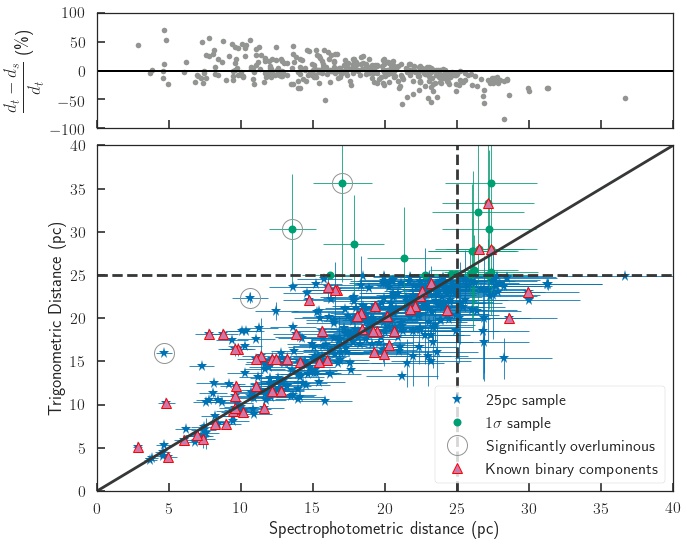}
\caption{Spectrophotometric distance estimates compared to trigonometric distance measurements. \emph{(Top)} Fractional percentage errors between trigonometric ($d_t$) and spectrophotometric ($d_s$). \emph{(Bottom)} The $25\,pc$ sample is shown in green and the $1\sigma$ sample is shown in blue. The black solid line delineates the one-to-one correspondence between trigonometric and photometric distances. Sources significantly above the line and beyond three standard deviations are likely unresolved binaries. In particular, the sources encircled in grey are 2MASS J1733+1655 ($d_t = 16.03\pm0.10$\,pc), NLTT 40017 ($d_t = 22.4\pm0.7$\,pc), SDSS J1221+4632 ($d_t = 30.3\pm6.4$\,pc), and SDSS J0911+2248 ($d_t = 35.7\pm11.5$\,pc). None of these objects have mentions of binarity in the literature.\label{fig:trigspecphot}}
\end{figure*}

We calculated spectrophotometric distances using the adopted literature spectral type and the absolute magnitude empirical relations from~\citet{2012ApJS..201...19D}. Distances were calculated for the NIR filters $J$, $H$, and $K_s$, and averaged, weighted by their uncertainties. We adopt trigonometric distances if available (for $93\%$ of the sample), and use spectrophotometric distances for 29 sources that do not have a parallax measurement. Distances are reported in Table~\ref{tab:dists}. Figure~\ref{fig:distunc} summarizes the distance uncertainties for these measurements, and Figure~\ref{fig:trigspecphot} compares trigonometric to  spectrophotometric distances for the $25\,pc$ and $1\sigma$ samples. Trigonometric and spectrophotometric distances agree within 6.9\% of each other, except for obviously overluminous sources.

Using the best distance measure, a strict cut on 25\,pc was applied to select our volume-limited sample with 410 sources whose measured literature optical or NIR spectral types lie within M7$-$L5, \emph{and} whose distance was within 25\,pc, i.e. excluding objects with only a photometric estimation of their spectral type. We assess Poisson uncertainties as described in~\citet{2017ApJS..228...18G} for our sample size in subsequent analysis. Sources whose $1\sigma$ uncertainties placed them within 25\,pc, amounting to 60 objects, were added to an expanded $25\,pc+1\sigma$ sample of 470 objects.


\section{Spectral Sample}\label{sec:specsamp}




Two hundred and forty $25\,pc$ sample members had SpeX spectra in the SpeX Prism Library (SPL;~\citealt{2014ASInC..11....7B}) prior to 2015. We observed an additional 286 sources with SpeX between UT 2015 February 24 and 2018 November 22 as part of NASA IRTF programs 2015A074, 2015B087, 2016A079, 2016B114, 2017A102, 2018B120 (PI: Bardalez Gagliuffi), and 2016A038 (PI: Burgasser), over a total of 15 nights. The observations log is summarized in Table~\ref{tab:obstable}. The latitude, equatorial mount, and location of IRTF allow for observation of declinations in the $-50^{\circ}<\delta<+67^{\circ}$ range. Ninety percent of the $25\,pc$ sample lies within these declinations, and between existing work and our contributions, we obtained spectra for $89\%$ of these sources, or $81\%$ of the $25\,pc$ sources overall.  Sources were observed in prism mode, which completely samples wavelengths $0.75-2.5\mu$m at a dispersion of $20-30\mathrm{\AA}$ pixel$^{-1}$ in a single observation. Most stars were observed with the $0\farcs5$ slit, 10 sources were observed with the $0\farcs8$ slit if the seeing rose above $1\farcs2$. The slit was aligned with the parallactic angle. Integration times ranged between $60-150$\,s per exposure, depending on the brightness of the source and atmospheric conditions. Observations were carried out in an ABBA dither pattern along the slit, with additional AB cycles if more counts were needed to achieve S/N$\sim100$. Bright A0 stars were observed close in time at a similar airmass and used for flux calibration of the raw science spectra and correction for telluric absorption. Internal flat fields and Ar arc lamps were observed with each flux standard for pixel response and wavelength calibration, respectively. All data were reduced with SpeXtool package v4.1~(\citealt{2004PASP..116..362C,2003PASP..115..389V}) using standard settings.

\section{Sample Characterization}\label{sec:characterization}

\subsection{Spatial Distribution}\label{sec:spacedist}

\begin{figure*}
\figurenum{7}
\centering
\includegraphics[width=\textwidth]{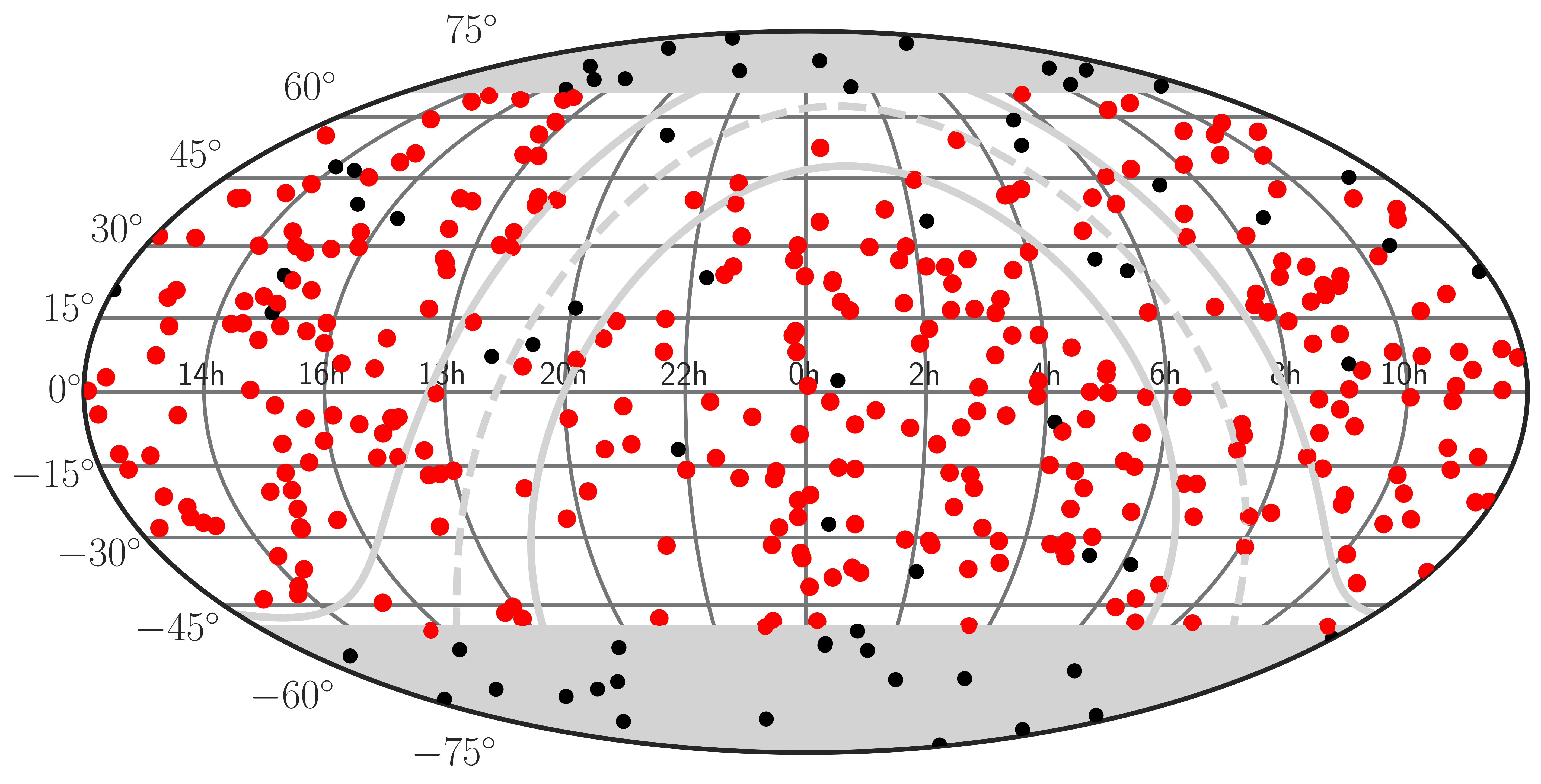}
\caption{Spatial distribution of $25\,pc$ targets in the M7$-$L5 25\,pc sample. The sample is shown as black dots, objects for which we have SpeX spectra are shown as red dots. The sky regions inaccessible by IRTF are shaded in grey. The galactic plane ($b=0^{\circ}$) is shown as a dashed light gray line, and the $\pm15^{\circ}$ parallels from the galactic plane are shown as solid light gray lines.\label{fig:map}}
\end{figure*}

Figure~\ref{fig:map} shows the spatial distribution of all our targets. The $25\,pc$ literature sample is evenly distributed across the sky, with the exception of the Galactic plane. Since 25\,pc is a relatively small radius compared to the radius of the Milky Way ($R_{MW}\sim25\,$kpc) and its vertical scale height ($\sim300$\,pc;~\citealt{1991ApJ...378..131K,2010AJ....139.2679B}), we assume an isotropic distribution of sources within this volume. There are 217 sources at northern declinations and 193 at southern declinations. In Galactic coordinates, there are 228 sources above the plane of the galaxy and 182 below it. We convert the 381 sources with measured parallaxes in our $25\,pc$ from equatorial to galactic X, Y, Z right-handed coordinates centered at the Sun. In the $\vec{X}$ direction we find 161 objects between the Sun and the Galactic center, and 220 between the Sun and the outer edge of the Galaxy. In the $\vec{Y}$ direction we find 206 objects  in the direction of the Sun's motion, and 175 objects trailing behind it. In the $\vec{Z}$ direction, we find 207 objects above the plane of the Sun, and 174 below it. All of these values are within $3\sigma$ of each other, considering Poisson uncertainties, yet not consistent at the $1\sigma$ level.  ~\citet{2016AandA...589A..26B} have suggested an inhomogeneity in the spatial distribution of brown dwarfs compared to stars, most likely an effect of small number statistics and incomplete coverage of observations. The slight preference for northern sources is due to the larger number of panchromatic survey observations in the northern hemisphere (in particular SDSS). The Galactic plane looks sparse due to overcrowding and background source contamination, and this region is excluded from our space density analysis below  (c.f.~\citealt{2007MNRAS.374..445K,2003AandA...403..929K}).



\subsection{Spectral Classification}\label{sec:sptclass}

We compared our SpeX spectra to NIR spectral standards defined in~\citet{2010ApJS..190..100K}, following the method described therein, which compares the $0.9-1.4\mu$\,m spectrum of an object to standards using a $\chi^2$ minimization routine. The resulting distribution of spectral types is shown in Figure~\ref{fig:spthist}.

\begin{figure*}
\figurenum{8}
\centering
\includegraphics[width=\textwidth]{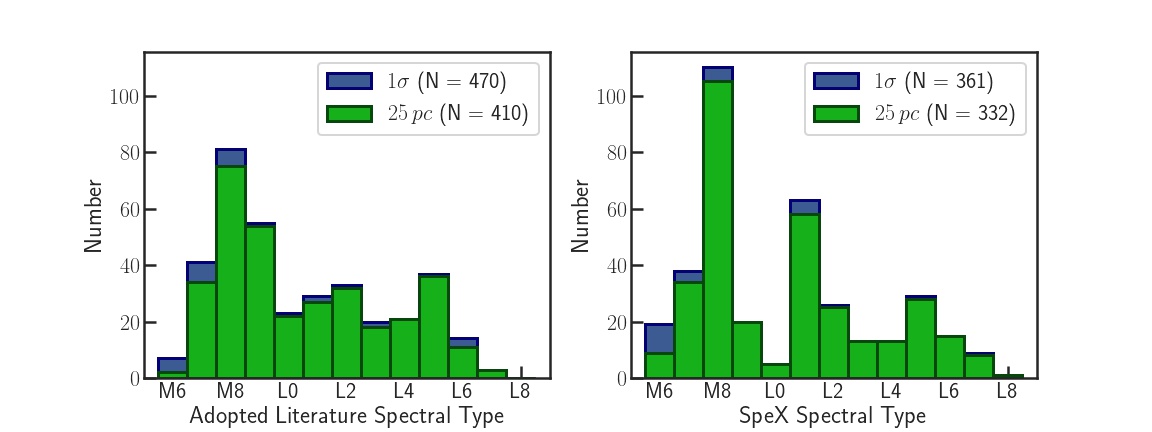}
\caption{\emph{(Left)} Adopted literature spectral type distribution of $25\,pc$ and $1\sigma$ samples. \emph{(Right)} Spectral type distribution of $25\,pc$ and $1\sigma$ samples according to their SpeX classification. Objects outside of the M7$-$L5 range have at least one spectral classification within that range.\label{fig:spthist}}
\end{figure*}

After classifying the spectra, we compared their literature and measured spectral types. For most objects, we measured a NIR spectral type within one subtype of the published literature type. Objects with only a photometric estimate from the literature and whose SpeX spectral type placed them outside of the M7$-$L5 range are in the $1\sigma$ sample. 

\begin{figure*}
\figurenum{9}
\centering
\includegraphics[scale=0.55]{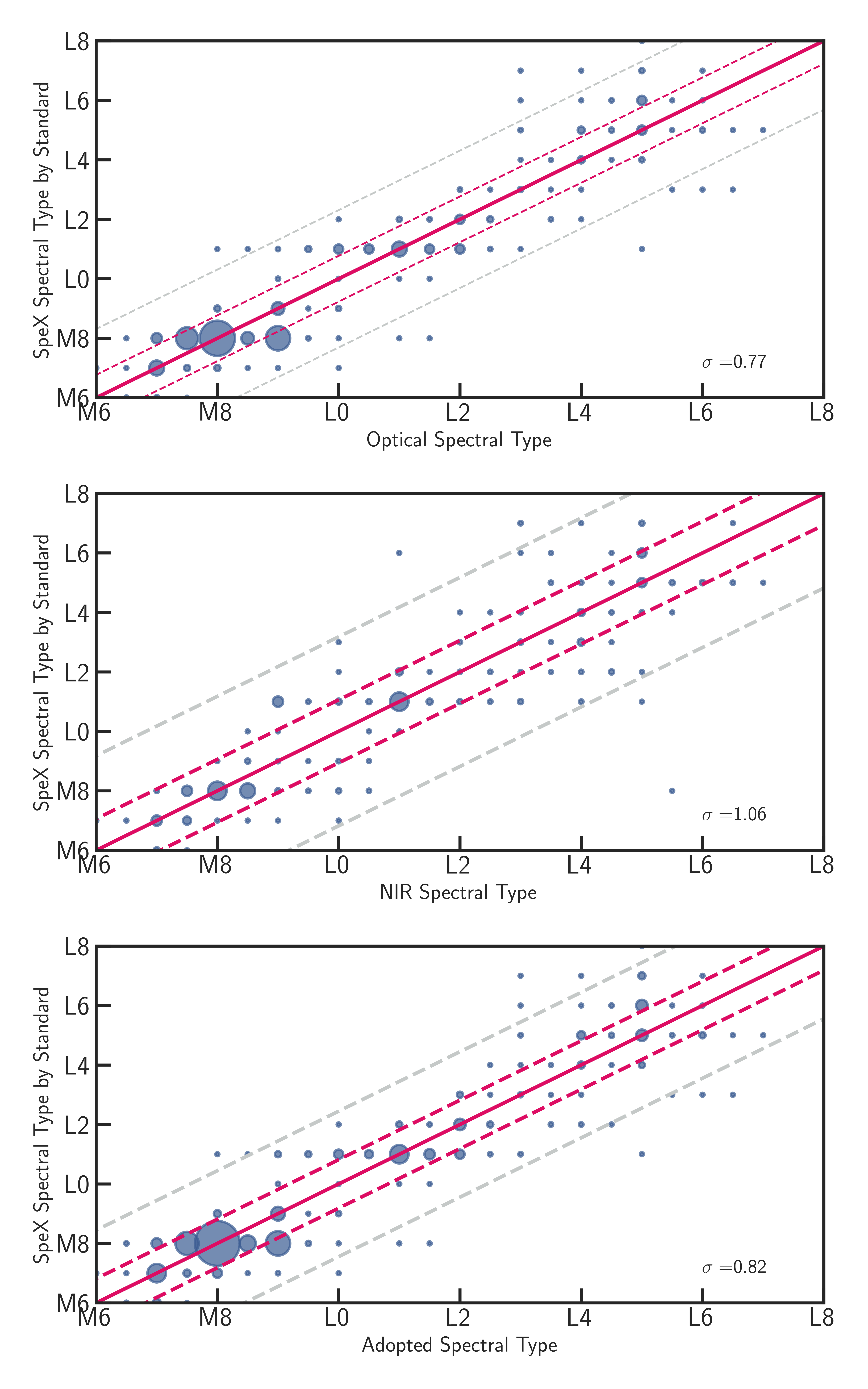}
\caption{Literature optical and NIR spectral types compared to SpeX spectral types with~\citet{2010ApJS..190..100K} NIR standards. Circle sizes are proportional to the number of sources in a given optical-NIR spectral type pair. The solid line indicates equal classification, and the pink and grey dashed lines are the $1\sigma$ and $3\sigma$ limits, respectively.\label{fig:sptstd}}
\end{figure*}

Figure~\ref{fig:sptstd} compares the literature adopted optical or NIR classifications to the SpeX classification. The scatter for the optical-SpeX comparison is $\sigma = 0.77$ subtypes, the scatter for the NIR-SpeX comparison is $\sigma = 1.06$ subtypes, and the scatter in the adopted-SpeX comparison is $\sigma = 0.82$ subtypes. The larger scatter between NIR-SpeX classifications may be due to poorly defined prior NIR types, sensitivity to surface gravity, metallicity, clouds; and variance in the spectral region used for NIR classification.

\begin{figure*}
\figurenum{10}
\centering
\includegraphics[scale=0.5, trim=0 240 0 0, clip]{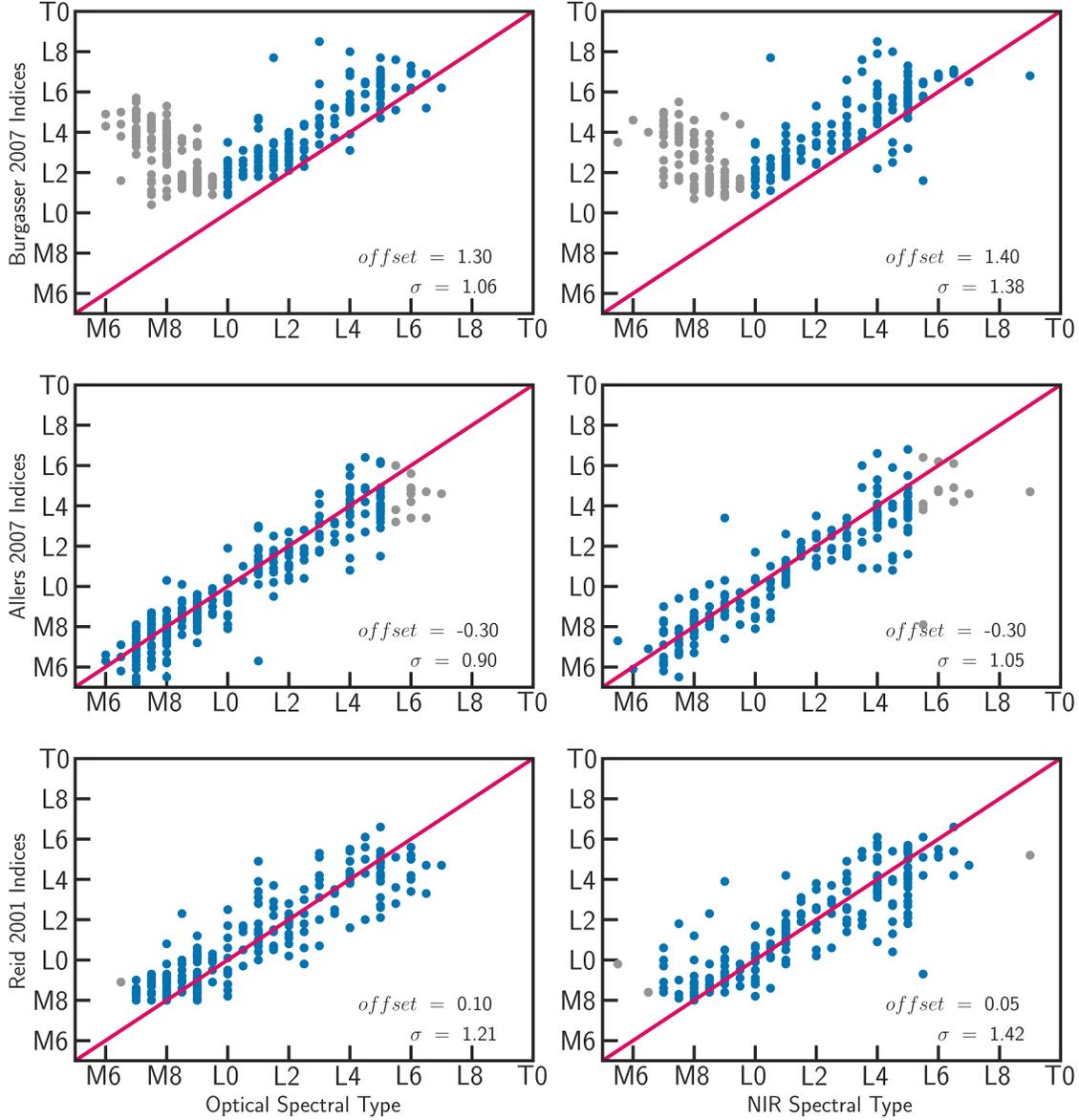}
\caption{Literature optical and NIR spectral types compared against measured spectral types with the index sets of~\citet{2007ApJ...659..655B},~\citet{2007ApJ...657..511A} and~\citet{2001AJ....121.1710R}. Points outside the spectral type ranges defined for each index classification are plotted in grey and do not enter the $\sigma$ calculation.\label{fig:sptindices}}
\end{figure*}

We also classified our SpeX spectra using spectral indices from~\citet{2007ApJ...659..655B},~\citet{2007ApJ...657..511A}, and~\citet{2001AJ....121.1710R}. These indices are applicable in the L0$-$T8, M5$-$L5, and M7$-$L8 spectral type ranges, respectively. Figure~\ref{fig:sptindices} shows the comparisons from these index-classification systems against optical and NIR spectral types reported in the literature. The points outside of the allowed classification ranges are plotted in light grey and are not included in the median offset and scatter calculations. The indices from~\citet{2007ApJ...659..655B} have a systematic offset of $+1.30$ and $+1.40$ subtypes compared to optical and NIR types, respectively, and overestimate the spectral type of our sources. The~\citet{2007ApJ...657..511A} indices are the most accurate at predicting optical spectral types with $\sigma=0.90$ subtypes. The scatter is larger for NIR types ($\sigma=1.05$ subtypes), with a slight tendency to predict spectral types earlier than measured in the literature (offset = $-0.30$ in both cases). For both optical and NIR types, the~\citet{2001AJ....121.1710R} indices have the smallest offset ($0.10$ and $0.05$ subtypes for optical and NIR spectral types, respectively) but slightly larger scatters than~\citet{2007ApJ...657..511A}, at $\sigma=1.21$ and $\sigma=1.42$ subtypes, respectively.  All spectral types for sample sources are summarized in Table~\ref{tab:spts}.

\subsection{Gravity Classification and Young Moving Group Membership}\label{sec:grav}

Young brown dwarfs ($\tau\lesssim200$\,Myr) are undergoing cooling and contraction, and are both larger in radius and less massive than their older counterparts at a similar spectral type. These physical properties translate into lower surface gravities, affecting spectral features such as reduced collision-induced absorption, and narrower alkali lines~(\citealt{2007ApJ...657..511A,2010ApJS..190..100K}). Due to their low surface gravity and typically dusty atmospheres, young brown dwarfs share physical properties with directly-imaged exoplanets, making the former ideal analogs to the latter~(\citealt{2013MmSAI..84..955F,2016ApJS..225...10F}).

We obtained gravity classifications of our SpeX spectra, following the NIR scheme of~\citet{2013ApJ...772...79A}, defined for the spectral type range M5$-$L7, except that spectral types were determined from H$_2$O indices without a visual comparison of the $J$-band with NIR standards. 

Additionally, we obtain 7 very low gravity (VL-G) and 64 intermediate gravity (INT-G) candidate classifications from our spectra in the combined $25\,pc$ and 1$\sigma$ samples (Table~\ref{tab:grav}). All low-gravity candidates were examined for visual signatures of low gravity, comparing the spectra band-by-band to low-gravity standards (see~\citealt{2015ApJS..219...33G} and~\citealt{2018AJ....155...34C}), leading to the rejection of 26 INT-G classifications. We labeled 11 sources with conflicting signatures as peculiar, such as blue $J-K_S$ colors, indicating low metallicity effects rather than low gravity~\citep{2016AJ....151...46A}.  Most VL-G sources are previously known, but we have identified 2MASS J1739+2454 as a new very low-gravity source. Thirteen of the 26 INT-G sources are first reported in this paper. The unresolved spectrum of the M8+M8 binary system 2MASS J0027+2219AB~\citep{2005AandA...435L...5F} was also classified as an INT-G source. Since both components have the same spectral type, and since the system is coeval, we assume that both components would be independently classified as INT-G, leading to a final number of INT-G objects of 26 plus one more including the $1\sigma$ sample.

While 2MASS J1022+5825~\citep{2008AJ....136.1290R}, 2MASSW J2148+4003~\citep{2008ApJ...686..528L} and 2MASS J0512$-$2949~\citep{2003AJ....126.2421C} were previously classified as having field gravity (FLD-G;~\citealt{2013ApJ...772...79A,2016ApJS..225...10F}), our spectra yield INT-G classifications. Similarly, SDSS J0443+0002 was classified as a VL-G in~\citet{2013ApJ...772...79A}, but our spectra yields an INT-G classification. These discrepancies may be due to instrumental or reduction differences. 

We used BANYAN~$\Sigma$~\citep{2018ApJ...856...23G} on our low-gravity candidates to assess possible membership in 27 young moving groups, using new kinematic data from \textit{Gaia} DR2~\citep{2018AandA...616A...1G}, and report the probabilities for young moving group membership in Table~\ref{tab:grav}. The~\citet{2013ApJ...772...79A} gravity classification scheme is a spectroscopic test for youth, while BANYAN $\Sigma$ uses kinematic information to determine membership in a young moving group. Many of our low-gravity sources are classified as 0\% probability members of any young group by BANYAN $\Sigma$, which implies that these objects might be young and unassociated, field interlopers, or belonging to moving groups other than the 27 known associations included in BANYAN~$\Sigma$, possibly as a result of ejection.

\begin{figure*}
\figurenum{11}
\centering
\includegraphics[width=\textwidth]{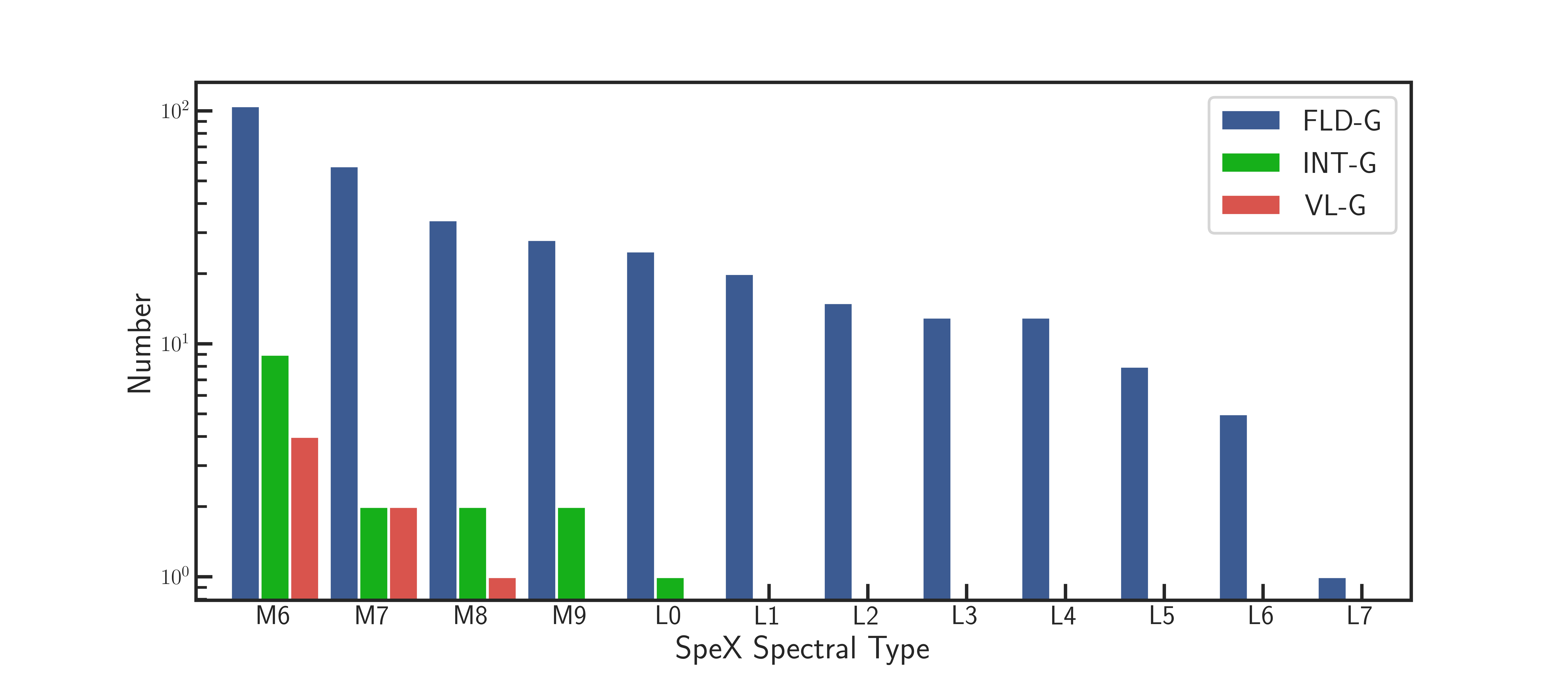}
\caption{Distribution of spectral types as classified by field spectral standard for different gravity types. Objects with gravity classifications of very-low gravity (VL-G) or intermediate gravity (INT-G) are plotted in red and green, respectively.\label{fig:gravhist}}
\end{figure*}

Figure~\ref{fig:gravhist} shows the distributions of gravity types from our SpeX spectra by spectral type, as classified by field standards. We find the very-low-gravity and intermediate-gravity fractions for our $25\,pc$ sample to be $2.1^{+0.9}_{-0.8}\%$ and $7.8^{+1.7}_{-1.5}\%$, respectively, with uncertainties based on Poisson statistics.

\begin{figure*}
\figurenum{12}
\centering
\includegraphics[width=1.1\textwidth]{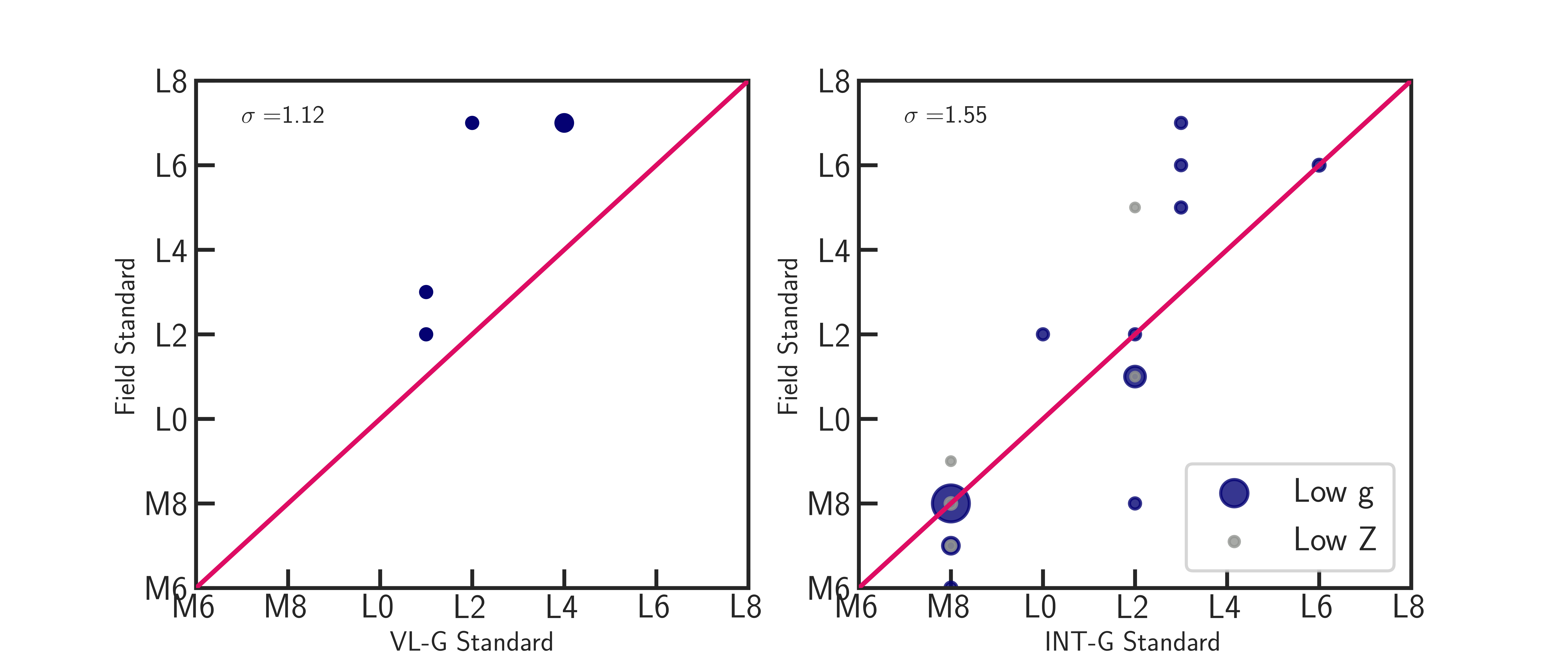}
\caption{\emph{(Left)} Comparison between spectral classification by very low gravity and field gravity standards for the 4 objects classified as having very low gravity by the prescription of~\citet{2013ApJ...772...79A}. Size of markers is proportional to the number of equally-classified sources. The magenta line represents a one-to-one match between classifications. \emph{(Right)} Same comparison between intermediate gravity and field gravity standards. Objects with an INT-G classification most likely not young, but metal-poor instead, are shown in grey, with a lower proportionality of number of sources to marker size.\label{fig:vlgintgstd}}
\end{figure*}

The spectral types of our low-gravity objects were further refined using VL-G and INT-G spectral standards from~\citet{2013ApJ...772...79A}. The comparison between classifications is shown on Figure~\ref{fig:vlgintgstd}. The 7 VL-G sources in our sample have much earlier types (by 1-3 subtypes) when classified with a VL-G standard than with a field standard, although this is too small of a sample to precisely quantify the bias.  Figure~\ref{fig:vlgfldgs} shows the 7 VL-G sources classified with a field standard and VL-G standard.

\begin{figure*}
\figurenum{13}
\centering
\includegraphics[scale=0.5]{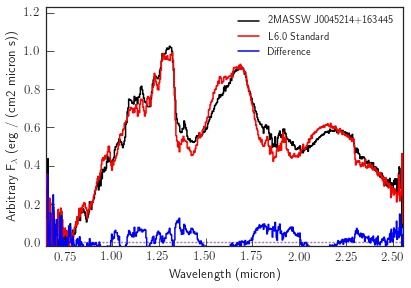}\includegraphics[scale=0.5]{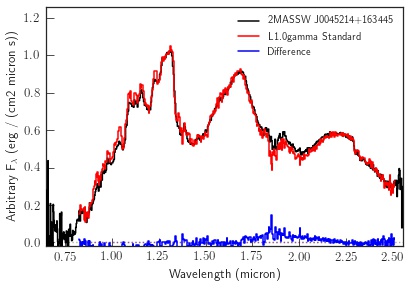}\\
\includegraphics[scale=0.5]{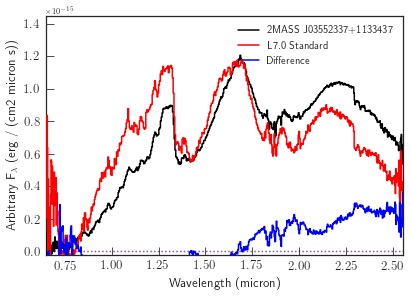}\includegraphics[scale=0.5]{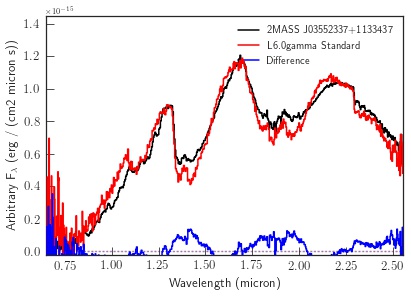}\\
\includegraphics[scale=0.5]{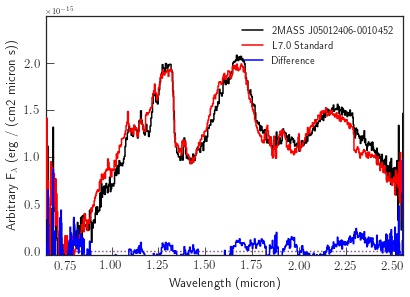}\includegraphics[scale=0.5]{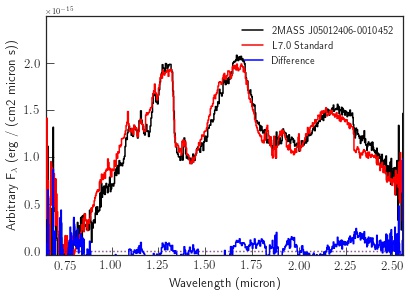}\\
\includegraphics[scale=0.5]{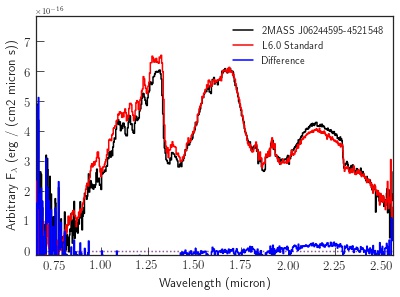}\includegraphics[scale=0.5]{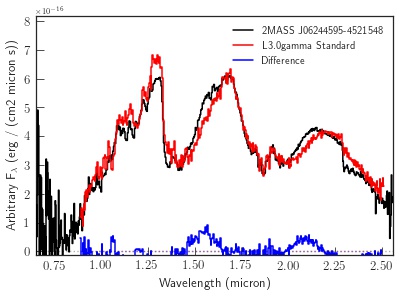}\\
\caption{Sources classified as very-low gravity (VL-G) compared against field \emph{(Left)} and VL-G \emph{(Right)} standards. Spectra (black) are consistently redder than their field standards (red). The positive difference between spectra and standards (blue) is clear, emphasizing the need to fit spectra to appropriate gravity standards.\label{fig:vlgfldgs}}
\end{figure*}

\begin{figure*}
\figurenum{12}
\centering
\includegraphics[scale=0.5]{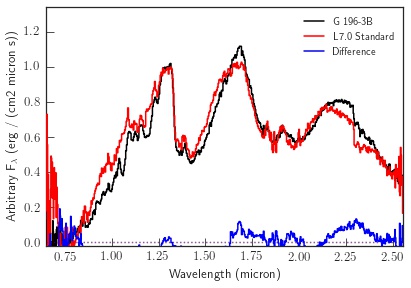}\includegraphics[scale=0.5]{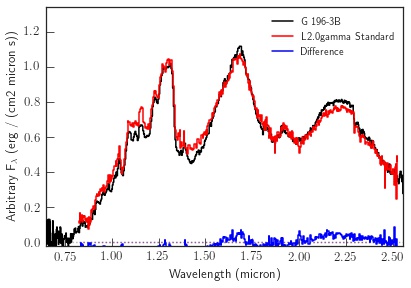}\\
\includegraphics[scale=0.5]{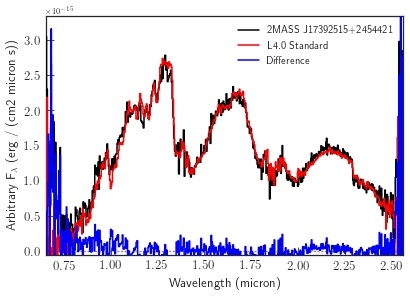}\includegraphics[scale=0.5]{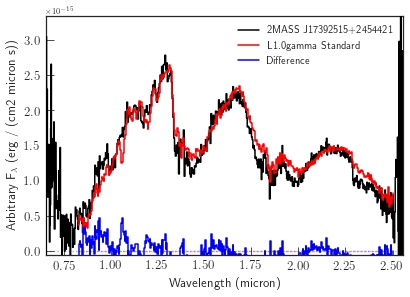}\\
\includegraphics[scale=0.5]{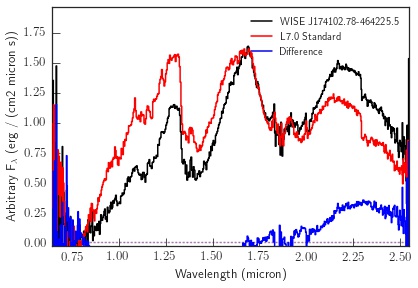}\includegraphics[scale=0.5]{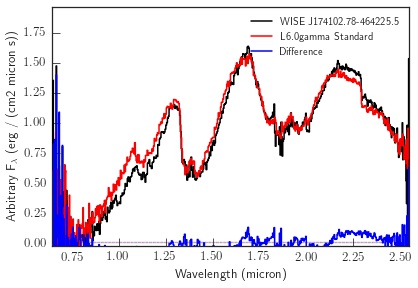}
\caption{Continued.\label{fig:vlgfldgs}}
\end{figure*}

For INT-G sources, there is a better correlation but larger scatter ($\sigma=1.67$), particularly among L dwarfs, which are expected to show stronger gravity features even as INT-G. These differences highlight the strong role of gravity-sensitive features and reinforce the importance of comparing low gravity sources to equivalent standards.

\subsection{Color Outliers}

Red and blue $J-K_S$ color outliers are empirically-defined subpopulations. Their unusual color is likely a proxy for physical properties such as age, low or high surface gravity, atmospheric cloud content, opacity, and metallicity~(\citealt{2006ApJ...651.1166M,2008ApJ...674..451B,2008ApJ...686..528L,2009AJ....137....1F}).

Clouds play a key role in $J-K_S$ color evolution from late-M to L-type, as increased opacity originating from condensates and possibly clouds reddens spectral energy distributions~(e.g.,~\citealt{1996AandA...308L..29T, 2006asup.book....1L}). This is intrinsic reddening, as objects in the 25\,pc sample should be minimally reddened by interstellar dust. The thickness of clouds may be an independent parameter (e.g.,~\citealt{2001ApJ...556..872A,2016ApJ...830...96H}), or may correlate with youth (e.g.,~\citealt{2013AJ....145....2F}), and/or metallicity~(~e.g.,~\citealt{2003ApJ...592.1186B}). Color outliers may also indicate the presence of an unresolved companion~(e.g.,~\citealt{2014ApJ...794..143B}). Unusually blue objects and subdwarfs have enhanced collision-induced H$_2$ opacity~(\citealt{1994ApJ...424..333S,2003ApJ...592.1186B}) due to their metal-poor atmospheres.

\begin{figure*}
\figurenum{14}
\centering
\includegraphics[width=\textwidth]{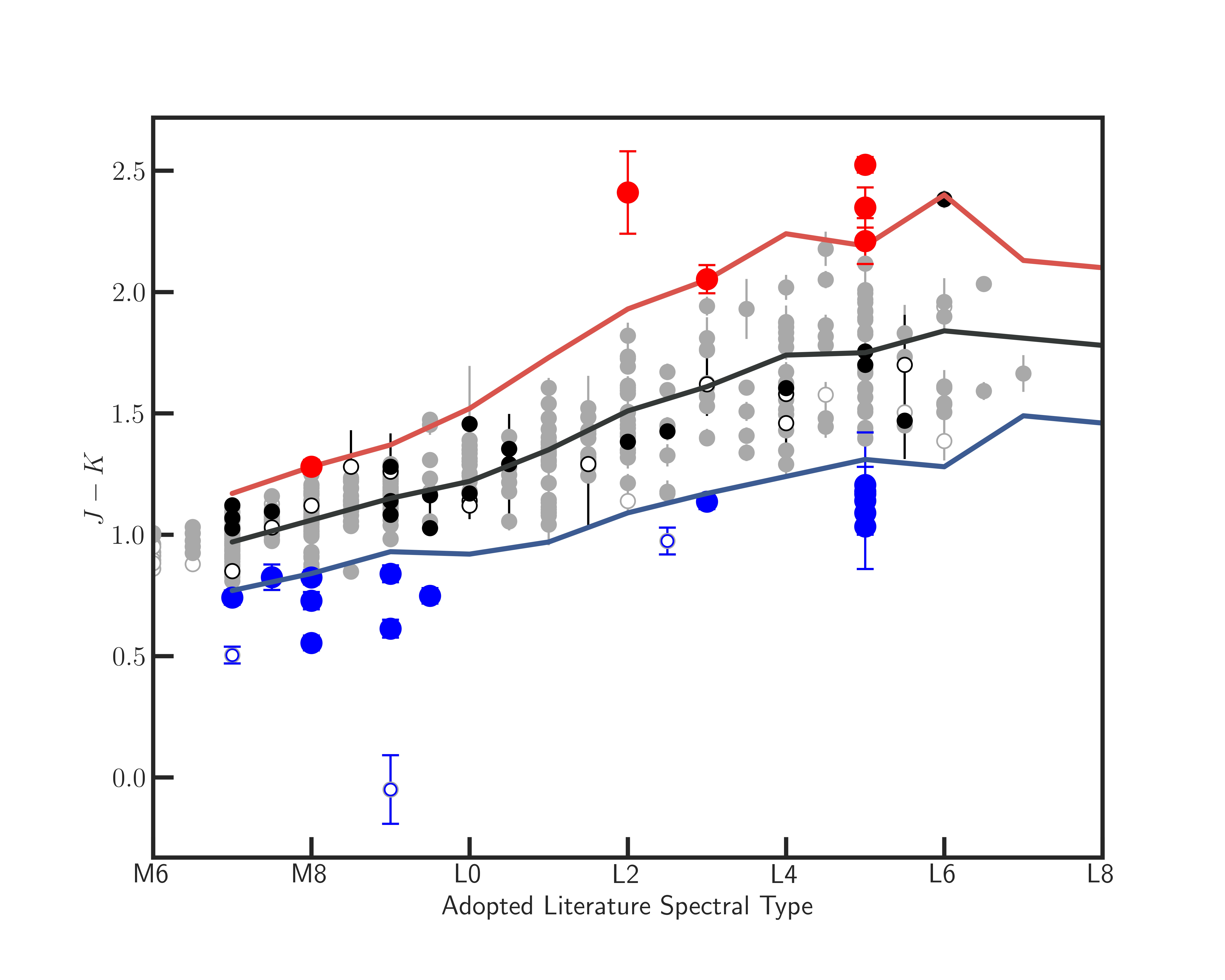}
\caption{$J-K$ color outliers per spectral type. $25\,pc$ sources with 2MASS photometry are filled grey circles, and 1$\sigma$ sources are open grey circles. Black filled and open circles are sources where the adopted magnitudes are in the MKO system for the $25\,pc$ and $1\sigma$ samples, respectively. Red and blue circles are color outliers for their spectral type, as defined by the color averages of~\citet{2016ApJS..225...10F}. The average $J-K_S$ color is the dark grey line, and the $2\sigma$ limits are the red and blue lines. The red outlier at L2 is the binary Kelu-1A.\label{fig:JKcolor}}
\end{figure*}

To isolate the color outliers of our sample, we compared their $J-K_S$ colors to the average colors and standard deviations as a function of spectral type from~\citet{2016ApJS..225...10F}, defined over the M7$-$L8 range. We identified outliers as $2\sigma$ deviants, shown in Figure~\ref{fig:JKcolor}. From the 387 objects in the $25\,pc$ whose adopted spectral type is within M7$-$L5\footnote{Objects with an adopted spectral type outside of the M7$-$L5 range have an optical spectral type also outside the range, but either a NIR or photometric type estimation within the range.}, and with both $J$ and $K_S$ photometry\footnote{GJ~1116AB only has unresolved photometry, so we do not count the B component in this calculation. Gl~779B only has $K_S$ photometry, so it is excluded as well.}, 188 have $J-K$ positive excesses, while 184 have negative color excesses, and 15 do not have a color excess. This even distribution of sources indicates that our sample does not have a NIR color bias, despite widely used 2MASS color selections~\citep{2015AJ....149..158S}, for which redder selection criteria were necessary to excise background population. 


The individual outliers are listed in Table~\ref{tab:JKcolor}. In our $25\,pc$ sample, 15 objects were found to have unusually blue $J-K_S$ colors and 6 have unusually red $J-K_S$ colors. In the $1\sigma$ sample we find 2 more unusually blue objects. Given the numbers of color outliers from the $25\,pc$ sample, we infer fractions of $1.4^{+0.6}_{-0.5}$\% for red and $3.6^{+1.0}_{-0.9}$\% for blue M7$-$L5 dwarfs in the Solar neighborhood (with Poisson uncertainties). Among the 5 red outliers, 2MASS J0355+1133, G 196$-$3B, and 2MASS J1741$-$4642 have been reported as young in the literature (\citealt{2015ApJS..219...33G,2016ApJS..225...10F}), while LHS2397aA and Kelu$-$1A are classified as having field gravity, but are also known binaries~(\citealt{2003IAUS..211..261F,2008arXiv0811.0556S}). From all the sources with \emph{Gaia} kinematics, we explored a reduced proper motion diagram and found no potential subdwarfs, i.e. sources with high proper motion, high reduced proper motion, and blue $G-G_{RP}$ colors.

Five blue sources were also classified as INT-G, cementing their status as metal-poor objects (see Section~\ref{sec:grav} and~\citealt{2016AJ....151...46A}). Two unusually blue sources, G 203$-$50B and 2MASS J1721+3344, are also rejected spectral binary candidates, as blue sources tend to be contaminants in the identification of spectral binaries~\citep{2014ApJ...794..143B}\footnote{When identifying spectral binaries via spectral indices alone, objects with a bluer spectral slope are often false positives that are rejected by visual inspection of their binary template fits.}.

\begin{figure*}
\figurenum{15}
\centering
\includegraphics[scale=0.5]{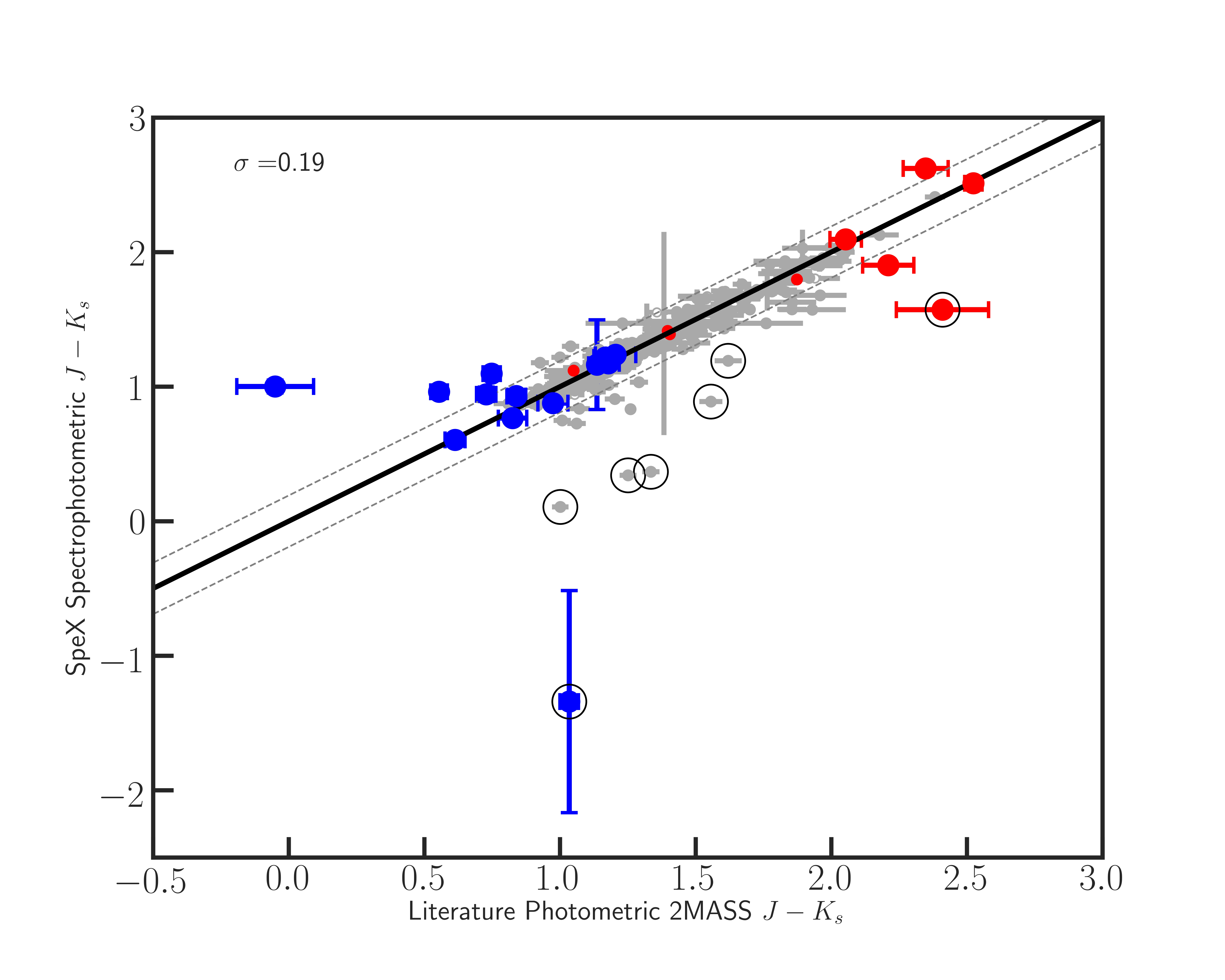}
\caption{Photometric 2MASS $J-K_S$ color from the literature compared to spectrophotometric $J-K_S$ color from our SpeX observations. Same color-coding as Figure~\ref{fig:JKcolor}. Objects inside open black circles are $>2\sigma$ outliers.\label{fig:JKcalib}}
\end{figure*}

Additionally, we calibrated our SpeX spectra to 2MASS $J$ and $K_s$ magnitudes to find spectrophotometric $J-K_S$ colors. These were compared against 2MASS $J-K_S$ colors, and found to have a scatter of 0.18\,mag. $2\sigma$ outliers or higher are highlighted in Figure~\ref{fig:JKcalib}, and could be due to intrinsic atmospheric variability (e.g.,~\citealt{2012ApJ...750..105R}). These sources are: LHS 5166B, 2MASS J1152+2438, 2MASS J1200+2048, Kelu-1 A (unusually red), 2MASS J1416+1348A (unusually blue), and 2MASS J1438+6408. Kelu-1 has a variability detection in 410\AA~with a peak-to-peak amplitude of 11.9$\pm$0.8\,mmag~\citep{2003MNRAS.341..239C}, reported before the discovery of its nearby companion~\citep{2005ApJ...634..616L}.~\citet{2013AJ....145...71K} reported marginal variability in $J$-band for 2MASS J1416+1348A. The remaining outliers have not been targeted in variability surveys.

\subsection{Spectral Binaries}

Spectral binaries of ultracool dwarfs are systems composed of a late-M/L-type primary and a hidden T-dwarf secondary, identifiable only by their peculiar blended-light spectrum in NIR wavelengths (\citealt{2004ApJ...604L..61C,2010ApJ...710.1142B,2014ApJ...794..143B}). Identifying these potentially closely-separated binaries allows us to probe the very low mass binary separation distribution at all scales and select potential systems for orbital measurement (see~\citealt{2015AJ....150..163B}).



We applied the spectral binary technique of~\citet{2014ApJ...794..143B}\footnote{The boundaries of the parameter spaces were modified in~\citet{2015AJ....150..163B} to include the M9+T5 spectral binary WISE~J072003.20$-$084651.2~\citep{2014AandA...561A.113S,2015AJ....149..104B}.} to the SpeX spectral sample.  The spectral binary technique consists of two parts: spectral index selection and binary template fitting, the second of which incurs a hypothesis test to determine whether binary template fits are statistically better fits to a candidate than single templates. Spectral binary candidates are listed in Table~\ref{tab:SB}. Forty-two objects were selected by the index-index parameter spaces as candidates, but rejected by the low confidence from hypothesis testing. Seven objects were rejected despite passing the spectral binary fitting due to their blue colors, as blue objects are known contaminants of the spectral binary technique~\citep{2014ApJ...794..143B}. 

We found five previously identified and confirmed spectral binaries in our $25\,pc$ sample: 2MASSW J0320284$-$044636~\citep{2008ApJ...678L.125B,2008ApJ...681..579B}, WISE~J072003.20$-$084651.2~\citep{2015AJ....149..104B}, 2MASS J08053189+4812330~(\citealt{2007AJ....134.1330B,2016ApJ...827...25B,2012ApJS..201...19D}), 2MASS J13153094$-$2649513~\citep{2011ApJ...739...49B}, and 2MASS J22521073$-$1730134~\citep{2006AJ....132..891R}. We recover the L4+T3 spectral binary 2MASS~J0931+2802~\citet{2014ApJ...794..143B} outside our $25\,pc$ sample. We identify two previously unreported spectral binary candidates in our spectral sample, both of which lie formally outside our 25\,pc distance limit:

\indent \emph{2MASS~J14111847+2948515.} Its spectrum shows a deep $H$-band dip at $1.62\,\mu$m, and an angled $J$-band peak at $1.25\,\mu$m, both signs of a hidden T-dwarf companion. The $K_s$-band of the object is slightly fainter compared to the binary template, which could be an indication of slightly blue L dwarf, known contaminants to the spectral binary technique. However, the best single fits to its SpeX spectrum fail to reproduce the dip in the $H$-band, and are fainter in $J$ and $K_s$-bands in comparison to 2MASS~J1411+2948. Its component spectral types are likely to be L4+T4. No parallax has been measured for this source, whose distance would be larger than the estimated spectrophotometric distance of $49\pm6$\,pc if it is a binary.

\indent \emph{2MASS~J14211873$-$1618201.} The spectrum of this source shows an angled $J$-band peak and a small dip in the $H$-band. Its inferred component spectral types are M8+T5, similar to 2MASS~J0320$-$0446~(\citealt{2008ApJ...678L.125B,2008ApJ...681..579B}) and 2MASS~J0006$-$0852~\citep{2012ApJ...757..110B}, and WISE J0720$-$0846~(\citealt{2014AandA...561A.113S,2015AJ....149..104B}). Our strict distance cut left this source outside of the $25\,pc$ sample, yet it rests right at the 25\,pc limit ($d_t = 25.15\pm0.14$\,pc;~\citealt{2018AandA...616A...1G}).

To calculate the frequency of spectral binary systems, we used the definition of~\citet{1993AandA...278...81R} (See Section~\ref{sec:binaries}), where the binary fraction is the number of binaries over the total number of systems. For this calculation, we only consider systems with a measured SpeX spectrum, since otherwise we would not be able to assess spectral binarity\footnote{I.e. spectra of secondaries are not counted in the calculation since we are only concerned with the number of systems, neither do spectra of UCD components of higher order systems.}. Since 2MASS~J1421$-$1618 lies at our limit distance, we calculate two spectral binary fractions, assuming 24\,pc (5 spectral binaries/282 spectra) and 26\,pc (6 spectral binaries/312 spectra) volumes. The fractions are $1.7^{+0.9}_{-0.7}$ and $1.9^{+0.8}_{-0.7}$ for 24\,pc and 26\,pc, respectively, or an average of $1.8^{+0.6}_{-0.5}\%$ assuming Poisson errors. This fraction is significantly lower than the total fraction of resolved binaries in the sample ($7.5^{+1.6}_{-1.4}\%$, see Section~\ref{sec:binaries}), but this is likely because spectral binary systems encompass a specific range of component spectral types to be selected. We analyze the spectral binaries in this sample and their implication for the brown dwarf binary fraction in a companion paper.

\begin{figure*}
\figurenum{16}
\centering
\includegraphics[width=0.45\textwidth]{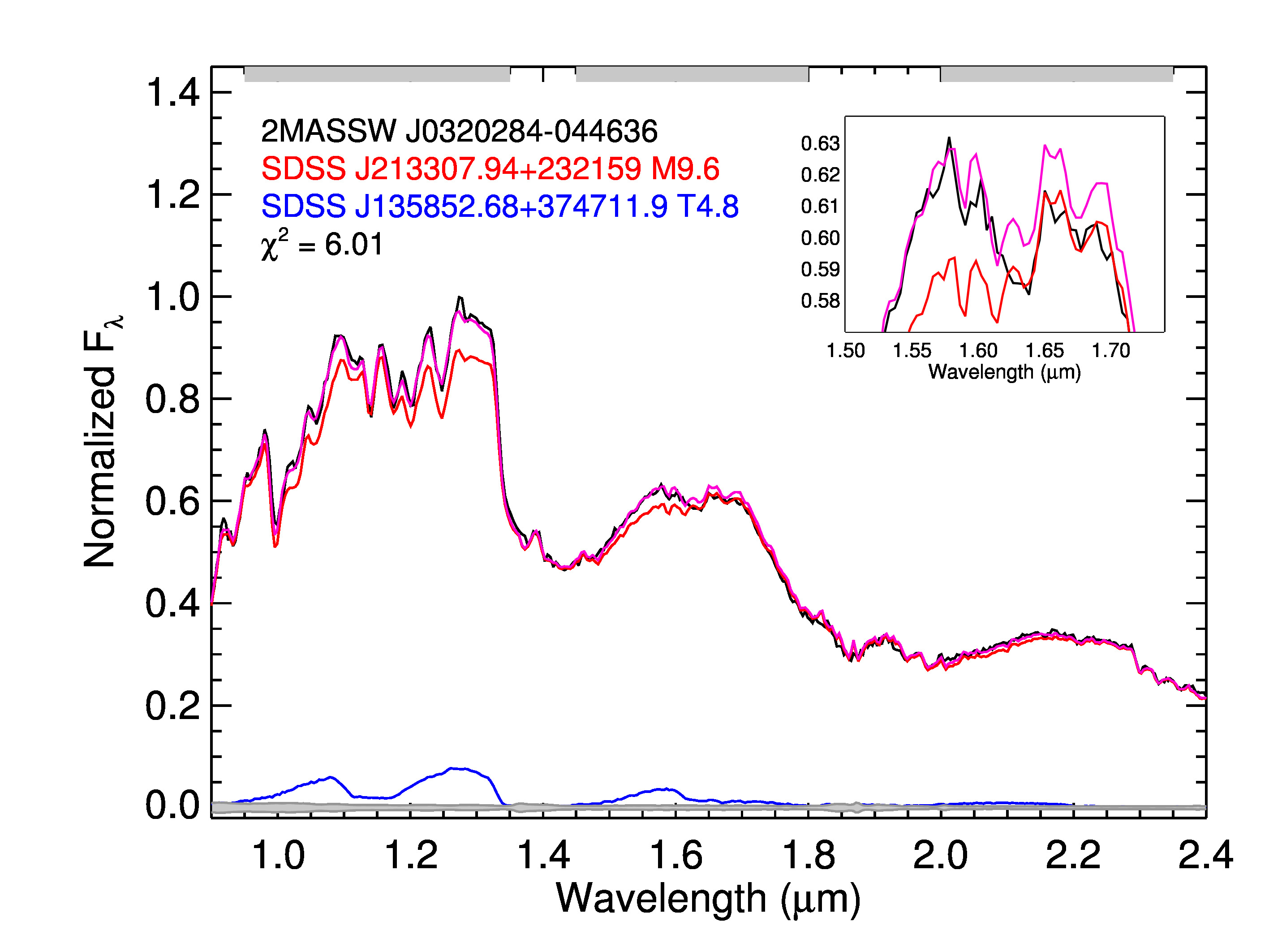}\includegraphics[width=0.45\textwidth]{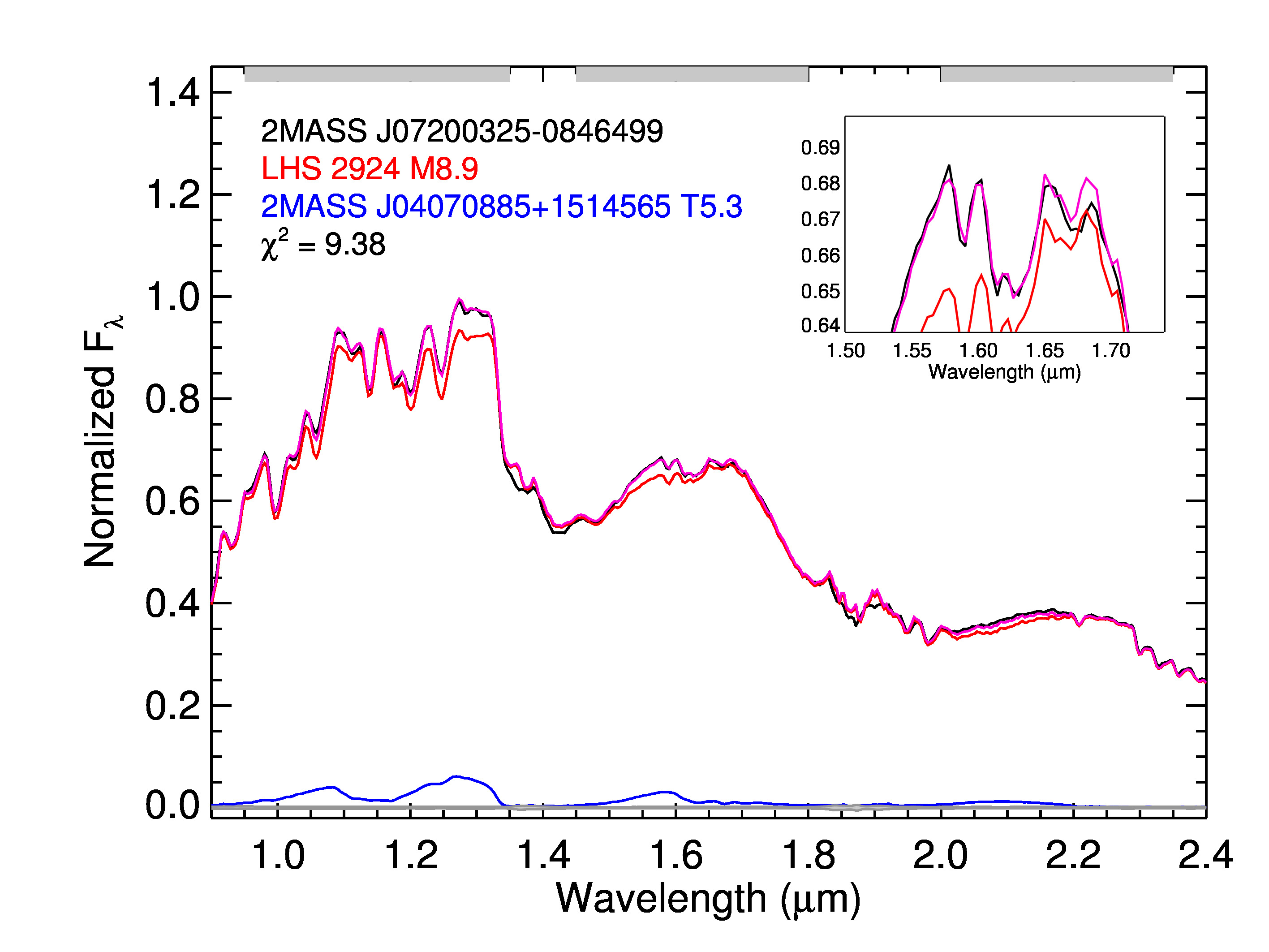}
\includegraphics[width=0.45\textwidth]{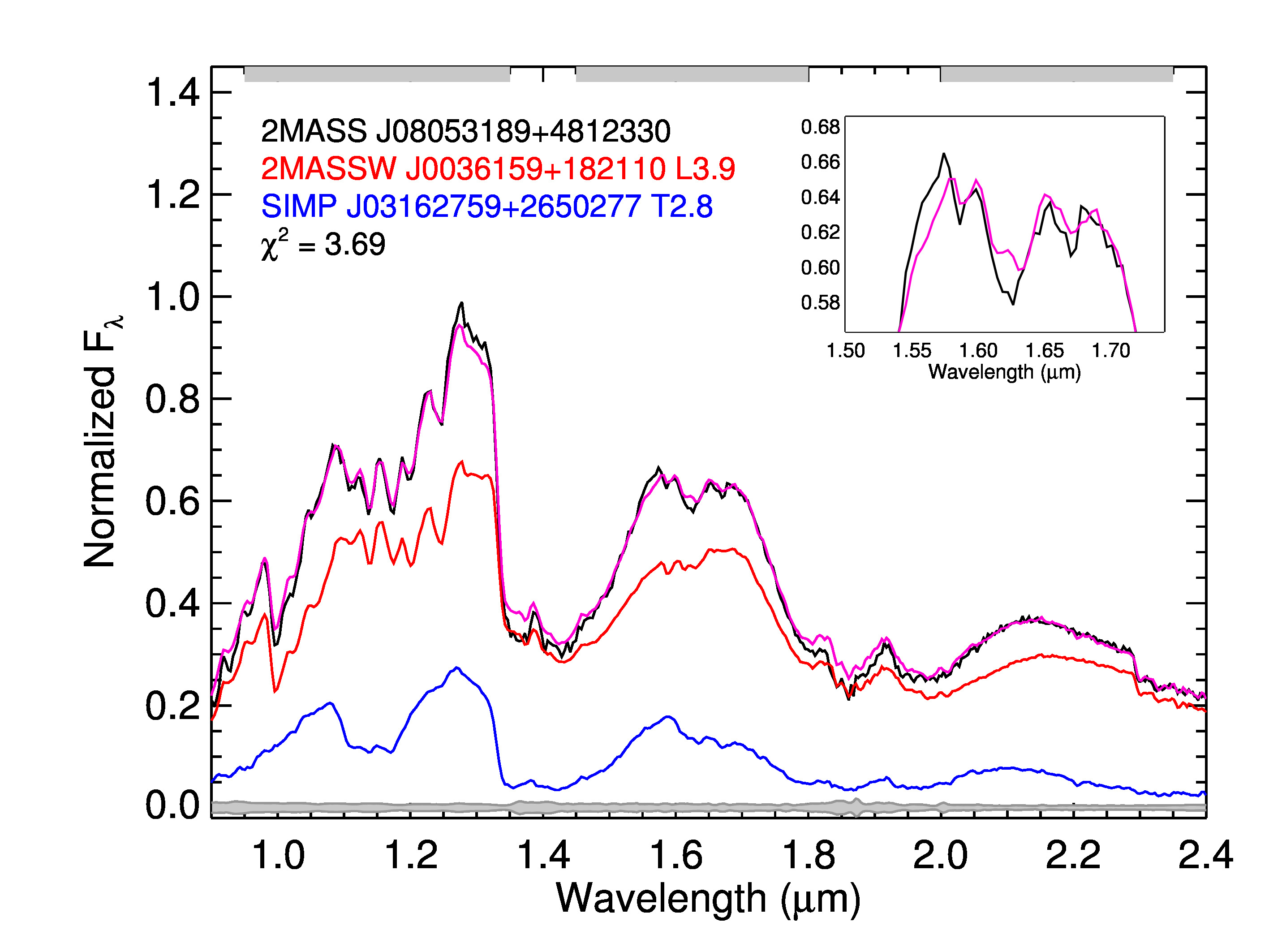}\includegraphics[width=0.45\textwidth]{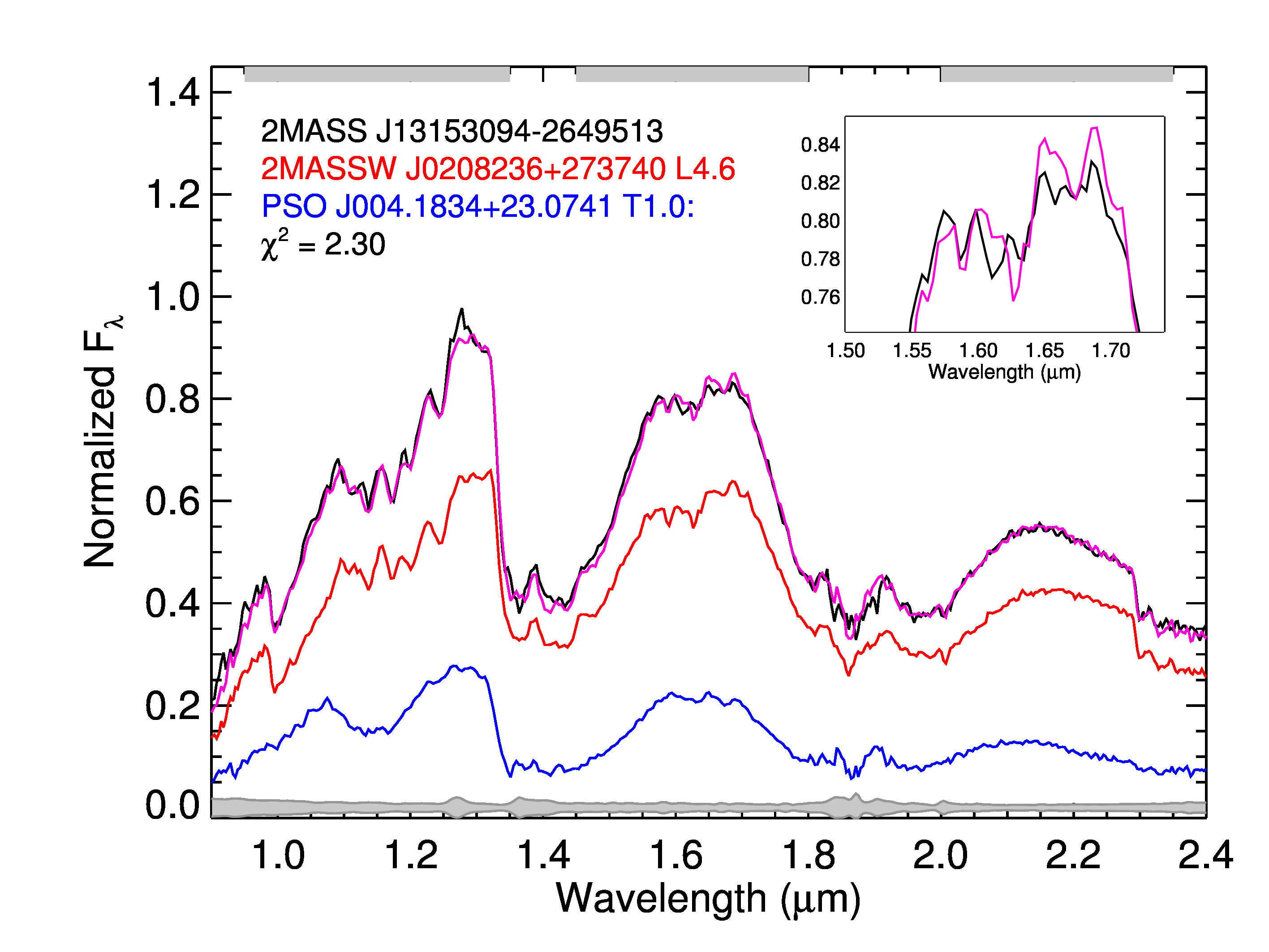}
\includegraphics[width=0.45\textwidth]{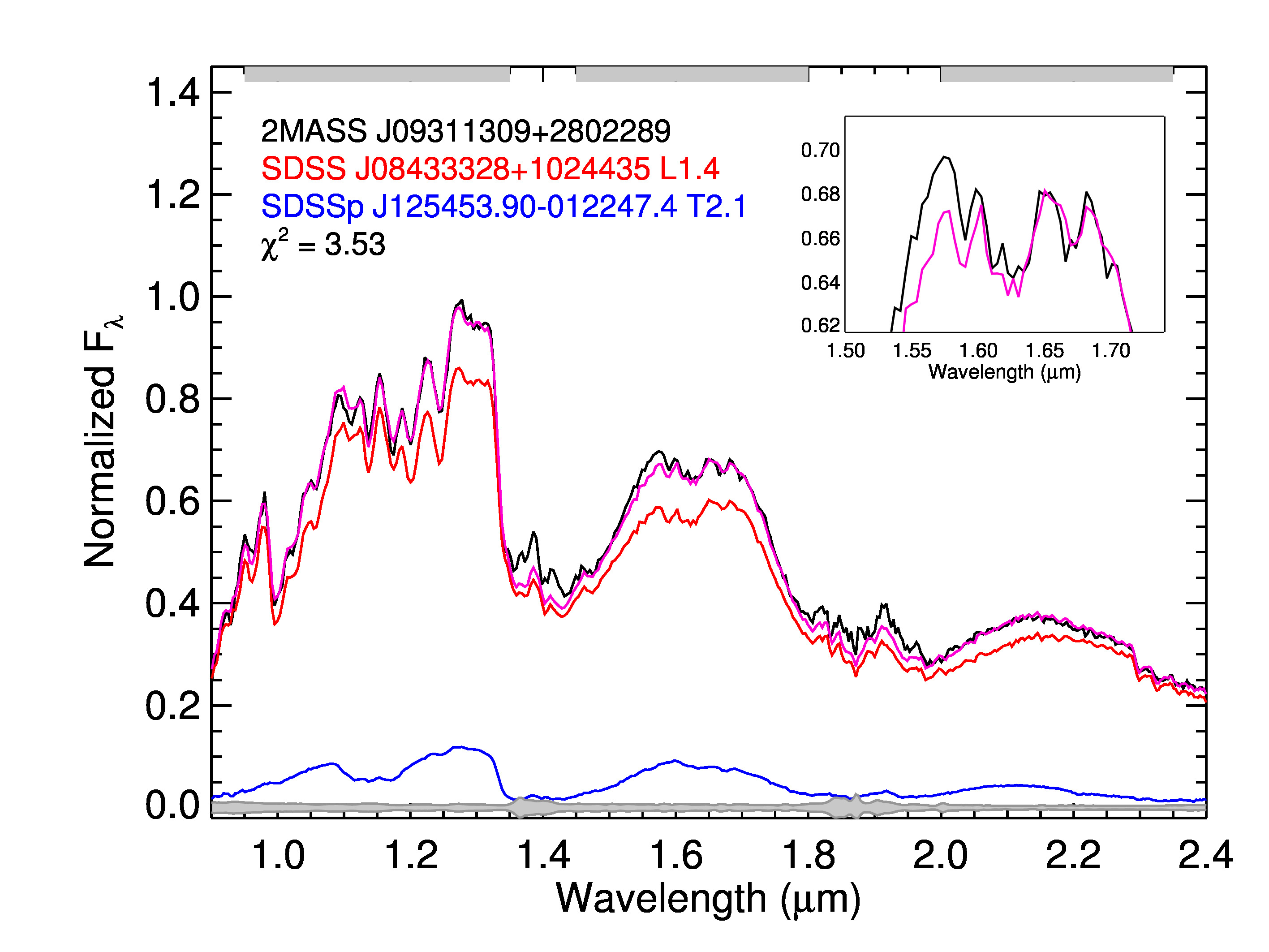}\includegraphics[width=0.45\textwidth]{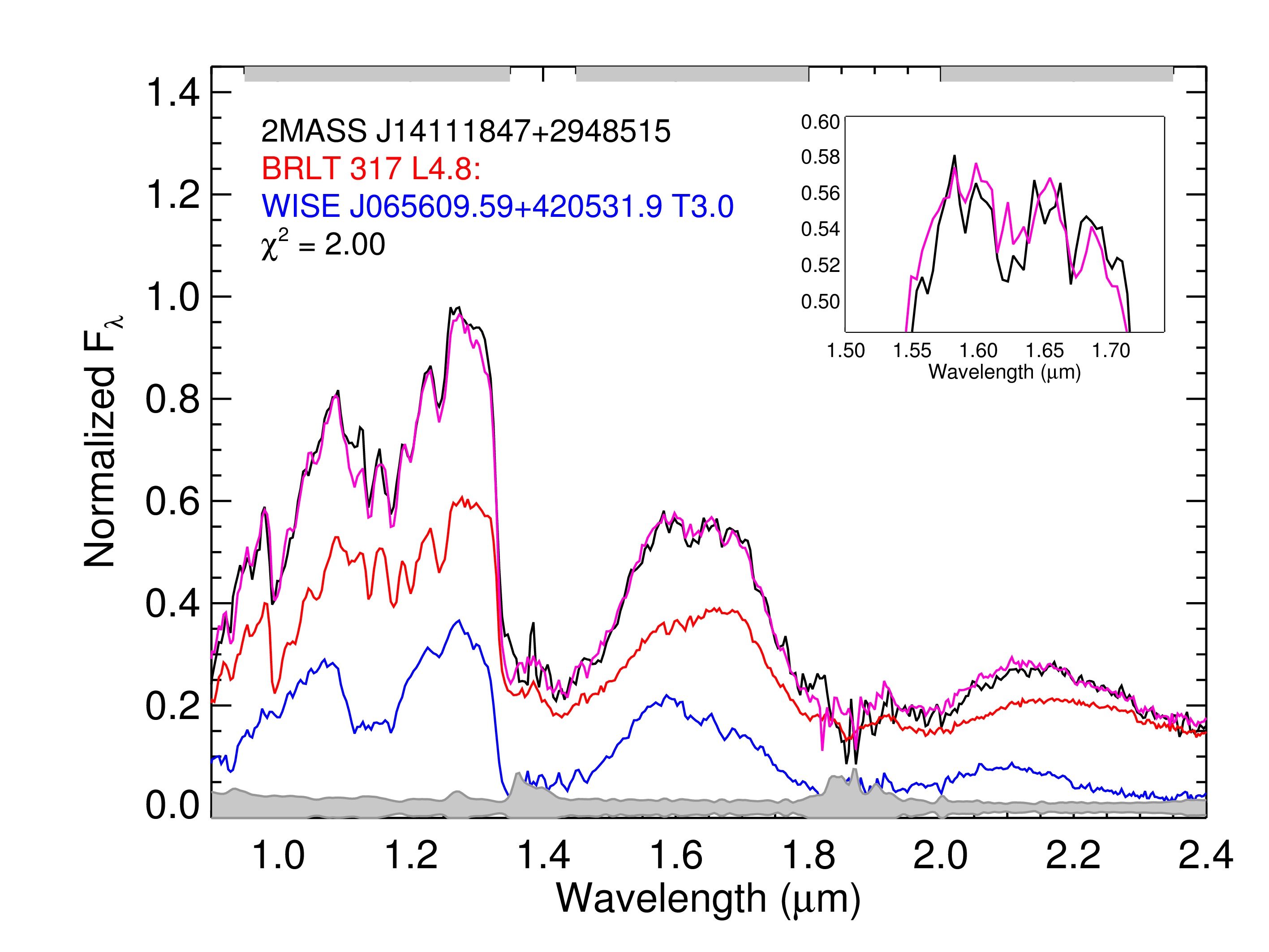}
\includegraphics[width=0.45\textwidth]{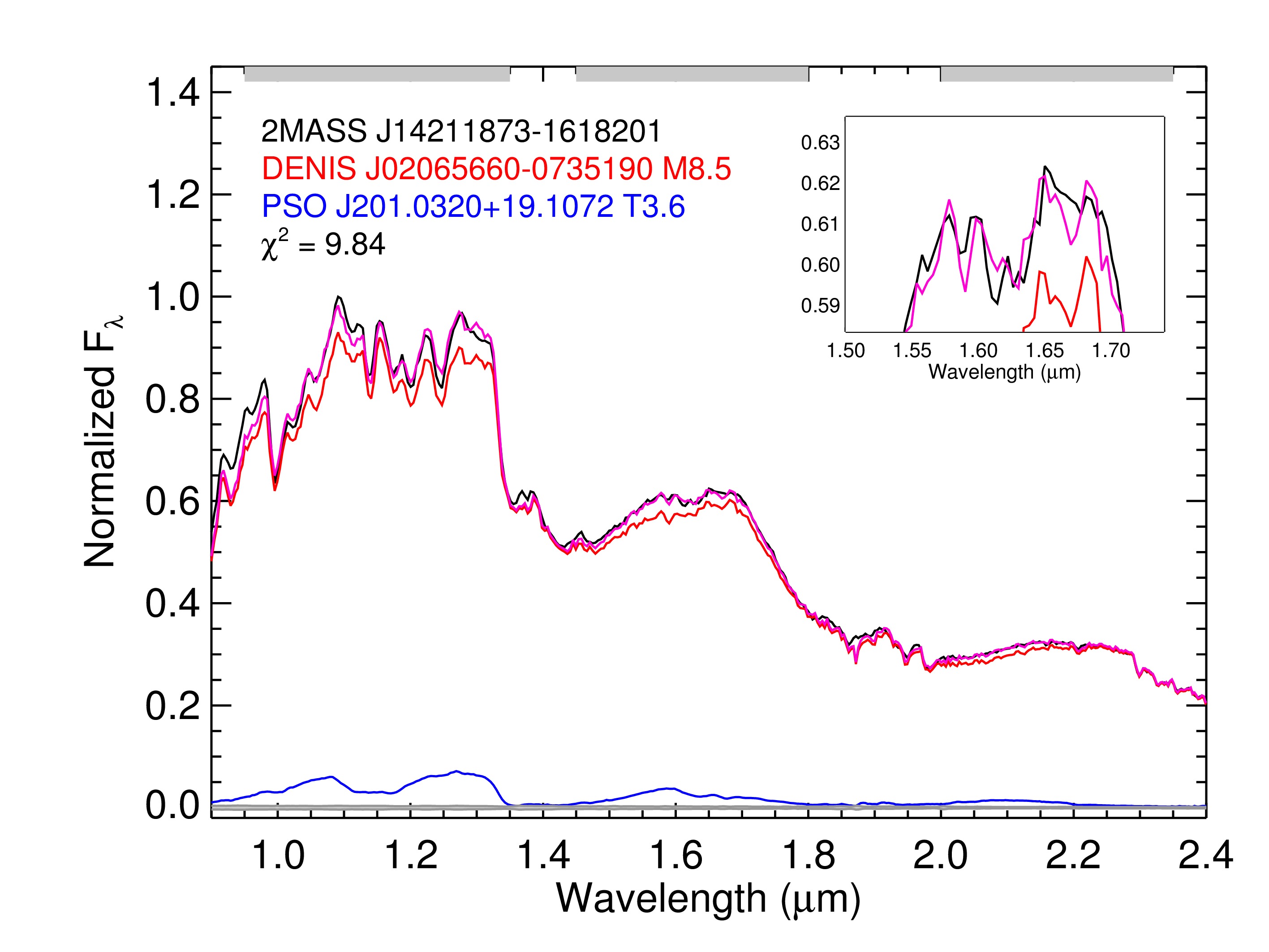}\includegraphics[width=0.45\textwidth]{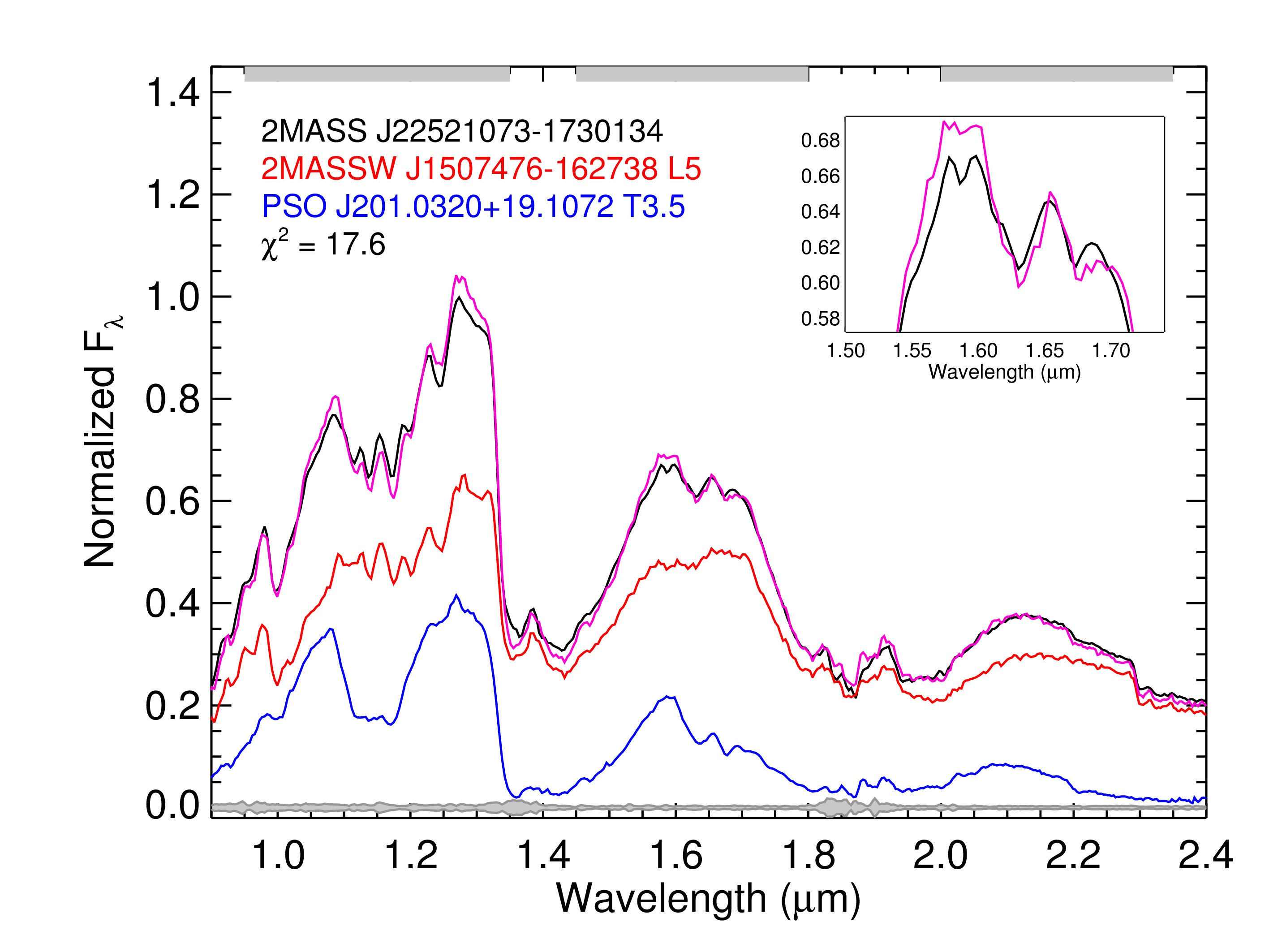}\\
\caption{Best fit templates to spectral binaries with M7$-$L5 primaries with a confidence $>90\%$. 2MASSW J0320$-$0446, WISE~J0720$-$0846,  2MASS J0805+4812, 2MASS J1315$-$2649, 2MASS J2252$-$1730 are all within 25\,pc, whereas 2MASS J0931+2802, 2MASS J1411+2948, and 2MASS J1421$-$1618 are outside 25\,pc. All the spectral binary candidates in the 25\,pc sample have already been confirmed as true binaries.\label{fig:SBfits25pc}}
\end{figure*}


\subsection{Binary systems containing UCDs in the 25\,pc volume}\label{sec:binaries}


Binaries and multiple systems reported in the literature were identified in our sample through crossmatches with the Washington Double Star Catalog\footnote{Eight matches to the WDS were ruled out in the notes from the Sixth Catalog of Orbits of Visual Binary Stars, found at~\url{https://ad.usno.navy.mil/wds/orb6/orb6notes.html}. 2MASS J0355+1603 was refuted as a binary in~\citet{2013AJ....145....2F}, and 6 other sources are only binary candidates, so are not considered in our binary statistics.}~(WDS;~\citealt{2001AJ....122.3466M}), SIMBAD~\citep{2000AandAS..143....9W}, and \url{vlmbinaries.org}.
Table~\ref{tab:binaries} lists the UCD binaries with primary components between M7$-$L5 found in our sample previously reported in the literature, as well as UCD companions to main sequence stars. Our $25\,pc$ sample contains 410 objects in 393 systems, 341 single systems, 42 binary systems, and 10 triple systems. Only 28 binaries and no triples have a primary with a spectral type M7 or later. Including the $1\sigma$ sample, we find 4 more binaries and one quintuple system, HD 114762, comprised of Aa, Ab, and Ac components F9+F8+F4 stars, an 11$\pm$0.1\,M$_{J}$~\citep{2011ApJ...735L..41K}  brown dwarf orbiting the F9 star~\citep{1989Natur.339...38L}, and an M6:: dwarf as the B component 130\,AU away from the F triple system~\citep{2002ApJ...581..654P}. 

We calculate several statistics to represent the multiplicity of the sample: the multiplicity fraction, which provides the probability that a given source is a multiple system; the companion star fraction, which is the probability for an object to be in a multiple system; the pairing factor, which is the mean number of companions per primary; and the companion frequency that indicates the mean number of companions per object. These equations are defined and explained in detail in~\citet{1993AandA...278...81R} and~\citet{2004AandA...423..169G}. Since we have no triple systems with primaries M7 or later, our multiplicity fraction is effectively a binary fraction. We determine the binary fraction of the $25\,pc$ sample to be $7.5^{+1.6}_{-1.4}\%$, including both spectral binaries and RV variable systems. The companion star fraction for this sample is $14.1^{+2.1}_{-1.9}\%$, the pairing factor is $1\pm0.3$, since there are no triple systems with primaries $\geq$M7, and the companion frequency is $0.14\pm0.02$ companions per object (following the definition of~\citealt{2004AandA...423..169G}).

\begin{figure}
\figurenum{17}
\centering
\includegraphics[scale=0.5]{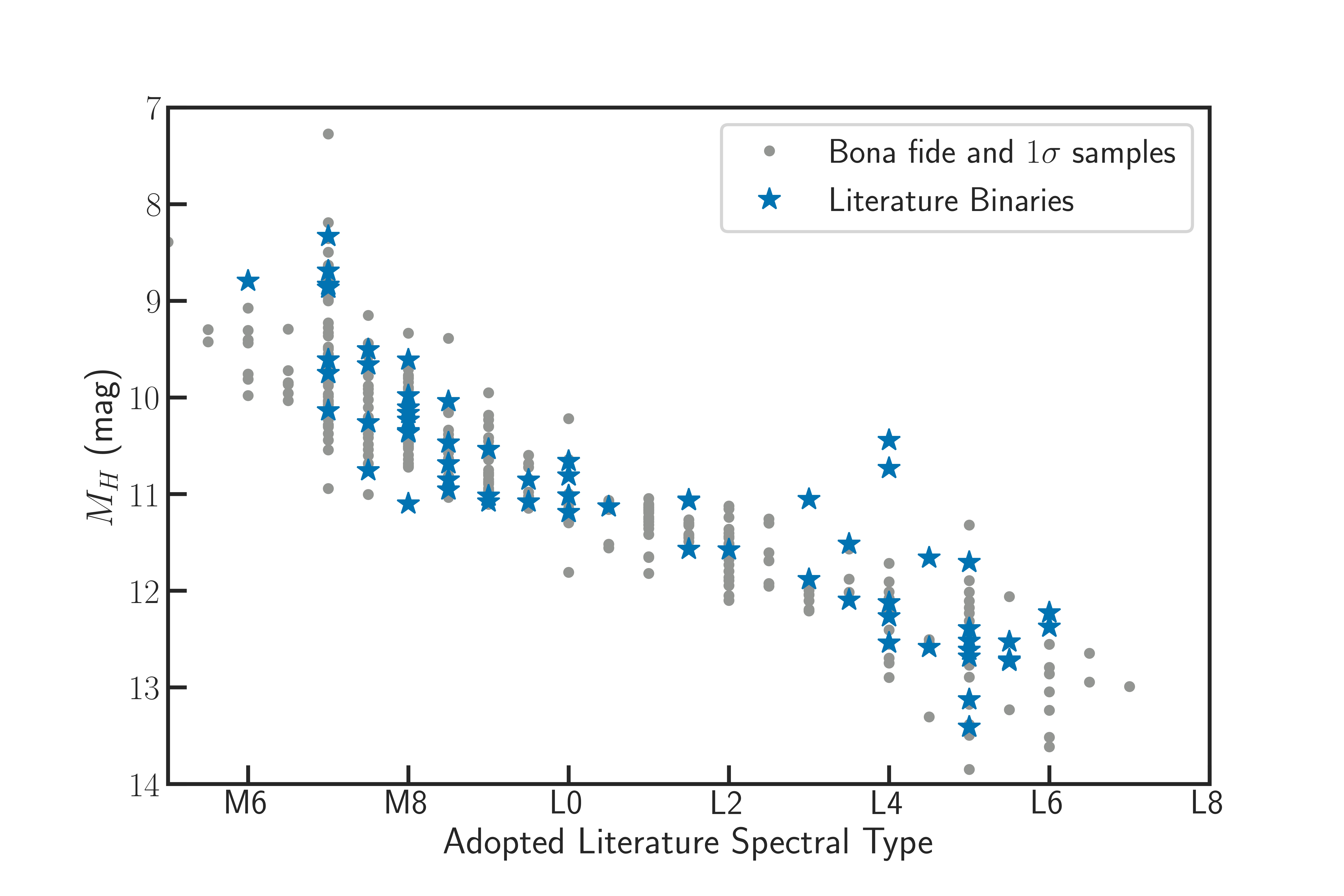}
\caption{Adopted literature spectral type vs.\ 2MASS $H$ absolute magnitude for our extended $1\sigma$ sample highlighting the UCD binary systems reported in the literature. Most binaries in this plot have resolved absolute magnitudes, and thus their individual components look normal. The two L4 dwarfs well above the sequence are HD~130948B and C are companions to the young F9 variable star~\citep{2002ApJ...567L..59G}, known to be overluminous on color-magnitude diagrams~\citep{2016ApJS..225...10F}. \label{fig:Habs}}
\end{figure}

\begin{figure}
\figurenum{18}
\centering
\includegraphics[scale=0.5]{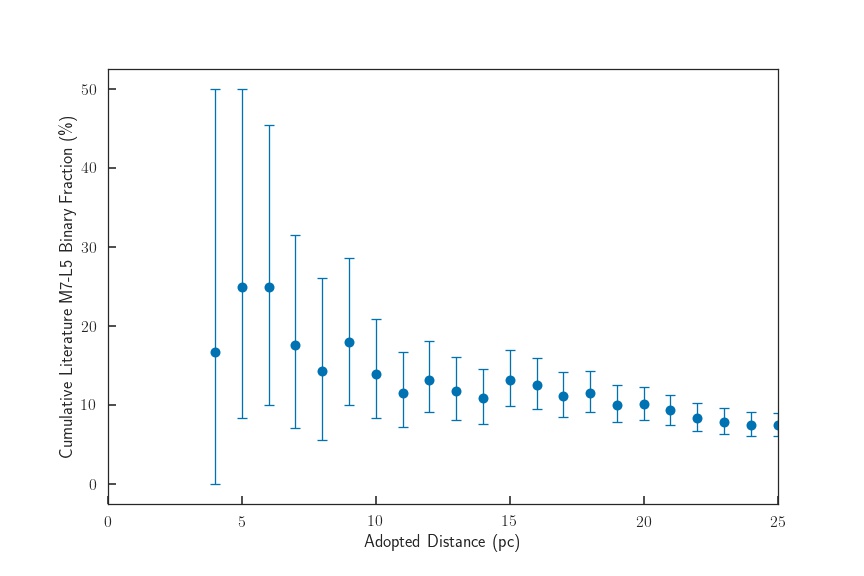}
\caption{Cumulative binary fraction as a function of distance.\label{fig:bfdist}}
\end{figure}

\begin{figure}
\figurenum{19}
\centering
\includegraphics[scale=0.5]{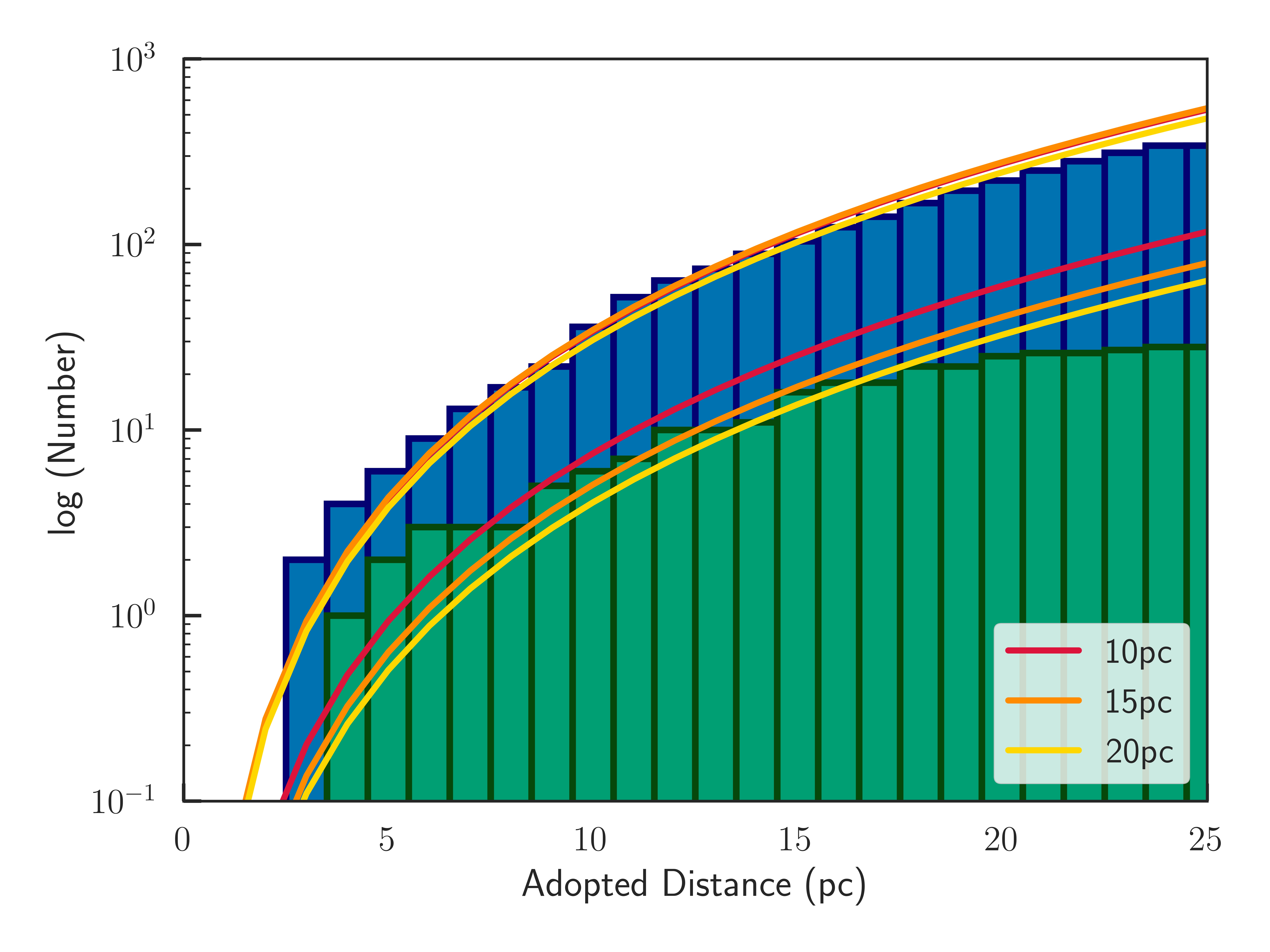}
\caption{Cumulative histogram of sources per unit adopted distance. The full $25\,pc$ sample is shown in blue, and the binaries with primaries M7 or later are shown in green. Three curve fits are shown for each histogram, assuming completeness between $5-10\,$pc (red), $5-15\,$pc (orange), and $5-20\,$pc (yellow).\label{fig:bincomp}}
\end{figure}

Figure~\ref{fig:bfdist} shows the cumulative binary fraction as a function of distance. Out to a distance of 9\,pc, the binary fraction oscillates around $13-25\%$, and at larger distances it begins to drop and settle around $\sim7\%$. The resolved UCD binary fraction has been thoroughly studied (e.g.,~\citealt{2003AJ....126.1526B,2007ApJ...659..655B}) leading to $\sim10-20\%$ for separations $>1$\,AU, while sub-AU systems comprise $1-4\%$ of the population~(\citealt{2010ApJ...723..684B,2007ApJ...668..492A}). However, this is the first time the UCD binary fraction has been calculated in a volume-limited sample\footnote{In their 6.5\,pc volume literature study,~\citet{2016AandA...589A..26B} identify 48.5\% of stars and 15.4\% of brown dwarfs as part of multiple systems, but no combined UCD fraction.}, and as seen in Figure~\ref{fig:Habs}, there may be a significant fraction of overluminous binaries that have not been confirmed by high resolution imaging, astrometry, or RV monitoring yet. Additionally, in the previous Section we found that 5 out of the 25 binaries within 25\,pc are spectral binaries. Since spectral binaries require specific combinations of spectral types to be identified as such, we do not expect them to dominate the binary detection yield. Yet in this study, $\sim20\%$ of our binaries are spectral binaries, supporting our hypothesis that the population of binaries in the 25\,pc sample literature is incomplete. The incompleteness of binaries is shown in Figure~\ref{fig:bincomp} as a cumulative histogram over distance which flattens beyond 20\,pc compared to the general $25\,pc$ sample. Fitting curves to the $5-10\,$pc, $5-15\,$pc, and $5-20\,$pc regions, and extrapolating to 25\,pc, we estimate a large binary incompleteness of 76\%, 65\%, or 56\%, respectively.

\section{Selection and Luminosity Functions}\label{sec:selfunc}

The luminosity function measures the number density of sources as a function of luminosity, or equivalently, absolute magnitude, temperature, or spectral type. For main sequence stars there is generally a one-to-one mapping between luminosity and mass functions; for UCDs, because brown dwarfs cool as they age, there is not a one-to-one mapping between a brown dwarf luminosity function and a brown dwarf mass function. However, the luminosity function is the initial crucial measurement towards a fundamental understanding of low mass star and brown dwarf formation through a field present-day mass function. The luminosity function of UCDs covering the M7$-$L5 spectral type range, has been most notably measured by~\citet{2007AJ....133..439C}, hence here we provide an updated reevaluation. 

\subsection{Area Coverage}\label{sec:area}

The area covered by our spectral survey is limited by the declinations accessible from IRTF, roughly $-50^{\circ} <\delta<+67^{\circ}$. Additionally, our survey suffers from an inherent incompleteness of sources in the Galactic plane. We therefore restrict our analysis to the area of sky outside $-15^{\circ} < b < +15^{\circ}$ and within $-50^{\circ} <\delta<+67^{\circ}$ which corresponds to an area of 26,051.54\,$deg^2$, or 63.2\% of the sky. 


Bright stars reduce the total available sky area by obscuring patches of sky where a UCD could otherwise be found. To account for this effect, we drew one million sources from our sample and reassigned them to random coordinates within our observable area. This list was crossmatched with the 2MASS catalog using TOPCAT with a $5\farcs0$ radius, returning 22,126 matches. Of these, 2,345 stars were as bright or brighter than the simulated input targets within the search radius, thus effectively obscuring nearby UCD. Accounting for this effect reduces the effective observable sky by 0.15\% to 25990.45\,$deg^2$.~While we note that 0.5\% of the sky is obscured by bright stars and excluded from the 2MASS survey\footnote{See 2MASS Explanatory Supplement,~\url{https://old.ipac.caltech.edu/2mass/releases/second/doc/}}, we do not take it into account in our calculations, since our sources also come from optical and mid-infrared surveys.

\subsection{Volume Completeness}\label{sec:volcomp}

\begin{figure*}
\figurenum{21}
\centering
\includegraphics[scale=0.5]{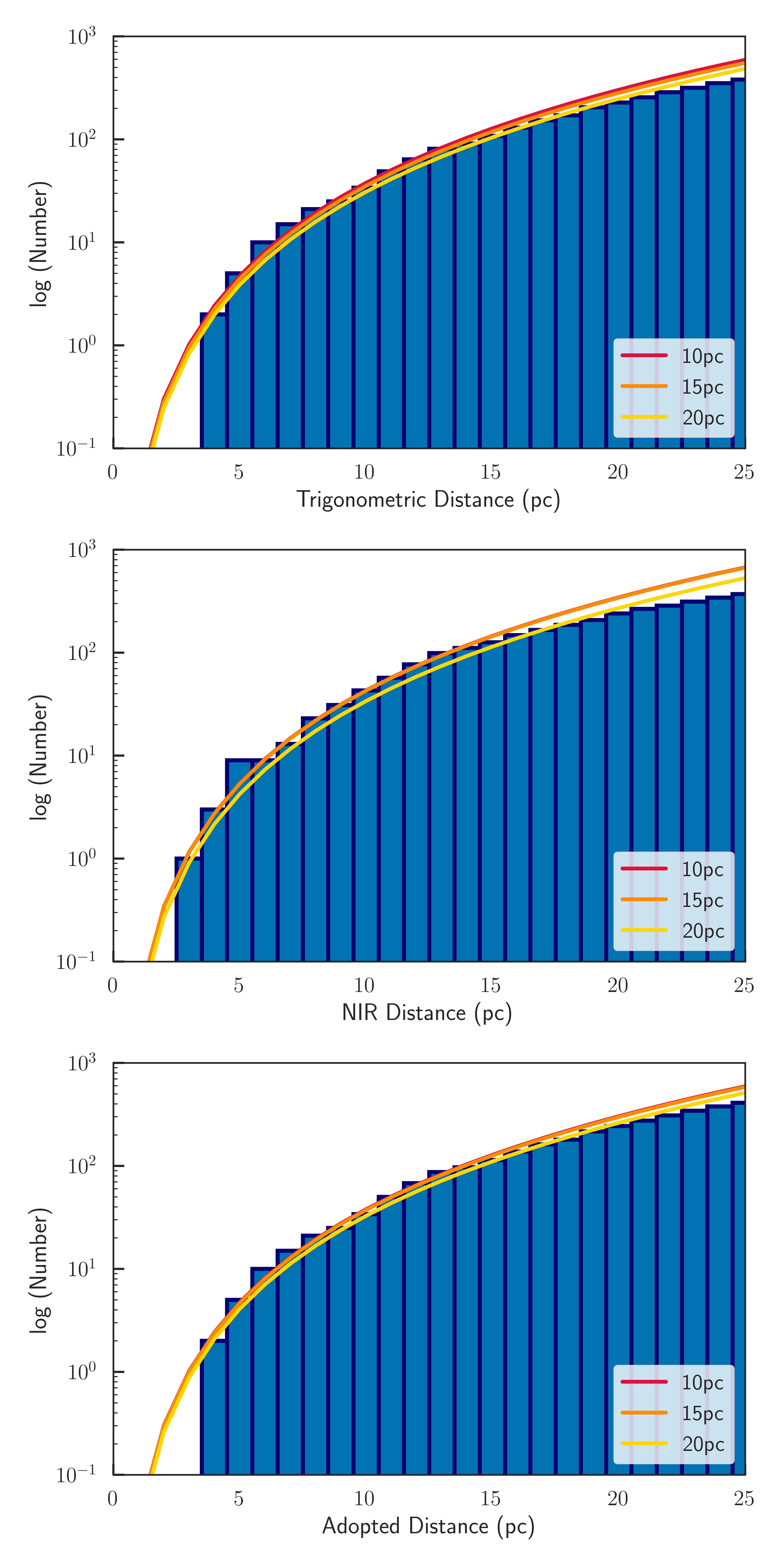}
\caption{Cumulative distance histograms for trigonometric, spectrophotometric, and adopted distances. The red, orange and yellow curves show the cube fit to the histograms in blue up to 10\,pc, 15\,pc and 20\,pc, including their Poisson uncertainties.\label{fig:distcumhists}}
\end{figure*}

A volume within 25\,pc around the Sun is well embedded within the thin disk of the Galaxy (scale height $\sim300$\,pc;~\citealt{1991ApJ...378..131K,2010AJ....139.2679B}), and therefore should be relatively uniform in density. Assuming a uniform distribution of sources, the cumulative number of objects should increase with distance following an $r^3$ relation. We estimate our volume completeness in trigonometric, spectrophotometric, and adopted distances by fitting power law curves to the cumulative distribution of sources between $5-10$\,pc, $5-15$\,pc and $5-20$\,pc, assuming completeness in those ranges, considering Poisson uncertainties (Figure~\ref{fig:distcumhists}), and extrapolating expected numbers to 25\,pc. The ratio of number of objects in our sample to expected number is used to estimate our completeness. These values are summarized in Table~\ref{tab:comp}.
\begin{figure*}
\figurenum{20}
\centering
\includegraphics[scale=0.6]{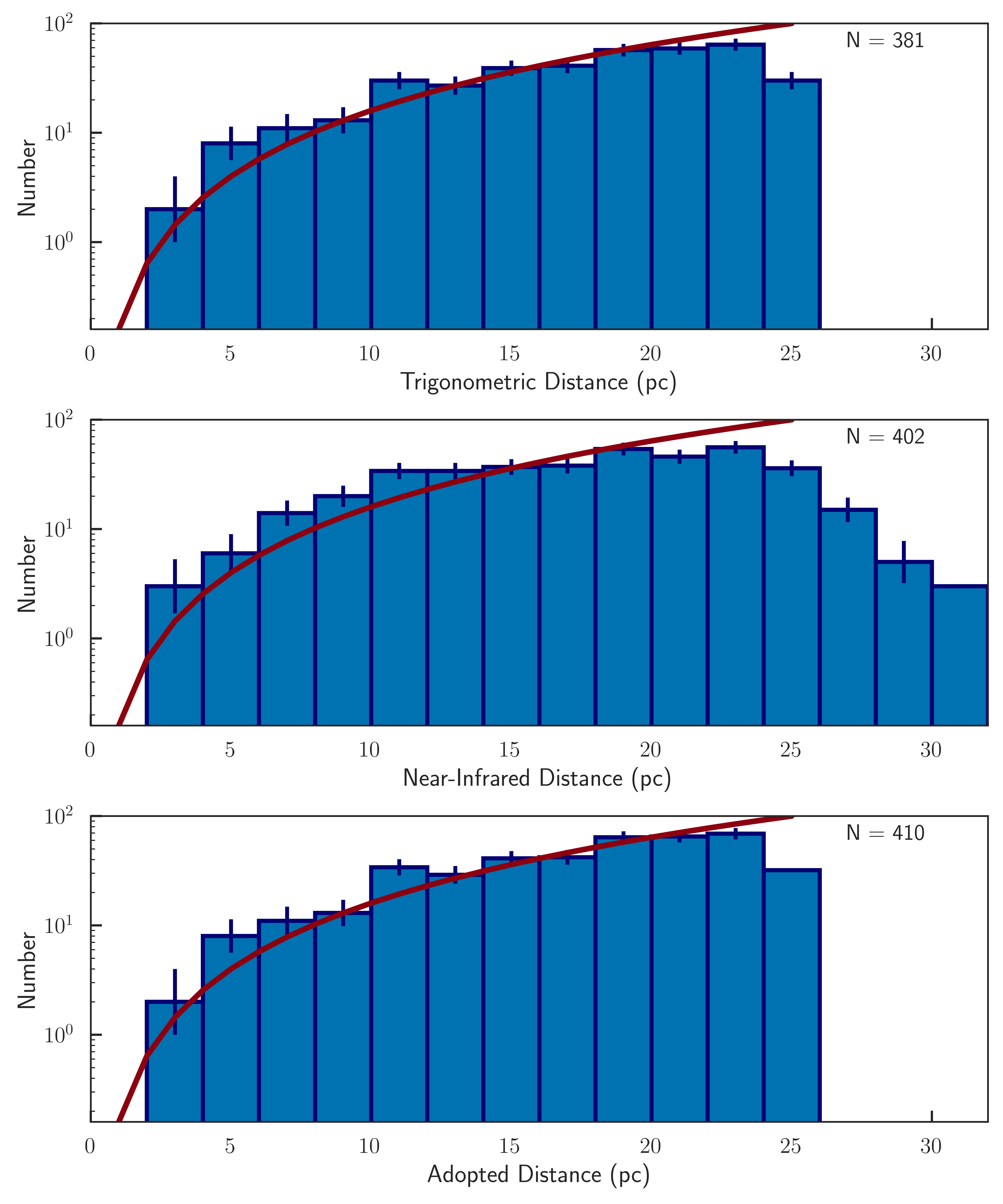}
\caption{Distributions of trigonometric \emph{(top)}, spectrophotometric \emph{(middle)} and adopted distances \emph{(bottom)}. Solid line is an $r^2$ fit normalized at the 25\,pc bin. Note the drop off in the largest distance bins, which reflects incompleteness likely due to brightness limits and selection biases.\label{fig:disthists}}
\end{figure*}

\begin{deluxetable}{c|cc|cc}
\tabletypesize{\scriptsize}
\tablenum{12}
\tablecaption{Estimated volume completeness.}\label{tab:comp}
\tablehead{ & \multicolumn{2}{c}{Predicted Numbers} & \multicolumn{2}{c}{Completeness}\\
\cline{2-3}
\cline{4-5}
\colhead{Fit Range (pc)} & 
\colhead{Trigonometric} &  
\colhead{Adopted Distance} &
\colhead{Trigonometric} &  
\colhead{Adopted Distance}}
\startdata
\sidehead{\emph{$25\,pc$ sample (N = 410)}}
$5-10$ & 592 & 592 &  $64^{+8}_{-7}\%$  & $69^{+9}_{-8}\%$\\ 
$5-15$ & 552 & 583 & $69^{+9}_{-8}\%$  & $70^{+9}_{-8}\%$\\
$5-20$ & 484 & 511 & $79^{+9}_{-8}\%$  & $80^{+9}_{-8}\%$\\
\hline
\sidehead{\emph{$25\,pc$ M dwarfs (N = 223)}}
$5-10$ & \nodata  & 509 & \nodata  & $44^{+7}_{-6}\%$\\
$5-15$ & \nodata & 357 & \nodata & $62^{+8}_{-7}\%$\\
$5-20$ & \nodata  & 283 & \nodata  & $78^{+9}_{-8}\%$\\
\hline
\sidehead{\emph{$25\,pc$ L dwarfs (N = 187)}}
$5-10$ & \nodata  & 83  & \nodata  & $226^{+16}_{-15}\%$\\
$5-15$ & \nodata  & 226  & \nodata & $83^{+10}_{-9}\%$\\
$5-20$ & \nodata  & 228  & \nodata & $82^{+10}_{-9}\%$\\
\enddata
\end{deluxetable}

The completeness of late-M dwarfs is lower than that of L dwarfs. Using the $5-15$\,pc fit, which is a good trade-off between completeness and sample size, our sample contains $62^{+8}_{-7}\%$ of the late-M dwarfs within 25\,pc, and $83^{+10}_{-9}\%$ of L dwarfs. Late-M dwarfs may have been missed in previous surveys, due to color-selection biases designed to exclude more numerous and brighter mid-M dwarfs, as indicated by~\citet{2015AJ....149..158S}. While most L dwarfs in the Solar neighborhood have already been identified in previous searches, many may be hidden in crowded areas like the Galactic plane (e.g. the L8 dwarf recently identified at 11\,pc;~\citealt{2018ApJ...868...44F}). From the trigonometric distances, we estimate our total sample completeness to be between $64-79\%$. Including spectrophotometric distances when parallaxes are not available, the sample completeness is between $69-80\%$, but we adopt the value for the $5-15$\,pc fit,  $70^{+9}_{-8}\%$. This completeness value is used in Section~\ref{sec:LF} to scale the corrected number of sources in the 25\,pc volume when measuring the luminosity function (see Equation~\ref{eq:Ncorr}). We expect most of the incompleteness to come from missing sources beyond 20\,pc, as seen in Figure~\ref{fig:disthists}, possibly including sources in the Galactic plane, the southern hemisphere, or UCD candidates recently identified in~\citet{2018arXiv180908244R} in need of spectroscopic validation.


\begin{figure*}
\figurenum{22}
\centering
\includegraphics[scale=0.5]{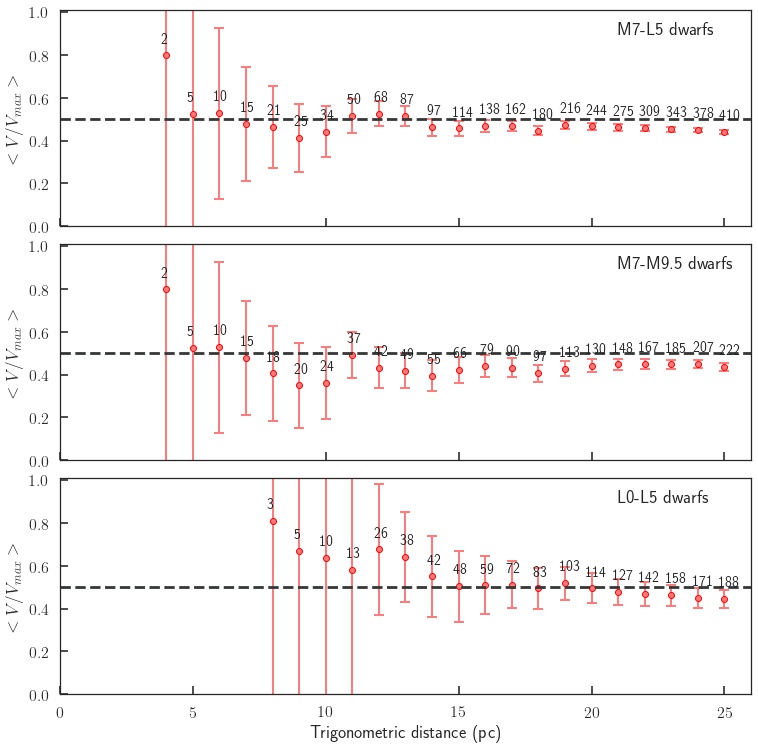}
\caption{Average $\langle V/V_{max}\rangle$ values for our $25\,pc$ sample, and also for subsamples of M and L dwarf with uncertainties calculated as described in~\citet{2019ApJS..240...19K}. The numbers indicate the cumulative number of sources counted up to that distance. We used the adopted distances for this calculation.\label{fig:vvmax}}
\end{figure*}

Additionally, we estimate $\langle V/V_{max}\rangle$ averages suggested by~\citet{1968ApJ...151..393S} to evaluate the homogeneous spatial distribution of our sample. $\langle V/V_{max}\rangle$ measures the number of sources in each half of a given volume, approaching 0.5 for a uniformly distributed sample with equal counts on each half-volume. Figure~\ref{fig:vvmax} shows the distribution of $\langle V/V_{max}\rangle$ values. Uncertainties are calculated as $0.5-\frac{n/2 - a_{max}}{n}$, where $a_{max}$ is the distance at which the value of $\langle V/V_{max}\rangle$ last equals 0.68 (4\,pc for the full sample, and M dwarfs only, and 8\,pc for L dwarfs), corresponding to one Gaussian standard deviation. For M dwarfs, the largest distance at which $\langle V/V_{max}\rangle$ approximates 0.5 is 13\,pc, suggesting incompleteness of M7$-$M9.5 dwarfs at larger distances. Conversely, L dwarfs have $\langle V/V_{max}\rangle$ consistent with 0.5 up until 25\,pc, indicating a homogeneous distribution of L0$-$L5 dwarfs in our sample. 

\subsection{Sample Simulation}

Compiling a sample of objects starting from past literature compilations leads to a complicated selection function. Rather than determining the selection function of each selection process separately, we simulate a sample of UCDs in a volume larger than 25\,pc, including unresolved binaries, and apply selections based on our spectral type and distance cuts, from both parallaxes and spectrophotometric estimates. This procedure aims to measure systematic effects in the sample construction.

We simulate $10^6$ UCDs, assigning distances drawn from a uniform spatial distribution out to 50\,pc. We calculate ``true'' parallaxes by inverting the distances. An underlying spectral type distribution was derived by population simulations (c.f.~\citealt{2004ApJS..155..191B}) using the~\citet{2005ASSL..327...41C} IMF, a uniform age distribution, the~\citet{2001RvMP...73..719B} evolutionary models, and the effective temperature to spectral type empirical relations from~\citet{2013ApJS..208....9P}, which cover the full stellar and substellar spectral type range from O3 to Y2. From this distribution, $10^6$ ``true'' spectral types between M5$-$L7 were randomly drawn and assigned to our simulated UCD sources. 


We calculate absolute magnitudes empirically, from the simulated spectral types, using the following linear relations:
\begin{eqnarray}
M_{J} =& 0.37\times SpT + 4.29,~rms = 0.35\\
M_{H} =& 0.32\times SpT + 4.61,~rms = 0.29\\
M_{K_{S}} =& 0.29\times SpT + 4.67,~rms = 0.29
\end{eqnarray}
determined from a subset of 230 single M7$-$L5 dwarfs with parallax measurements, 2MASS magnitudes, not classified as VL-G, INT-G, unusually red, or unusually blue from our $25\,pc$ sample. The scatter in these relations is slightly smaller than in other empirical relations covering broader spectral type ranges~(e.g.~\citealt{2012ApJS..201...19D}; $\sigma=0.4$\,mag). To simulate the intrinsic brightness distribution of the population, we add offsets to these empirical absolute magnitudes, drawn from a Gaussian distribution centered at zero and scaled by the scatter in the empirical relations.
 


Parallax-limited and magnitude-limited samples are subject to different biases affecting the total number of included sample members. The Lutz-Kelker bias affects parallax-limited samples by allowing objects from outside a distance limit into the observed volume~\citep{1973PASP...85..573L}. For an observed parallax $\pi_0$, there is a range of true parallaxes $\pi_0\pm\delta\pi$ for normally-distributed measurement uncertainties. Assuming a uniform number density of stars, the number of objects per parallax bin will be proportional to $N_{*}\propto1/\pi^4$, implying that the number of stars increases as the parallax decreases, i.e. there are more objects in the volume outside a given distance than within. Subsequently, this means that more stars will appear to have smaller true parallaxes than their observed parallaxes, and that the average distance for sample members will be farther than the distance limit~\citep{1973PASP...85..573L}.  

In magnitude-limited samples, intrinsically brighter sources (i.e. on the high end of the absolute magnitude distribution) and unresolved binaries will be selected in larger numbers than intrinsically fainter sources, again due to the larger volume sampled by the brighter sources, an effect known as the Malmquist bias~\citep{1922MeLuF.100....1M}. Depending on the relative uncertainty in distance and magnitude measurements, and intrinsic scatter in the population, the effect from the Malmquist bias can be significantly larger than that of the Lutz-Kelker bias. Since our sample is defined by both trigonometric and spectrophotometric distances, both effects are significant in our calculations, although the Lutz-Kelker bias plays a more significant role given the large number of parallaxes in our sample (93\% of the sample).

\begin{figure}
\figurenum{23}
\centering
\includegraphics[width=\textwidth]{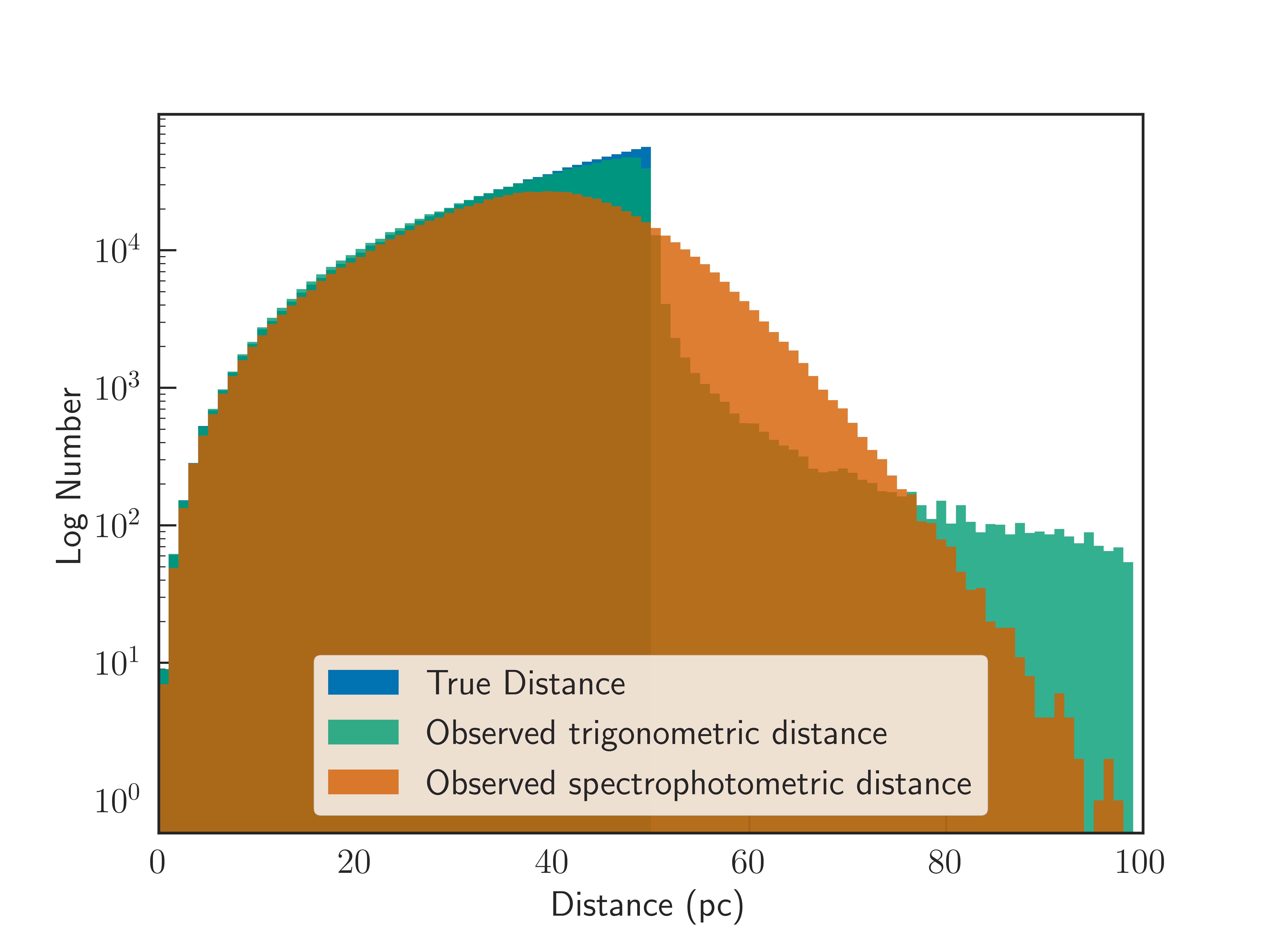}
\caption{True and observed distances from our simulation. The blue histogram shows the distribution of true distances, following an $r^3$ shape, defined up to 50\,pc. The green histogram shows the distribution of observed trigonometric distances, measured after a Gaussian uncertainty was added to the true parallax, with the scale of the distribution emerging from our sample's parallax uncertainty distribution. The orange histogram shows the distribution of the observed spectrophotometric distances, measured with spectral types, apparent magnitudes, and empirical absolute magnitude relations. This distribution is affected by the Malmquist bias, including sources located farther than the volume limit.\label{fig:distbias}}
\end{figure}

We model the Lutz-Kelker bias in our simulation by adding an uncertainty offset to our parallax measurements drawn from the uncertainty distribution of our observed parallaxes (see Figure~\ref{fig:distbias}). We excluded 2246 simulated sources with observed negative parallaxes.  We account for unresolved binarity by adding a magnitude offset to $20\%$ of stars in our simulated sample, the fraction based on estimates of the underlying UCD binary fraction~(\citealt{2003AJ....126.1526B,2003AJ....125.3302G,2007prpl.conf..427B}). We randomly assigned mass ratios from a power law distribution~($\propto q^{1.8}$;~\citealt{2007ApJ...668..492A}) to compute secondary masses. Effective temperatures, spectral types, and absolute magnitudes for the secondaries were obtained in the same manner as the primaries, resulting in combined system absolute magnitudes. Magnitude offsets ranged between $\Delta m = 0-0.75$\,mag\footnote{Systems with a magnitude offset larger than 0.75 (corresponding to an equal mass binary) occurred when the secondary was slightly brighter than the primary in any band as allowed by the added scatter, despite a larger primary mass. This is the case for 33,194 sources, 3.3\% of the simulated sample or 16.6\% of the simulated binaries. All of these systems were dropped from the simulation, resulting in 964,560 objects total.}. For simplicity, we assumed that the addition of flux to the simulated binaries does not affect the spectral type classification, which is likely true for late-M and early-L dwarf primaries but not necessarily for late-L+T dwarf systems~(\citealt{2004ApJ...604L..61C,2010ApJ...710.1142B}). The addition of magnitude offsets for simulated binaries, and uncertainties to the true absolute magnitudes for all simulated sources models the effects from the Malmquist bias.

To model observed spectral types, offsets were drawn from a Gaussian distribution with a standard deviation equal to 0.95 subtypes (see Section~\ref{sec:spts}). Apparent magnitudes were assigned based on the distance modulus and absolute magnitudes, adding an observational uncertainty drawn from a Gaussian distribution with a standard deviation following the same photometric error distribution from our literature sample. Observed parallaxes were modeled by adding a Gaussian uncertainty to the true parallaxes.



\subsection{Selection Function}

\begin{figure*}
\figurenum{24}
\centering
\includegraphics[width=0.49\textwidth]{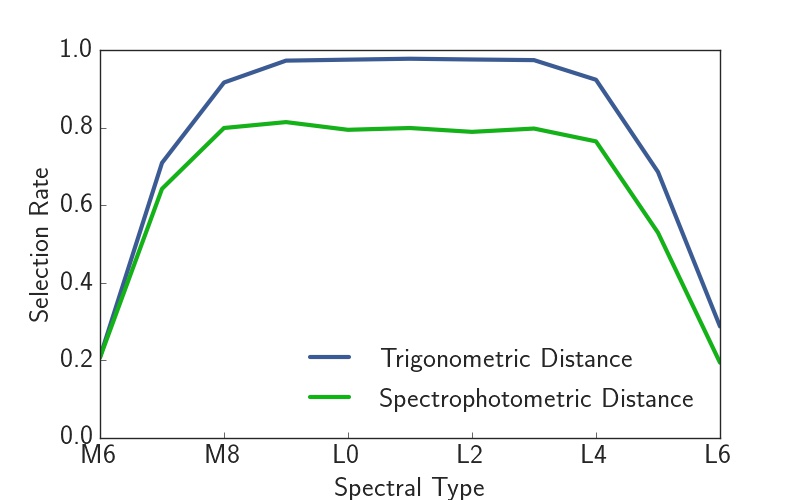}
\includegraphics[width=0.49\textwidth]{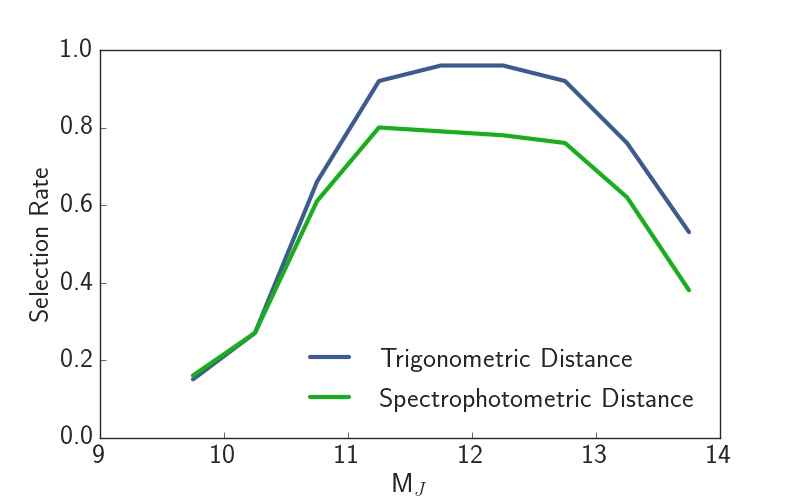}
\caption{Selection functions from trigonometric (blue) and spectrophotometric (green) distance cuts as a function of spectral type \emph{(Left)} and absolute magnitude in $J$-band \emph{(Right)}.\label{fig:SF}}
\end{figure*}

We quantify four selection functions, one for trigonometric and one for spectrophotometric distance selections as functions of spectral type and absolute magnitude. First, we define our ``intrinsic sample'' as those simulated sources whose true distances are $d\leq25\,pc$. We define ``observed samples'' by requiring observed trigonometric or observed spectrophotometric distances $d\leq25\,pc$. In each sample, we select objects with an observed spectral type between M7$-$L5, and organize them according to their true spectral type, given that we are concerned with modeling our observations, yet aware that the true subtype may be different from the observed one. For the selection function by absolute magnitudes, we organized this selected sample in bins of 0.5\,mag observed absolute magnitudes. Our trigonometric and spectrophotometric selection functions are the ratio of objects selected by observations over the number of objects selected by their true parameters.  These selection functions are summarized in Tables~\ref{tab:sfspt} and~\ref{tab:sfmag} and illustrated in Figure~\ref{fig:SF}.




Our trigonometric selection function is relatively high ($92-98\%$) for the central part of the M7$-$L5 spectral type range, except at the edges where the selection rate drops to 71\% for M7 and 69\% for L5. The spectrophotometric selection function runs parallel to the trigonometric one, following a similar shape at a lower rate, $77-82\%$ for M8$-$L4 and dropping to 65\% for M7 and 53\% for L5. The trigonometric selection function based on $J$-band absolute magnitudes steadily increases from 66\% at 10.75\,mag (roughly equivalent to M7) to 96\% at 12.25\,mag, then dropping to 92\% and 76\% in the subsequent fainter bins. The corresponding spectrophotometric selection function follows a similar shape at a lower rate as well, starting at 61\% for 10.75\,mag, reaching a peak of 80\% at 11.25\,mag, and decreasing towards fainter magnitudes down to 62\% at 13.25\,mag (roughly equivalent to L4). These results are presented in Tables~\ref{tab:sfspt} and~\ref{tab:sfmag}. As expected, the edges of our sample suffer from higher contamination than the bulk of it. Contamination from bright sources that do not belong in the $25\,pc$ M7$-$L5 sample is most noticeable in the low spectrophotometric selection rate of the brightest absolute magnitude bins. 



We also calculated the proportion of true negatives and false positives per spectral subtype and absolute magnitude bin. True negatives are true M7$-$L5 dwarfs with true distances within 25\,pc which are not selected by observed trigonometric or spectrophotometric cuts at 25\,pc, i.e. true sources missed by our selections. The true negative fraction is 2\% for any spectral subtype using a parallax cut, except for L0 where the missed fraction is 3\%. However, for a spectrophotometric cut, the true negative fraction rises with spectral type from 8\% to a maximum of 21\% at L2, then decreasing again to 17\% at L5. The true negative fraction by absolute magnitude bins is also 2\% for trigonometric cuts and 7-22\% for spectrophotometric cuts, with the maximum at 12.25\,mag. False positives are contaminants, either sources outside the M7$-$L5 spectral range within 25\,pc or true M7$-$L5 dwarfs outside 25\,pc selected by observations.  The false positive fraction for M7$-$L5 dwarfs varied between $6-9\%$ for spectral type bins selected by parallax, and $10-31\%$ if selected by spectrophotometric distance. The false positive rates by absolute magnitude bins are $2-8\%$ for trigonometric selections and $14-30\%$ for spectrophotometric selections. Thus, the true negative and false positive rates for trigonometric and spectrophotometric selections are comparable across spectral type and absolute magnitude bins. Tables~\ref{tab:plxFP} and ~\ref{tab:photFP} show the fraction of simulated sources outside 25\,pc with a given spectral type and their observed spectral type as selected by observed trigonometric and spectrophotometric distances. For example, on Table~\ref{tab:plxFP}, 4\% of observed M8 dwarfs are actually M9 dwarfs outside of 25\,pc. Overall, it appears that parallax selections are more resistant to scattering of earlier type objects. Diagonal elements indicate objects of matching true and observed spectral subtype, outside of 25\,pc but falsely selected to be within the volume, possibly very close to the 25\,pc limit (Lutz-Kelker bias) or brighter than most other objects of the same subtype (Malmquist bias).

\begin{table} 
\tabletypesize{\scriptsize}
\tablenum{15}
\caption{False positive fractions per spectral subtypes for observed trigonometric selection.\label{tab:plxFP}}
\centering
    \begin{tabular}{@{} cl*{9}c @{}}
        & & \multicolumn{9}{c}{Observed Spectral Type} \\[2ex]
        & & M7 & M8 & M9 & L0 & L1 & L2 & L3 & L4 & L5 \\
        \cmidrule{2-11}
       & M5 &  0.02 &  0.00 &  0.00 &  0.00 &  0.00 &  0.00 &  0.00 &  0.00 &  0.00 \\
       & M6 &  0.04 &  0.01 &  0.00 &  0.00 &  0.00 &  0.00 &  0.00 &  0.00 &  0.00 \\
       & M7 &  0.03 &  0.03 &  0.00 &  0.00 &  0.00 &  0.00 &  0.00 &  0.00 &  0.00 \\
       & M8 &  0.02 &  0.04 &  0.01 &  0.00 &  0.00 &  0.00 &  0.00 &  0.00 &  0.00 \\
       & M9 &  0.00 &  0.03 &  0.04 &  0.02 &  0.01 &  0.00 &  0.00 &  0.00 &  0.00 \\
       & L0 &  0.00 &  0.01 &  0.02 &  0.03 &  0.03 &  0.01 &  0.00 &  0.00 &  0.00 \\
       & L1 &  0.00 &  0.00 &  0.00 &  0.01 &  0.03 &  0.03 &  0.01 &  0.00 &  0.00 \\
       & L2 &  0.00 &  0.00 &  0.00 &  0.00 &  0.02 &  0.03 &  0.02 &  0.01 &  0.00 \\
        \rot{\rlap{True Spectral Type}}
         & L3 &  0.00 &  0.00 &  0.00 &  0.00 &  0.00 &  0.02 &  0.03 &  0.02 &  0.00 \\
       & L4 &  0.00 &  0.00 &  0.00 &  0.00 &  0.00 &  0.00 &  0.02 &  0.04 &  0.02 \\
       & L5 &  0.00 &  0.00 &  0.00 &  0.00 &  0.00 &  0.00 &  0.01 &  0.03 &  0.04 \\
       & L6 &  0.00 &  0.00 &  0.00 &  0.00 &  0.00 &  0.00 &  0.00 &  0.00 &  0.02 \\
         \cmidrule[1pt]{2-11}
    \end{tabular}
\end{table}

\begin{table} 
\tabletypesize{\scriptsize}
\tablenum{16}
\caption{False positive fractions per spectral subtypes for observed spectrophotometric selection.\label{tab:photFP}}
\centering
    \begin{tabular}{@{} cl*{9}c @{}}
        & & \multicolumn{9}{c}{Observed Spectral Type} \\[2ex]
        & & M7 & M8 & M9 & L0 & L1 & L2 & L3 & L4 & L5\\
        \cmidrule{2-11}
	& M5 &  0.25 &  0.06 &  0.00 &  0.00 &  0.00 &  0.00 &  0.00 &  0.00 &  0.00 \\
	& M6 &  0.42 &  0.17 &  0.01 &  0.00 &  0.00 &  0.00 &  0.00 &  0.00 &  0.00 \\
	& M7 &  0.08 &  0.20 &  0.06 &  0.01 &  0.00 &  0.00 &  0.00 &  0.00 &  0.00 \\
	& M8 &  0.01 &  0.10 &  0.08 &  0.03 &  0.00 &  0.00 &  0.00 &  0.00 &  0.00 \\
	& M9 &  0.00 &  0.02 &  0.08 &  0.11 &  0.08 &  0.01 &  0.00 &  0.00 &  0.00 \\
	& L0 &  0.00 &  0.00 &  0.01 &  0.06 &  0.16 &  0.10 &  0.01 &  0.00 &  0.00 \\
	& L1 &  0.00 &  0.00 &  0.00 &  0.00 &  0.07 &  0.15 &  0.06 &  0.01 &  0.00 \\
	& L2 &  0.00 &  0.00 &  0.00 &  0.00 &  0.01 &  0.06 &  0.11 &  0.06 &  0.01 \\
        \rot{\rlap{True Spectral Type}}        
	& L3 &  0.00 &  0.00 &  0.00 &  0.00 &  0.00 &  0.01 &  0.08 &  0.14 &  0.06 \\
	& L4 &  0.00 &  0.00 &  0.00 &  0.00 &  0.00 &  0.00 &  0.01 &  0.09 &  0.14 \\
	& L5 &  0.00 &  0.00 &  0.00 &  0.00 &  0.00 &  0.00 &  0.00 &  0.01 &  0.09 \\
	& L6 &  0.00 &  0.00 &  0.00 &  0.00 &  0.00 &  0.00 &  0.00 &  0.00 &  0.01 \\
         \cmidrule[1pt]{2-11}
    \end{tabular}
\end{table} 



\subsection{Luminosity Function}\label{sec:LF}

\subsubsection{Luminosity Function with respect to Spectral Types}\label{sec:LFspt}


Luminosity functions are a result of the underlying mass function and stellar birth rates. Calculating a luminosity function of UCDs in the 25\,pc volume around the Sun is the first step towards building a field IMF across the stellar/substellar boundary. To measure our luminosity function with respect to spectral types, we prioritize literature optical, SpeX, and literature NIR spectral types in that order, since optical classifications are more precise than NIR ones\footnote{We made an exception to prioritize literature NIR over SpeX classification for 2MASS J22521073$-$1730134A, which has a literature NIR spectral type of L4, no literature optical spectral type, and an unresolved SpeX spectral type of T0.}. Since our study is concerned with the areas accessible by SpeX and outside of $\pm15^{\circ}$ from the galactic plane, we excluded literature sources outside of these areas, reducing our sample to 331 sources. However, 4 sources do not have unresolved $J$-band magnitude (see Section~\ref{sec:photdata}), hence our effective sample includes 327 objects. From these, we find 306 sources in our $25\,pc$ sample with prioritized spectral types within M7$-$L5 within declinations accessible by SpeX ($-50^{\circ}\leq\delta\leq+67^{\circ}$), and outside galactic latitudes $\pm15^{\circ}$ from the galactic plane. 

To estimate the expected total number of objects in our $25\,pc$ sample per spectral type bin, we scale our counts by our selection functions and completeness. We proportionally apply the trigonometric and spectrophotometric selection functions ($SF_{plx}$ and $SF_{phot}$, respectively) to each spectral type bin by splitting our counts, $N_{bin} = N_{plx} + N_{phot}$ according to their type of adopted distance (trigonometric or spectrophotometric), and then scaled by the completeness percentage for the 5-15\,pc fit from Section~\ref{sec:volcomp}, i.e.,

\begin{equation}\label{eq:Ncorr}
N_{corrected} = \Big(\frac{N_{plx}}{SF_{plx}} + \frac{N_{phot}}{SF_{phot}}\Big)\cdot\Big(\frac{1}{\mathrm{completeness}}\Big)
\end{equation}

\begin{figure*}
\figurenum{27}
\centering
\includegraphics[scale=0.5]{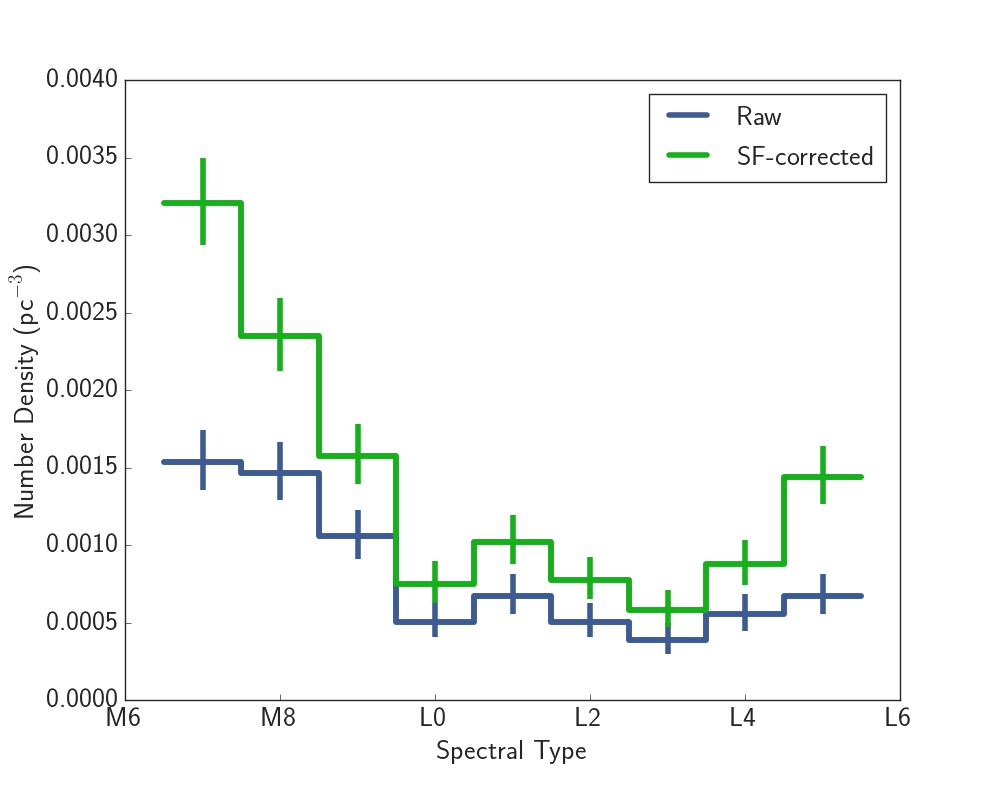}
\caption{Raw and selection-function corrected number densities per subtype for our $25\,pc$ sample.\label{fig:mynumdens}}
\end{figure*}

These corrected counts were divided over the volume estimated in Section~\ref{sec:area} to obtain our luminosity function with respect to spectral types. Our number densities are listed in Table~\ref{tab:numdens}, and shown in Figure~\ref{fig:mynumdens} with and without selection function and completeness corrections. 

\begin{figure*}
\figurenum{28}
\centering
\includegraphics[scale=0.5]{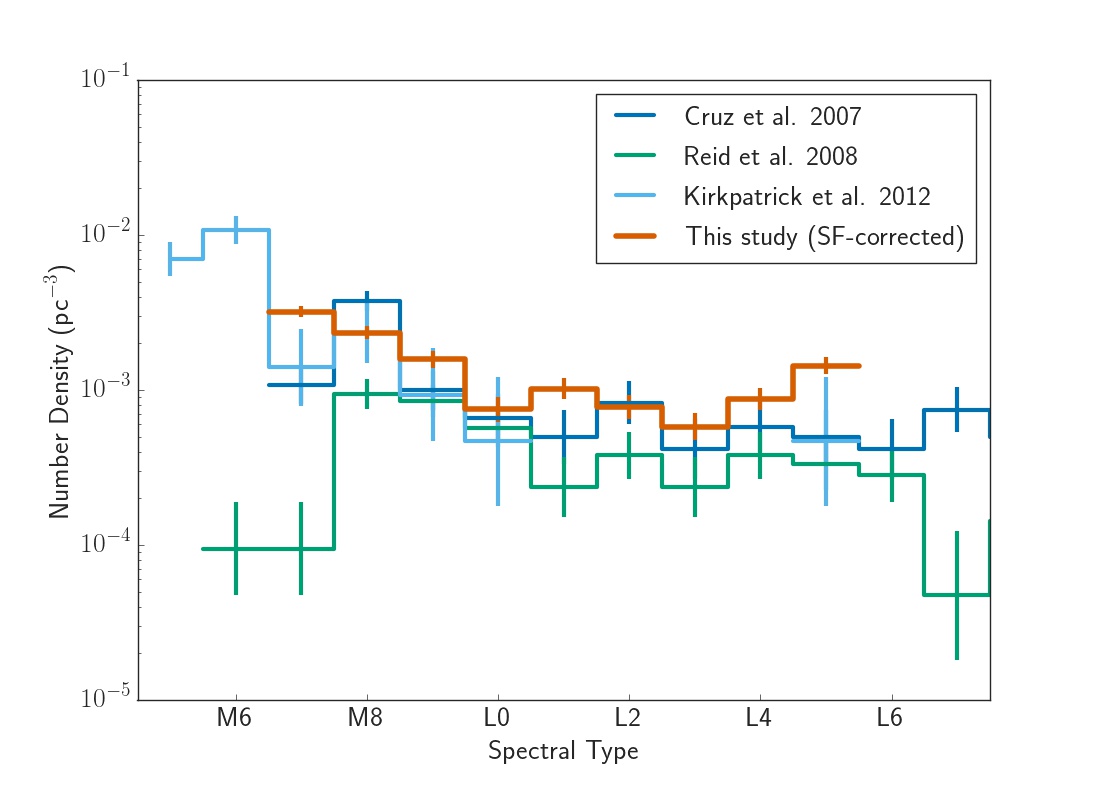}
\caption{Number densities per subtype for the surveys of~\citet{2007AJ....133..439C},~\citet{2008AJ....136.1290R},~\citet{2012ApJ...753..156K}, and this study.\label{fig:numdens}}
\end{figure*}

Figure~\ref{fig:numdens} compares our number densities to other UCD field studies, including the 20\,pc samples of~\citet{2007AJ....133..439C} and~\citet{2008AJ....136.1290R}, and the 8\,pc sample of~\citet{2012ApJ...753..156K}, extended into the substellar regime. Our number densities areconsistently higher than those of~\citet{2008AJ....136.1290R}, particularly on the M dwarfs, although their study does not claim completeness on spectral types earlier than L0. Except for the M7 and L5 edges, our number densities are comparable within 2$\sigma$ to those of~\citet{2007AJ....133..439C} for all spectral types, albeit they claim only a lower limit on L dwarf densities. However, out densities are on average slightly higher than those of~\citet{2007AJ....133..439C}, except for the M8 bin.~\citet{2007AJ....133..439C} found 99 objects between M7$-$L8 in 20\,pc with a sky coverage of 36\%, which scales to 244 sources at 25\,pc for our sky coverage of 63.5\% and 69\% completeness, yet we count 327 sources within a shorter spectral type range. This $\geq34\%$ difference can be attributed to new discoveries, improvements in source color selection (i.e.~\citealt{2015AJ....149..158S}), and broader availability of parallaxes. The 8\,pc sample of~\citet{2012ApJ...753..156K} is sparse on the L dwarf regime, with only one L5 within that volume, and while they include 11 M7$-$M9.5 dwarfs, they claim no completeness on the M dwarf range.  We identify 19 M7$-$M9.5 sources in the literature within the 8\,pc volume and therefore have larger number densities than~\citet{2012ApJ...753..156K}, including a few new discoveries since then.


Table~\ref{tab:numdens} also shows number densities for the M7$-$M9.5, L0$-$L5, and M7$-$L5 ranges. We find that the late-M dwarf raw number density agrees within 20\% of~\citet{2007AJ....133..439C}, but our number density corrected by the selection function and incompleteness is $\sim45\%$ higher, largely driven by the latter. Our L dwarf densities cover a smaller spectral type range than~\citet{2007AJ....133..439C}, and raw and corrected densities follow the same proportions as for the M-dwarf regime. Taking the full range of M7$-$L5 spectral subtypes, we find 40\% higher densities than~\citet{2007AJ....133..439C}, with a raw density of $(7.3\pm0.4)\times10^{-3}~pc^{-3}$ and a corrected density of $(12.6\pm0.6)\times10^{-3}~pc^{-3}$. Our volume density implies that the total number of M7$-$L5 dwarfs within the 25\,pc volume could be as high as $\sim820$. 


\subsubsection{Luminosity Function with respect to Absolute Magnitudes}

\begin{figure*}
\figurenum{25}
\centering
\includegraphics[width=1\textwidth]{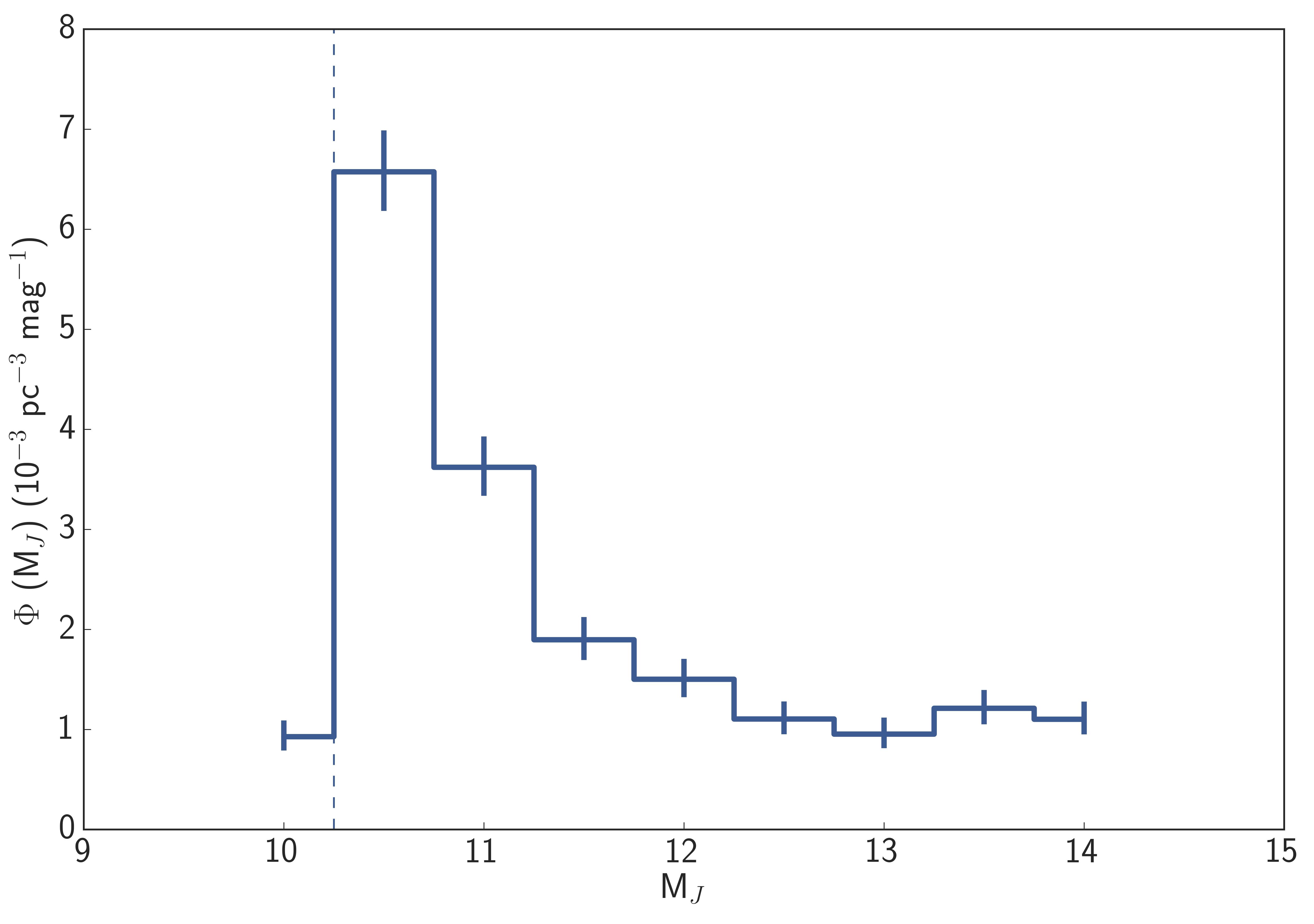}
\caption{Measured luminosity function for M7$-$L5 ultracool dwarfs with Poisson error bars, corrected by the selection function and completeness. We do not claim completeness at magnitudes brighter than the dashed line.}\label{fig:myLFmag}
\end{figure*}

We follow a similar procedure to calculate the luminosity function with respect to absolute magnitude in $J$. We use the subsample of 306 sources described in Section~\ref{sec:LFspt}, yet we organize it in absolute magnitude bins. Our luminosity function is described in Table~\ref{tab:lfmag}. Figure~\ref{fig:myLFmag} shows the resulting luminosity function, including Poisson error bars. Using the~\citet{2015ApJ...810..158F} empirical relations, we determine that the $10.3-14.2$\,mag range in $J$-band encompasses the M7$-$L5 dwarf range, including the $1\sigma$ (0.4\,mag) relation uncertainties. Our luminosity function peaks at the $10.25-10.75$\,mag bin, which roughly corresponds to the peak at the M7$-$M8 spectral class, matching our spectral type distribution from Figure~\ref{fig:spthist}. Our luminosity function then tapers off to a plateau after the 12.25\.mag bin.
 



\begin{figure*}
\figurenum{26}
\centering
\includegraphics[width=1\textwidth]{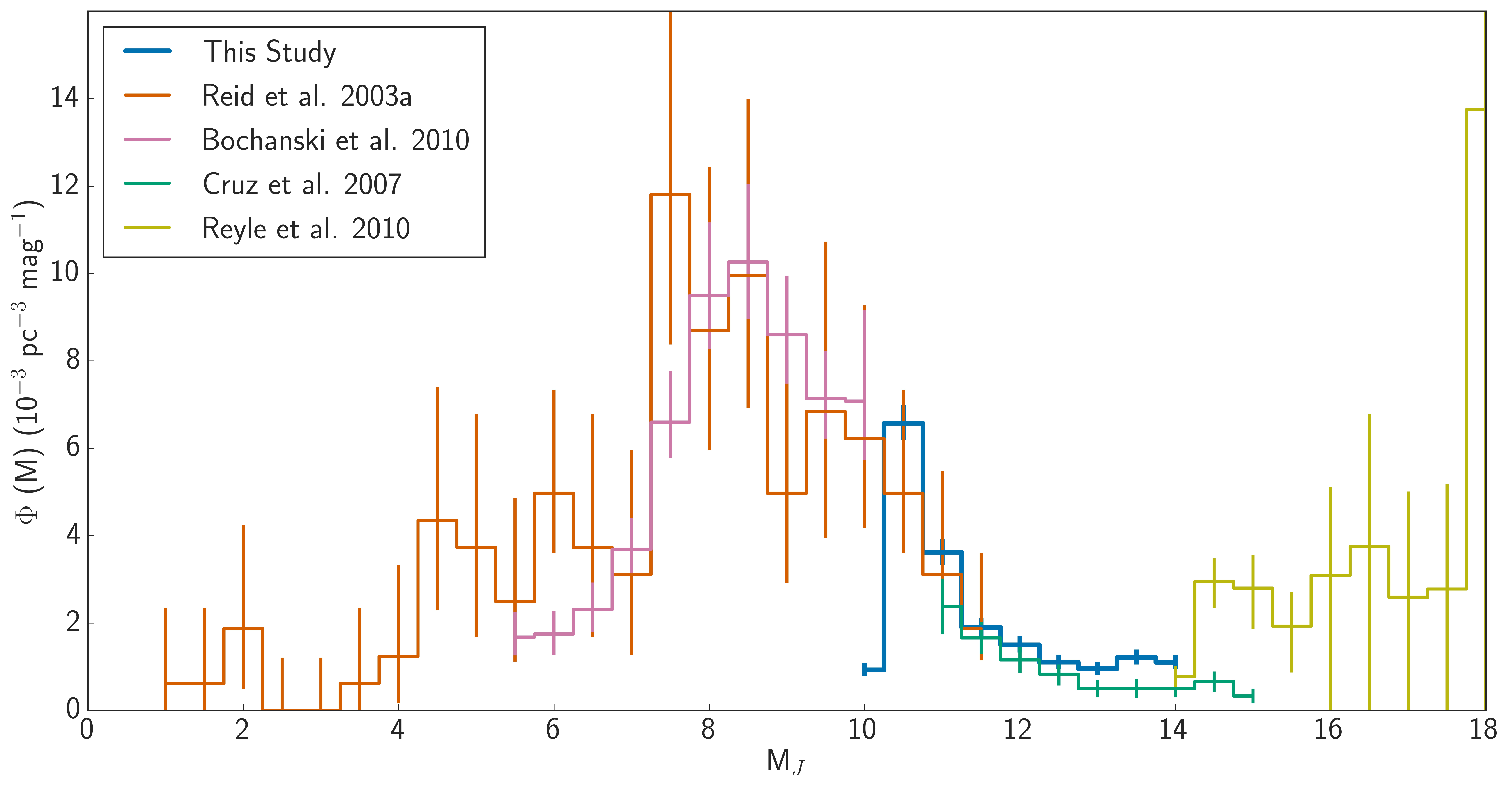}
\caption{Luminosity functions for ultracool dwarfs according to our study (dark blue),~\citet{2003AJ....125..354R} (orange),~\citet{2010AJ....139.2679B} (pink),~\citet{2007AJ....133..439C} (green), and~\citet{2010AandA...522A.112R} (yellow).}\label{fig:allLFmag}
\end{figure*}




Our luminosity function follows from the faint end of the~\citet{2003AJ....125..354R} luminosity function, as seen in Figure~\ref{fig:allLFmag}, matching it well within uncertainties. Throughout the $10.75-13.75$\,mag range, our luminosity function resembles the downward slope of the~\citet{2007AJ....133..439C} corresponding function.

\subsection{Towards building a substellar IMF}

The IMF is a direct outcome of the formation process. Measurements of the IMF across the hydrogen burning limit have revealed that brown dwarfs are not a significant contributor to dark matter~\citep{2003AJ....125..354R}, yet brown dwarfs could be as abundant as stars~(e.g.,~\citealt{2017MNRAS.471.3699M}). The efficiency of the star formation process at low masses, and the minimum mass allowed by the gravitational fragmentation of a molecular cloud can be determined by quantifying the IMF. Constraining the IMF at low masses is a necessary step towards determining the prevalence of different brown dwarf formation mechanisms~(\citealt{2001AJ....122..432R,2002ApJ...576..870P,2004AandA...427..299W,2007MNRAS.382L..30S}), and whether they depend on environmental conditions or not~(e.g.,~\citealt{2007prpl.conf..459W,2019MNRAS.484.2341B}).

Mass functions are typically derived from luminosity functions, a straightforward operation for main sequence stars. For ultracool dwarfs, however, the mapping is no longer one-to-one due to the long lifetimes of very low mass stars and the mass-age-luminosity degeneracy of brown dwarfs. Substellar IMFs can be directly measured in clusters and young moving groups where age is known for all members (e.g., Taurus,~\citealt{2000ApJ...544.1044L}; TW Hydrae,~\citealt{2011PhDT.......245L,2017ApJS..228...18G}). Measuring the substellar field IMF requires assumptions about the age distribution~\citep{2004ApJS..155..191B}. Nevertheless, the field luminosity function presented here is an important step towards measuring an accurate mass function across the hydrogen-burning limit in the field, and the overall formation history and evolution of UCDs in the Milky Way.

This sample also has the potential to reveal ultracool dwarf hosts to habitable zone terrestrial planets like those orbiting TRAPPIST-1~(\citealt{2016Natur.533..221G,2017Natur.542..456G}). Currently, this source is the only example of a planetary system around an ultracool dwarf, and the only planetary system known with 3 potentially habitable terrestrial worlds. With this volume-limited ultracool sample, planetary population studies around the lowest mass stars and brown dwarfs can be approached in a systematic way~(e.g., SPECULOOS,~\citealt{2018SPIE10700E..1ID}).

\section{Summary}\label{sec:conclusions}

We have compiled a volume-limited sample of M7$-$L5 ultracool dwarfs out to 25\,pc, with targets originating from various surveys in the literature. The variety of selection criteria that goes into defining these surveys makes for a potentially complicated selection function with biases difficult to quantify. Nevertheless, we estimate the compiled sample to be $70^{+9}_{-8}\%$ complete to 25\,pc, and highly complete for L dwarfs.



The main results of this study are summarized as follows:
\begin{enumerate}
\item We find 410 UCDs in 394 systems in the 25\,pc volume surrounding the Sun, with 60 more sources in the $1\sigma$ periphery of 25\,pc. Thanks to \textit{Gaia} DR2, our sample is largely volume-limited, with $93\%$ of the sample having parallaxes. 
\item We obtained low-resolution, NIR, SpeX prism spectra for $89\%$ of the observable sample, and uniformly classified them with spectral and gravity standards. 
\item We identify 7 very low gravity sources and 26 intermediate gravity sources in our $25\,pc$ spectral sample, corresponding to fractions of $2.1^{+0.9}_{-0.8}\%$ and $7.8^{+1.7}_{-1.5}\%$, respectively.  One new very low gravity source,  2MASS J1739+2454, is identified in this study. Thirteen new intermediate gravity sources are also reported. Eleven other sources identified as having intermediate gravity also have blue $J-K_S$ colors, suggesting instead low metallicity effects.
\item We calculate $J-K_S$ infrared colors and use them to determine the color distribution of our sample, and identify the red and blue color outlier fractions of $1.4^{+0.6}_{-0.5}$\% for red and and $3.6^{+1.0}_{-0.9}$\% for blue, from 5 and 15 red and blue color outliers, respectively. We do not identify a color bias in our sample given approximately equal numbers of sources with positive and negative $J-K_S$ color excesses.
\item We identify 5 previously confirmed spectral binaries in the 25\,pc volume, and 2 new additional candidates outside the 25\,pc volume. The resulting spectral binary fraction is $1.8^{+0.6}_{-0.5}\%$. In a future paper, we will explore the significance of this fraction with respect to the true ultracool binary fraction of M7$-$L5 dwarfs.
\item We also identified 25 resolved binaries and 13 ultracool companions to main sequence stars in the literature. The literature binary fraction from this sample is $7.5^{+1.6}_{-1.4}\%$. We expect that the identification of overluminous binaries and potentially hidden low gravity and small separation systems will increase this fraction closer to an ultracool resolved binary fraction of $10-20\%$.
\item Our sample is $70^{+9}_{-8}\%$ complete for all sources, mostly incomplete for late-M dwarfs. The completeness for M7$-$M9.5 is $62^{+8}_{-7}\%$, while for L0$-$L5 dwarfs it is $83^{+11}_{-10}\%$.
\item We have produced a $J$-band luminosity function for the 25\,pc sample that closely agrees with previous work but with smaller statistical uncertainties. 
\item We have calculated space densities per subtype and find a 40\% increase in our densities compared to~\citet{2007AJ....133..439C}. Our predicted number density of M7$-$L5 dwarfs is $(12.6\pm0.6)\times10^{-3}~pc^{-3}$, or $\sim820$ objects within 25\,pc.
\end{enumerate}

This homogeneous, volume-limited sample of ultracool dwarfs, with uniformly determined spectral types, measured distances, and masses that span the hydrogen burning limit, has important potential for future statistical studies of UCDs, such as the incidence of magnetic activity, binarity, color outliers, young sources, low metallicity sources, and searches for planetary systems around UCDs. 

\acknowledgments

The authors thank telescope operators Brian Cabreira, Dave Griep, and Tony Matulonis at IRTF for their support during observations. DBG and AJB acknowledge funding support from the National Aeronautics and Space Administration under Grant No. NNX15AI75G. This research has made heavy use of the VizieR catalogue access tool and SIMBAD database, operated at CDS, Strasbourg, France. The original description of the VizieR service was published in~\citet{2000AandAS..143...23O}, and the SIMBAD astronomical database was published in~\citet{2000AandAS..143....9W}. This publication makes use of data from the SpeX Prism Spectral Libraries, maintained by Adam Burgasser at \url{http://www.browndwarfs.org/spexprism}. This research has made use of the Washington Double Star Catalog maintained at the U.S. Naval Observatory. This research was worked on at the NYC \textit{Gaia} DR2 Workshop at the Center for Computational Astrophysics of the Flatiron Institute in 2018 April. This research has benefitted from the Ultracool RIZzo Spectral Library, maintained by Jonathan Gagn\'{e} and Kelle Cruz. The authors acknowledge being on the traditional territory of the Lenape Nations and recognizes that Manhattan continues to be the home to many Algonkian peoples. We give blessings and thanks to the Lenape people and Lenape Nations in recognition that we are carrying out this work on their indigenous homelands. The authors wish to recognize and acknowledge the very significant cultural role and reverence that the summit of Mauna Kea has always had within the indigenous Hawaiian community. We are most fortunate to have the opportunity to conduct observations from this mountain.\\

\software{Astropy \citep{2013AandA...558A..33A}, Astroquery~\citep{astroquery}, BANYAN $\Sigma$~\citep{2018ApJ...856...23G}, Matplotlib~\citep{Hunter:2007}, Pandas \citep{mckinney12}, SpeXtool~\citep{2004PASP..116..362C}, SpeX Prism Library Analysis Toolkit (SPLAT;~\citealt{2017ASInC..14....7B}), Tool for OPerations on Catalogues and Tables (TOPCAT;~\citealt{2005ASPC..347...29T})}.

%

\vspace{5mm}
\facilities{IRTF (SpeX)}

\startlongtable
\begin{deluxetable*}{llr}
\tabletypesize{\scriptsize}
\tablewidth{0pt}
\tablenum{1}
\tablecolumns{3}
\tablecaption{Literature sources providing UCD targets to the final sample\label{tab:litsources}}
\tablehead{
\colhead{Reference} & 
\colhead{Survey/Compilation} &
\colhead{UCD Targets}}
\startdata
\citet{2014ASInC..11....7B} & SpeX Prism Library (SPL) & 510 \\ 
\citet{2015ApJ...814..118B} & L/T Transition Dwarfs from Pan-STARRS1 & 5\\
\citet{2018ApJS..234....1B} & MLT Dwarfs from Pan-STARRS1 & 1041 \\ 
\citet{2011AJ....142...92B} & The Solar Neighborhood XXVIII & 119 \\
\citet{2016csss.confE.148C} & Carmencita, CARMENES Input Catalogue & 63 \\
\citet{2013ApJ...776..126C} & High Proper Motion L Dwarfs & 29 \\
\citet{2006AJ....131.2722C} & SDSS L and T Dwarfs &  71 \\
\citet{2010MNRAS.402..575C} & Southern ultracool dwarfs in young moving groups & 98 \\
\enddata
\end{deluxetable*} 
\begin{deluxetable}{llccccccc}
\tabletypesize{\scriptsize}
\tablecolumns{9}
\tablenum{3}
\tablewidth{0pt}
\setlength{\tabcolsep}{0.05in}
\tablecaption{Bona fide and $1\sigma$ samples of M7$-$L5 ultracool dwarfs in the 25\,pc volume\label{tab:finalsample}}
\tablehead{\colhead{Designation} & 
\colhead{SIMBAD Name} &
\colhead{Adopted SpT} &  
\colhead{SpT Flag} & 
\colhead{$J$ (mag)} & 
\colhead{$J-K_S$} & 
\colhead{Distance (pc)} & 
\colhead{Distance Type} & 
\colhead{Ref.}}
\startdata
\cutinhead{\emph{Bona fide sample}}
                    LP 584$-$4 &      J00020649+0115366 &    M9.0 &    NIR &  12.17$\pm$0.02 &  1.04$\pm$0.03 &   20.81$\pm$0.06 &        Trig &          2 \\
                   GJ 1001 C &     J00043484$-$4044058B &    L5.0 &    OPT &  13.76$\pm$0.04 &   1.7$\pm$0.06 &   12.18$\pm$0.06 &        Trig &          2 \\
                   GJ 1001 B &     J00043484$-$4044058C &    L5.0 &    OPT &    13.9$\pm$nan &    1.6$\pm$nan &   12.18$\pm$0.06 &        Trig &          3 \\
     2MASS J00044144$-$2058298 &    J000441442$-$20582984 &    M7.0 &    NIR &   12.4$\pm$0.02 &  1.01$\pm$0.03 &   15.08$\pm$0.04 &        Trig &          4 \\
     2MASS J00145575$-$4844171 &      J00145575$-$4844171 &    L2.5 &    OPT &  14.05$\pm$0.04 &  1.33$\pm$0.05 &   19.96$\pm$0.16 &        Trig &          5 \\
\enddata
\end{deluxetable}
\clearpage

\begin{deluxetable*}{lcc}
\tabletypesize{\scriptsize}
\tablecolumns{3}
\tablenum{4}
\tablewidth{0pt}
\setlength{\tabcolsep}{0.05in}
\tablecaption{Multi$-$wavelength photometry for bona fide and $1\sigma$ samples of M7$-$L5 ultracool dwarfs in the 25\,pc volume\label{tab:photometry}}
\tablehead{ 
\colhead{Column Number} &
\colhead{Label} & 
\colhead{Description}}
\startdata
1 & SIMBAD Name & Source name as listed in SIMBAD\\
2 & Adopted Spectral Type & Adopted spectral type following the preference order described in Section~\ref{sec:spts}.\\
3 & Spectral Type Flag & Source of spectral type\\
4, 5 & SDSS $r$ (mag) & SDSS $r$ magnitude and uncertainty\\
6, 7 & SDSS $i$ (mag) & SDSS $i$ magnitude and uncertainty\\
8, 9 & SDSS $z$ (mag) & SDSS $z$ magnitude and uncertainty\\
10, 11 & 2MASS $J$ (mag) & 2MASS $J$ magnitude and uncertainty\\
12, 13 & 2MASS $H$ (mag) & 2MASS $H$ magnitude and uncertainty\\
\enddata
\end{deluxetable*}
\clearpage

\begin{deluxetable}{lll}
\tabletypesize{\scriptsize}
\tablecolumns{3}
\tablenum{5}
\tablewidth{0pt}
\setlength{\tabcolsep}{0.05in}
\tablecaption{Bona fide and $1\sigma$ samples of M7$-$L5 ultracool dwarfs in the 25\,pc volume\label{tab:spts}}
\tablehead{ 
\colhead{Column Number} &
\colhead{Label} & 
\colhead{Description}}
\startdata
1 & Source Name & Name of source as shown in SIMBAD\\
 & \multicolumn{2}{l}{\emph{Literature spectral types}}\\
2 & Adopted Spectral Type &  Spectral type as reported in the literature adopted in this study\\
3 & Flag of Adopted Spectral Type & Type of spectral classification adopted\\
4 & Optical Spectral Type & Optical spectral classification as reported in the literature\\
5 & NIR Spectral Type & NIR spectral classification as reported in the literature\\
6 & SIMBAD Spectral Type & Spectral classification reported in SIMBAD, either optical or NIR\\
7 & SDSS Spectral Type & Spectral classification as reported in SDSS\\
\enddata
\end{deluxetable}
\clearpage

\startlongtable
\begin{deluxetable}{ccccccc}
\tabletypesize{\scriptsize}
\tablecolumns{7}
\tablenum{6}
\tablewidth{0pt}
\setlength{\tabcolsep}{0.05in}
\tablecaption{Trigonometric and spectrophotometric distances of bona fide and $1\sigma$ samples of M7$-$L5 ultracool dwarfs in the 25\,pc volume\label{tab:dists}}
\tablehead{ & \multicolumn{2}{c}{Spectral Type} & & \multicolumn{2}{c}{Distance} & \\
\cline{2-3}\cline{5-6}
\colhead{Source Name} & 
\colhead{Adopted} &  
\colhead{Flag} & 
\colhead{Parallax} & 
\colhead{Trigonometric} & 
\colhead{Spectrophotometric (NIR)} & 
\colhead{Ref.}}
\startdata
\cutinhead{\emph{Bona fide sample}}
                    LP 584$-$4 &    M9.0 &    NIR &   48.05$\pm$0.14 &   20.81$\pm$0.06 &     15$\pm$2 &           191 \\
                   GJ 1001 C &    L5.0 &    OPT &   82.09$\pm$0.38 &   12.18$\pm$0.06 &     11$\pm$1 &           191 \\
                   GJ 1001 B &    L5.0 &    OPT &   82.09$\pm$0.38 &   12.18$\pm$0.06 &  \nodata &           191 \\
     2MASS J00044144$-$2058298 &    M7.0 &    NIR &   66.33$\pm$0.16 &   15.08$\pm$0.04 &     23$\pm$3 &           191 \\
     2MASS J00145575$-$4844171 &    L2.5 &    OPT &   50.11$\pm$0.39 &   19.96$\pm$0.16 &     20$\pm$2 &           191 \\
        \enddata
\tablerefs{~(1)~This paper;~(4)~\citet{2018yCat.1345....0G};~(72)~\citet{2015AJ....149..104B};~(80)~\citet{2012ApJ...752...56F};~(129)~\citet{2009AJ....137.4109L};~(135)~\citet{2014AJ....147...94D};~(151)~\citet{2013AJ....146..161M};~(178)~\citet{2012ApJS..201...19D};~(191)~\citet{2018AandA...616A...1G};~(192)~\citet{2014ApJ...784..156D};~(193)~\citet{2016AJ....152...24W};~(194)~\citet{2017AJ....154..151B};~(195)~\citet{2016AandA...595A...4L};~(196)~\citet{2015AandA...577A..15S};~(197)~\citet{2007AandA...474..653V};~(198)~\citet{2005ApJ...630..528P}.}
\tablecomments{This table is available in its entirety in a machine-readable form in the online journal.}
\end{deluxetable}

\startlongtable
\begin{deluxetable*}{lcccccccc}
\tabletypesize{\scriptsize}
\tablewidth{0pt}
\tablenum{7}
\setlength{\tabcolsep}{3pt}
\tablecolumns{9}
\tablecaption{SpeX observing log\label{tab:obstable}}
\tablehead{
\colhead{Designation} & 
\colhead{2MASS $J$} &  
\colhead{2MASS $K_s$} & 
\colhead{Slit} & 
\colhead{Total t$_{exp}$~(s)} & 
\colhead{Airmass} & 
\colhead{Obs. Date} &
\colhead{Median S/N} &
\colhead{A0 Standard}}
\startdata
\sidehead{\emph{Within 25\,pc}}
     J00130931$-$0025521 &   12.167 &   11.319 &  0.5x15$''$ &   539 &    1.098 &  20151006 &   379.34 &     HD~1154  \\
       J00192626+4614078 &   12.603 &   11.502 &  0.5x15$''$ &   539 &    1.127 &  20151117 &   296.67 &   HD~222749 \\
     J00525468$-$2705597 &   13.611 &    12.54 &  0.5x15$''$ &   719 &    1.461 &  20150804 &   186.29 &  HD~ 222332  \\
     \nodata &   \nodata &   \nodata &  0.5x15$''$ &   717 &    1.474 &  20151116 &   163.85 &   HD~225200 \\
     J01004911$-$1933398 &   13.487 &   12.755 &  0.5x15$''$ &   478 &    1.348 &  20161007 &    85.90 &    HD~13433  \\
\enddata
\tablenotetext{a}{ Magnitudes are for combined system.}
\tablenotetext{b}{Background source.}
\tablecomments{This table is available in its entirety in a machine-readable form in the online journal.}
\end{deluxetable*}
\clearpage

\begin{deluxetable*}{lcccccccccc}
\tablewidth{0pt}
\setlength{\tabcolsep}{0.01in}
\tabletypesize{\scriptsize}
\tablenum{8}
\tablecolumns{11}
\tablecaption{Intermediate gravity and very low gravity sources in the M7$-$L5 25\,pc Sample\label{tab:grav}}
\tablehead{
& \multicolumn{2}{c}{Literature SpT} & \multicolumn{2}{c}{SpeX SpT}  & & & & \colhead{Trigonometric} & \colhead{BANYAN $\Sigma$} &  \\
\cline{2-5}
\colhead{Source Name} & 
\colhead{Optical} &  
\colhead{NIR} &
\colhead{Field} &   
\colhead{Low Gravity} & 
\colhead{$\mu_{\alpha}$ (mas/yr)} & 
\colhead{$\mu_{\delta}$ (mas/yr)} & 
\colhead{RV (km/s)} & 
\colhead{Distance (pc)} & 
\colhead{YMG Prob.} & 
\colhead{Ref.\tablenotemark{a}}}
\startdata
\cutinhead{\emph{Very Low Gravity Sources}}
    2MASSW J0045214+163445 &     L2$\beta$ &   L3.5 &   L3.0 &  L1.0$\gamma$ & 358.92$\pm$0.4 & -48.07$\pm$0.24 & 3.16$\pm$0.83 & 15.38$\pm$0.05 & Argus (99.9\%)  & 191;157;191;156,216 \\
    2MASS J03552337+1133437\tablenotemark{b} &    L5$\gamma$ &       L3$\gamma$ &   L7.0 &  L4.0$\gamma$ &   219.76$\pm$1.57 & -631.28$\pm$0.82 & 11.92$\pm$0.22 & 9.12$\pm$0.06 & AB Dor (99.9\%) & 191;218;191;59 \\
    2MASS J05012406$-$0010452 &    L4$\gamma$ &       L3$\gamma$ &   L7.0 &  L4.0$\gamma$ &  189.25$\pm$1.52 & -145.31$\pm$1.16 & 21.77$\pm$0.66 & 21.24$\pm$0.40 & Field (0\%) &  191;157;191;59,156 \\
    2MASS J06244595$-$4521548 &    L5: &     L5 &   L7.0 &  L4.0$\gamma$ & -35.75$\pm$0.89 & 376.67$\pm$1.24 & \nodata & 12.25$\pm$0.07 & Argus (95.1\%) &  191;\nodata;191;59 \\
    G 196$-$3B\tablenotemark{b}  &     L3$\beta$ &       L4 $\gamma$ &   L7.0 &  L2.0$\gamma$ &  -137.82$\pm$0.93 & -208.52$\pm$1.67 & \nodata & 22.55$\pm$0.41 & Field (45.3\%) & 191;\nodata;191;156 \\
\enddata
\tablerefs{~(1)~This paper;~(49)~\citet{2008AJ....135..785W};~(50)~\citet{2014AJ....147...34S};~(56)~\citet{2016ApJ...817..112S};~(59)~\citet{2015ApJS..219...33G};~(80)~\citet{2012ApJ...752...56F};~(102)~\citet{2014AJ....147...20N};~(115)~\citet{2012AJ....144...99D};~(156)~\citet{2013ApJ...772...79A};~(157)~\citet{2016ApJS..225...10F};~(191)~\citet{2018AandA...616A...1G};~(192)~\citet{2014ApJ...784..156D};~(216)~\citet{2014ApJ...783..121G};~(217)~\citet{2010AandA...512A..37S};~(218)~\citet{2003AJ....125..984M};~(219)~\citet{2015ApJS..220...18B};~(220)~\citet{2010ApJ...723..684B};~(221)~\citet{2008MNRAS.390.1517C};~(222)~\citet{2009ApJ...705.1416R};~(223)~\citet{2008MNRAS.384.1399J};~(224)~\citet{2010MNRAS.407.2269M};~(225)~\citet{1996AJ....112.2799H};~(226)~\citet{2017ApJ...845..110B}.}
\tablecomments{This table is available in its entirety in a machine-readable form in the online journal.}
\end{deluxetable*}
\clearpage

\startlongtable
\begin{deluxetable*}{llccccccc}
\tablecaption{Red and blue $J-K_s$ color outliers\label{tab:JKcolor}}
\tablecolumns{7}
\tablenum{9}
\tabletypesize{\scriptsize}
\tablewidth{0pt}
\setlength{\tabcolsep}{0.05in}
\tablehead{
\colhead{Source Name} &
\colhead{Adopted SpT} &  
\colhead{SpT Flag} & 
\colhead{$J-K_s$} & 
\colhead{$J-K_s$ Excess} & 
\colhead{YMG} &
\colhead{Ref.}}
\startdata
\cutinhead{\emph{Red Outliers}}
 2MASS J03552337+1133437\tablenotemark{b}  &    L5.0 &    OPT &  2.52$\pm$0.03 &      0.77 &   11;59 \\
                G 196$-$3B\tablenotemark{b}  &    L3.0 &    OPT &  2.05$\pm$0.06 &      0.44 &     51 \\
               LHS2397aA &    M8.0 &    OPT &  1.28$\pm$0.03 &      0.22 &  60 \\
                Kelu$-$1 A &    L2.0 &    OPT &  2.41$\pm$0.17 &      0.90 &       2 \\
2MASSW J1728114+394859A &    L5.0 &    NIR &  2.21$\pm$0.09 &      0.46 &      6 \\
 2MASS J17410280$-$4642218A\tablenotemark{b} &    L5.0 &    NIR &  2.35$\pm$0.08 &      0.60 &  50;157 \\
\hline
\cutinhead{\emph{Blue Outliers}}
 2MASS J09230296$-$2300415\tablenotemark{c} &    M8.0 &    NIR &  0.55$\pm$0.03 &     -0.51 &       2 \\
                 LHS 286 &    M8.0 &    OPT &  0.82$\pm$0.03 &     -0.24 &     2 \\
\enddata
\tablerefs{~(2)~\citet{2003yCat.2246....0C};~(6)~\citet{2000AJ....120..447K};~(11)~\citet{2008AJ....136.1290R};~(35)~\citet{2004AandA...421..763P};~(50)~\citet{2014AJ....147...34S};~(51)~\citet{1998Sci...282.1309R};~(59)~\citet{2015ApJS..219...33G};~(60)~\citet{1999ApJ...526L..25K};~(79)~\citet{2009AJ....137..304S};~(81)~\citet{2004AandA...428L..25S};~(91)~M. Gillon (priv. comm.);~(97)~\citet{2008ApJ...689..471R};~(98)~\citet{2011ApJ...743..109S};~(99)~\citet{2003yCat.1289....0Z};~(100)~\citet{2005AJ....130..337C};~(101)~\citet{2013AandA...557L...8B};~(104)~\citet{2012ApJ...760..152L};~(117)~\citet{2003AandA...397..575P};~(157)~\citet{2016ApJS..225...10F}.}
\tablenotetext{a}{Discovery Reference; Young Moving Group Reference.}
\tablenotetext{b}{2MASS J0355+1133 and 2MASS J1741$-$4642 are members of the AB Doradus young moving group, while G 196$-$3B is a young, unassociated source.}
\tablenotetext{c}{Also classified as INT-G, indicating low metallicity rather than low gravity.}
\tablenotemark{d}{Member of extended $25\,pc + 1\sigma$ sample.}
 \end{deluxetable*}
 \clearpage

\startlongtable
\begin{deluxetable*}{lccccccc}
\tabletypesize{\scriptsize}
\tablewidth{0pt}
\tablecolumns{8}
\tablenum{10}
\tablecaption{Spectral binary candidates with M7$-$L7 primary components.\label{tab:SB}}
\tablehead{
& \multicolumn{3}{c}{Spectral Type} & & & & \\
 \cline{2-4} 
\colhead{Designation} & 
\colhead{Combined} & 
\colhead{Primary}\tablenotemark{a} & 
\colhead{Secondary}\tablenotemark{a} & 
\colhead{2MASS $\Delta~J$} & 
\colhead{Confidence} & 
\colhead{Spectral Indices}\tablenotemark{b} &
\colhead{References\tablenotemark{*}}}
\startdata
\cutinhead{\emph{Within 25 pc}}
  2MASSW J0320284$-$044636  &  M8.0  &  M9.6$\pm$0.2 & T5.6$\pm$1.0 & 3.5$\pm$0.5 & $96\%$     & 12 &  202,213;191\\
  WISE~J072003.20$-$084651.2 &  M9.0 &  M8.9$\pm$0.0  & T5.1$\pm$0.7 & 3.5$\pm$0.2 & $100\%$    & 6     & 71;63 \\
  2MASS J08053189+4812330    &  L4.0  &  L4.3$\pm$0.4   & T5.0$\pm$1.1 & 1.5$\pm$0.3 & $>99\%$ &  6   & 214;188 \\
  2MASS J13153094$-$2649513 &  L5.5  &  L4.7$\pm$0.4  & T5.4$\pm$3.0 & 2.1$\pm$0.8 & $95\%$      & 12 &  168;158 \\
  2MASS J22521073$-$1730134     & T0.0  &  L4.8$\pm0.5$   &  T4.4$\pm$0.7 & 1.24$\pm$0.25 & $>99\%$ & 11 & \nodata;205\\  
\hline
\cutinhead{\emph{Outside 25 pc}}
  2MASS J09311309+2802289 &  L3.0 &   L1.4$\pm$0.1 & T2.3$\pm$0.8 & 2.3$\pm$0.1 & $>99\%$ &  11 &  \nodata \\
  2MASS J14111847+2948515 &  L3.5 &  L4.1$\pm$1.0 & T3.9$\pm$0.9 & 1.2$\pm$0.5	 & $>99\%$  & 6 &   \nodata \\
  2MASS J14211873$-$1618201 &  M7.5 &  M8.3$\pm$0.2 & T5.1$\pm$1.4	& 3.7$\pm$0.5 & $95\%$ & 5 & \nodata \\
\enddata
\tablenotetext{*}{Candidate; Confirmed.}
\tablenotetext{a}{Primary and secondary spectral types are a weighted average of the best binary template fits, inversely proportional to their ranked $\chi^2$.}
\tablenotetext{b}{Strong candidates have been selected by 8 or more index-index plots up to 12, weak candidates by 4-8.}
\tablerefs{~(63)~\citet{2015AJ....149..104B};~(79)~\citet{2008ApJ...689..471R};~(158)~\citet{2011ApJ...739...49B};~(159)~\citet{2010ApJ...710...45B};~(188)~\citet{2012ApJS..201...19D};~(191)~\citet{2008ApJ...681..579B};~(213)~\citet{2008ApJ...678L.125B};~(214)~\citet{2007AJ....134.1330B}.}
\end{deluxetable*}
\clearpage
\startlongtable
\begin{longrotatetable}
\begin{deluxetable*}{lccccccc}
\tablewidth{0pt}
\tablecolumns{6}
\tablenum{11}
\tabletypesize{\scriptsize}
\tablecaption{Ultracool Binaries with M7$-$L5 Primaries in the 25\,pc Sample\label{tab:binaries}}
\tablehead{& \multicolumn{3}{c}{Spectral type} & & \\
\cline{2-5}
\colhead{Name} &
\colhead{Combined} & 
\colhead{UCD Primary} &
\colhead{UCD Secondary} &
\colhead{Adopted Distance (pc)} &
\colhead{Binary Reference}}
\startdata
\cutinhead{\emph{Ultracool Binaries}}
 2MASS J00275592+2219328AB       &     M7.5        &  M7      &  M8 		&  15.27$\pm$0.89 	&           120 \\
 2MASSW J0320284$-$044636AB     &      M8          & M8       &  T5  	&   20.04$\pm$0.2 &           199 \\
 2MASS J04291842$-$3123568AB    &  \nodata       &   M7.5: &  L1 		&   12$\pm$1 &           126 \\
              V* V780 Tau                         &    M7            &  M7.0 &    unknown             &   10.25$\pm$0.29 &           200 \\
  2MASS J06523073+4710348AB      &    L4.5          &  L3.5   & L6.5 		&     9.12$\pm$0.04 &           182 \\
  2MASS J07003664+3157266AB      &      \nodata   & L3.5    &  L6 		&     12.2$\pm$0.3 &           182 \\
               LHS 1901AB                        &      \nodata   &  M7      &     M7 	&   12.85$\pm$0.5 &           200 \\
  2MASS J07200325$-$0846499AB   &   M8              & M9       & T5  	&    6.02$\pm$1.02 &            72 \\
\enddata
\end{deluxetable*}
\end{longrotatetable}
\clearpage 
\startlongtable
\begin{longrotatetable}
\begin{deluxetable*}{cc|ccc|cc||cc|cc||ccc|cc}
\tabletypesize{\scriptsize}
\tablewidth{0pt}
\tablenum{13}
\tablecolumns{16}
\tablecaption{Selection functions for trigonometric and spectrophotometric distance cuts per spectral type bin.\label{tab:sfspt}}
\tablehead{
\colhead{} & 
\colhead{} &
\colhead{} &
\multicolumn{2}{c}{Observed} & 
\multicolumn{2}{c}{Selection} & 
\multicolumn{2}{c}{Observed} & 
\multicolumn{2}{c}{Fraction} & 
\colhead{} &
\multicolumn{2}{c}{Observed} & 
\multicolumn{2}{c}{False Positive} \\
\colhead{SpT} & 
\colhead{Total} & 
\colhead{Intrinsic} &
\multicolumn{2}{c}{Selected} & 
\multicolumn{2}{c}{Function} & 
\multicolumn{2}{c}{Not Selected} & 
\multicolumn{2}{c}{Missed} & 
\colhead{Not Intrinsic} & 
\multicolumn{2}{c}{Selected} & 
\multicolumn{2}{c}{Fraction}\\
\cline{4-5}\cline{6-7}\cline{8-9}\cline{10-11}\cline{13-14}\cline{15-16}
\colhead{} & 
\colhead{} & 
\colhead{} & 
\colhead{Trig} & 
\colhead{Phot} & 
\colhead{Trig} & 
\colhead{Phot} & 
\colhead{Trig} &
\colhead{Phot} &
 \colhead{Trig} &
\colhead{Phot} &
\colhead{} & 
\colhead{Trig} & 
\colhead{Phot} & 
\colhead{Trig} & 
\colhead{Phot} \\
\colhead{(1)} & 
\colhead{(2)} & 
\colhead{(3)} & 
\colhead{(4)}& 
\colhead{(5)} & 
\colhead{(6)} & 
\colhead{(7)} & 
\colhead{(8)} & 
\colhead{(9)} & 
\colhead{(10)} & 
\colhead{(11)}& 
\colhead{(12)} & 
\colhead{(13)} & 
\colhead{(14)} & 
\colhead{(15)} & 
\colhead{(16)}\\
}
\startdata
      M5 &  189460 &      23612 &                      2129 &                       2123 &   0.09 &    0.09 &                   38 &                    44 &        0.00 &         0.00 &       165848 &                  174 &                  2400 &      0.01 &       0.12 \\
      M6 &  173166 &      21535 &                      4475 &                       4469 &   0.21 &    0.21 &                  106 &                   112 &        0.00 &         0.01 &       151631 &                  353 &                  4517 &      0.02 &       0.24 \\
      M7 &   75002 &       9459 &                      6704 &                       6113 &   0.71 &    0.65 &                  181 &                   772 &        0.02 &         0.08 &        65543 &                  490 &                  2417 &      0.06 &       0.30 \\
      M8 &   51619 &       6490 &                      5946 &                       5210 &   0.92 &    0.80 &                  160 &                   896 &        0.02 &         0.14 &        45129 &                  461 &                  1726 &      0.08 &       0.31 \\
      M9 &   80893 &      10141 &                      9861 &                       8282 &   0.97 &    0.82 &                  251 &                  1830 &        0.02 &         0.18 &        70752 &                  763 &                  2347 &      0.09 &       0.27 \\
      L0 &   83302 &      10572 &                     10305 &                       8427 &   0.97 &    0.80 &                  267 &                  2145 &        0.03 &         0.20 &        72730 &                  754 &                  2192 &      0.08 &       0.24 \\
\enddata
 \tablecomments{Columns: (1) Spectral type bins; (2) Number of objects per bin; (3) Number of true objects per spectral type bin within 25\,pc, i.e. intrinsic objects; (4) Number of intrinsic objects selected by an observed parallax cut at 25\,pc; (5) Number of intrinsic objects selected by an observed spectrophotometric distance cut at 25\,pc; (6) Selection function by parallax, i.e. fraction of intrinsic objects selected by their parallax cut at 25\,pc; (7) Selection function by spectrophotometric distance; (8) Number of intrinsic objects not selected by a trigonometric cut at 25\,pc; (9) Number of intrinsic objects not selected by a spectrophotometric cut at 25\,pc; (10) Fraction of intrinsic objects missed by a trigonometric cut at 25\,pc; (11) Fraction of intrinsic objects missed by a spectrophotometric cut at 25\,pc; (12) Number of objects per spectral type bin outside of 25\,pc; i.e. non-members (13) Number of non-members selected by an observed parallax cut at 25\,pc; (10) Number of non-members selected by an observed spectrophotometric distance cut at 25\,pc; (11) False positive rate for trigonometric selection; (12) False positive rate for spectrophotometric selection.}
\end{deluxetable*}
\end{longrotatetable}
\clearpage

\startlongtable
\begin{longrotatetable}
\begin{deluxetable*}{cc|ccc|cc||cc|cc||ccc|cc}
\tabletypesize{\scriptsize}
\tablewidth{0pt}
\tablenum{14}
\tablecolumns{16}
\tablecaption{Selection functions for trigonometric and spectrophotometric distance cuts per absolute magnitude bin.\label{tab:sfmag}}
\tablehead{
\colhead{} & 
\colhead{} &
\colhead{} &
\multicolumn{2}{c}{Observed} & 
\multicolumn{2}{c}{Selection} & 
\multicolumn{2}{c}{Observed} & 
\multicolumn{2}{c}{Fraction} & 
\colhead{} &
\multicolumn{2}{c}{Observed} & 
\multicolumn{2}{c}{False Positive} \\
\colhead{M$_J$} & 
\colhead{Total} & 
\colhead{Intrinsic} &
\multicolumn{2}{c}{Selected} & 
\multicolumn{2}{c}{Function} & 
\multicolumn{2}{c}{Not Selected} & 
\multicolumn{2}{c}{Missed} & 
\colhead{Not Intrinsic} & 
\multicolumn{2}{c}{Selected} & 
\multicolumn{2}{c}{Fraction}\\
\cline{4-5}\cline{6-7}\cline{8-9}\cline{10-11}\cline{13-14}\cline{15-16}
\colhead{} & 
\colhead{} & 
\colhead{} & 
\colhead{Trig} & 
\colhead{Phot} & 
\colhead{Trig} & 
\colhead{Phot} & 
\colhead{Trig} &
\colhead{Phot} &
 \colhead{Trig} &
\colhead{Phot} &
\colhead{} & 
\colhead{Trig} & 
\colhead{Phot} & 
\colhead{Trig} & 
\colhead{Phot} \\
\colhead{(1)} & 
\colhead{(2)} & 
\colhead{(3)} & 
\colhead{(4)}& 
\colhead{(5)} & 
\colhead{(6)} & 
\colhead{(7)} & 
\colhead{(8)} & 
\colhead{(9)} & 
\colhead{(10)} & 
\colhead{(11)} & 
\colhead{(12)} & 
\colhead{(13)} & 
\colhead{(14)} & 
\colhead{(15)} & 
\colhead{(16)}\\
}
\startdata
 9.75 &  161816 &      20008 &                 2962 &                  3155 &   0.15 &    0.16 &                    39 &                     98 &        0.00 &         0.00 &       141808 &                  272 &                  4129 &      0.02 &       0.23 \\
 10.25 &  186252 &      23385 &                 6363 &                  6346 &   0.27 &    0.27 &                   128 &                    336 &        0.01 &         0.01 &       162867 &                  451 &                  4531 &      0.02 &       0.22 \\
 10.75 &  117183 &      14691 &                 9628 &                  8925 &   0.66 &    0.61 &                   221 &                   1099 &        0.02 &         0.07 &       102492 &                  715 &                  3561 &      0.06 &       0.28 \\
 11.25 &  101656 &      12726 &                11672 &                 10136 &   0.92 &    0.80 &                   268 &                   1998 &        0.02 &         0.16 &        88930 &                  899 &                  3285 &      0.08 &       0.30 \\
 11.75 &   94099 &      11843 &                11358 &                  9325 &   0.96 &    0.79 &                   272 &                   2472 &        0.02 &         0.21 &        82256 &                  811 &                  2585 &      0.08 &       0.25 \\
 12.25 &   75482 &       9411 &                 9022 &                  7296 &   0.96 &    0.78 &                   218 &                   2080 &        0.02 &         0.22 &        66071 &                  568 &                  2228 &      0.07 &       0.27 \\
\enddata
 \tablecomments{Columns: (1) Bins of absolute magnitude in $J$-band; (2) Number of objects per bin; (3) Number of true objects per absolute magnitude bin within 25\,pc, i.e. intrinsic objects; (4) Number of intrinsic objects selected by an observed parallax cut at 25\,pc; (5) Number of intrinsic objects selected by an observed spectrophotometric distance cut at 25\,pc; (6) Selection function by parallax, i.e. fraction of intrinsic objects selected by their parallax cut at 25\,pc; (7) Selection function by spectrophotometric distance; (8) Number of intrinsic objects not selected by a trigonometric cut at 25\,pc; (9) Number of intrinsic objects not selected by a spectrophotometric cut at 25\,pc;  (10) Number of true objects per absolute magnitude bin outside of 25\,pc; i.e. extrinsic objects; (11) Number of extrinsic objects with observed trigonometric distances within 25\,pc; (12) Number of extrinsic objects with observed spectrophotometric distances within 25\,pc; (13) Number of extrinsic objects selected by an observed parallax cut at 25\,pc; (14) Number of extrinsic objects selected by an observed spectrophotometric distance cut at 25\,pc; (15) False positive rate for parallax selection; (16) False positive rate for spectrophotometric selection.}
\end{deluxetable*}
\end{longrotatetable}
\clearpage

\begin{deluxetable*}{cccc|cc||c}
\tabletypesize{\scriptsize}
\tablecolumns{7}
\tablenum{17}
\tablewidth{0pt}
\tablecaption{Number Densities by Spectral Subtype in units of $object~pc^{-3}$.\label{tab:numdens}}
\tablehead{
\colhead{Spectral Type} &
\colhead{\citet{2007AJ....133..439C}} &
\colhead{\citet{2008AJ....136.1290R}} & 
\colhead{\citet{2012ApJ...753..156K}} &
\colhead{N\tablenotemark{*}} &  
\colhead{Raw} & 
\colhead{SF-corrected}}
\startdata
   M5 &         \nodata                                           &     \nodata                                            & $6.99^{+2.05}_{-1.59}\times10^{-3}$  &  \nodata & \nodata   & \nodata \\
   M6 &          \nodata                                          &  $9.50^{+9.50}_{-4.70}\times10^{-5}$ &  $1.07^{+0.25}_{-0.20}\times10^{-2}$  &  \nodata &\nodata  &   \nodata \\
   M7 &     $1.08^{+0.34}_{-0.26}\times10^{-3}$ &  $9.50^{+9.50}_{-4.70}\times10^{-5}$ &  $1.40^{+1.07}_{-0.61}\times10^{-3}$  &  64         &  $1.54^{+0.21}_{-0.18}\times10^{-3}$  & $3.20^{+0.29}_{-0.27}\times10^{-3}$ \\
   M8 &     $3.73^{+0.60}_{-0.52}\times10^{-3}$ &  $9.47^{+2.37}_{-1.89}\times10^{-4}$ &  $2.33^{+1.30}_{-0.84}\times10^{-3}$  &  61        &  $1.47^{+0.20}_{-0.18}\times10^{-3}$  & $2.34^{+0.25}_{-0.23}\times10^{-3}$ \\
   M9 &     $9.95^{+3.32}_{-2.49}\times10^{-4}$ &  $8.53^{+2.26}_{-1.79}\times10^{-4}$ &  $9.33^{+9.33}_{-4.66}\times10^{-4}$ &  44         &  $1.06^{+0.17}_{-0.15}\times10^{-3}$  & $1.58^{+0.21}_{-0.18}\times10^{-3}$ \\
   L0 &     $6.63^{+2.80}_{-1.97}\times10^{-4}$  &  $5.68^{+1.89}_{-1.42}\times10^{-4}$ &  $4.66^{+7.54}_{-2.88}\times10^{-4}$ &  21         &  $5.04^{+1.23}_{-0.99}\times10^{-4}$ & $7.50^{+1.47}_{-1.23}\times10^{-4}$ \\
   L1 &     $4.97^{+2.49}_{-1.66}\times10^{-4}$  &  $2.37^{+1.32}_{-0.85}\times10^{-4}$ &  \nodata                                              &  28         &  $6.72^{+1.40}_{-1.16}\times10^{-4}$ & $1.02^{+0.17}_{-0.15}\times10^{-3}$ \\
\enddata
\tablenotetext{*}{\textbf{Number of sources within 25\,pc, declinations accessible by SpeX ($-50^{\circ}\leq\delta\leq+67^{\circ}$), and galactic latitudes outside of $\pm15^{\circ}$ from the plane.}}
\tablecomments{The units for all values are pc$^{-3}$. These number densities take into account the sky coverage in each survey, but not the survey incompleteness. For the 20\,pc sample of~\citet{2007AJ....133..439C} the volume coverage is 36\%, for the~\citet{2008AJ....136.1290R} survey of the same volume, the coverage is~63\%. For the~\citet{2012ApJ...753..156K} 8\,pc sample, the volume coverage is 100\%. For this study, the volume coverage is $63.6\pm0.59\%$. Uncertainties are calculated from Poisson statistics.}
\end{deluxetable*} 
\begin{deluxetable*}{cccccccc}
\tabletypesize{\scriptsize}
\tablewidth{0pt}
\tablenum{18}
\tablecolumns{8}
\tablecaption{Luminosity Function.\label{tab:lfmag}}
\tablehead{
\colhead{$M_J$} & 
\colhead{N} & 
\colhead{N$_{trig}$} & 
\colhead{N$_{phot}$} & 
\colhead{SFplx} & 
\colhead{SFphot} & 
\colhead{N$_{corrected}$} & 
\colhead{Density (mag pc$^{-3}$)}
}
\startdata
  9.75 &      4 &     4 &      0 &    0.15 &     0.16 &   38.65 &   $9.28_{-1.39}^{+1.62}\times10^{-4}$ \\
 10.25 &     51 &    41 &     10 &    0.27 &     0.27 &  273.75 &   $6.57_{-0.39}^{+4.14}\times10^{-3}$ \\
 10.75 &     68 &    60 &      8 &    0.66 &     0.61 &  150.76 &   $3.62_{-0.29}^{+0.31}\times10^{-3}$ \\
 11.25 &     50 &    49 &      1 &    0.92 &     0.80 &   79.00 &   $1.90_{-0.20}^{+0.23}\times10^{-3}$ \\
 11.75 &     41 &    39 &      2 &    0.96 &     0.79 &   62.55 &   $1.50_{-0.18}^{+0.20}\times10^{-3}$ \\
 12.25 &     30 &    28 &      2 &    0.96 &     0.78 &   45.99 &   $1.10_{-0.15}^{+0.18}\times10^{-3}$ \\
 12.75 &     25 &    24 &      1 &    0.92 &     0.76 &   39.71 &   $9.54_{-1.41}^{+1.65}\times10^{-4}$ \\
 13.25 &     26 &    24 &      2 &    0.76 &     0.62 &   50.44 &   $1.21_{-0.16}^{+0.18}\times10^{-3}$ \\
 13.75 &     16 &    14 &      2 &    0.53 &     0.38 &   45.91 &   $1.10_{-0.15}^{+0.18}\times10^{-3}$ \\
  \enddata
\end{deluxetable*}
\clearpage 

\end{document}